\documentclass[twocolumn,dvipsnames, usenames]{aastex63}
\usepackage{aas_macros}
\usepackage{graphicx}
\usepackage{dcolumn}
\usepackage{bm}
\usepackage{amsmath}
\usepackage{float}
\usepackage{lipsum}
\usepackage{xcolor}
\usepackage{textcomp}
\usepackage{latexsym}
\usepackage{mathtools}
\usepackage{url}
\usepackage{comment}
\usepackage[utf8]{inputenc}
\usepackage[normalem]{ulem}
\usepackage{xspace}
\usepackage{acronym}

\newcolumntype{C}[1]{>{\centering\arraybackslash}p{#1}}

\newcommand{\chieff}{\ensuremath{\chi_{\mathrm{eff}}}\xspace}

\newcommand{\red}[1]{#1}

\newcommand{\CIERA}{\affiliation{Center for Interdisciplinary Exploration and Research in Astrophysics (CIERA), Northwestern University, Evanston, IL 60201, USA}}
\newcommand{\Princeton}{\affiliation{Department of Physics, Princeton University, Princeton, NJ 08544, USA}}

\newcommand{\hellingerDistanceGWTCThreeSMR}{0.080\xspace}
\newcommand{\hellingerDistanceGWTCThreeRMR}{0.063\xspace}
\newcommand{\hellingerDistanceGWTCThreeISO}{0.041\xspace}
\newcommand{\hellingerDistanceMixedSimpleXi}{0.57\xspace}
\newcommand{\hellingerDistanceMixedSimpleXiVM}{0.058\xspace}
\newcommand{\hellingerDistanceMixedFullXi}{0.53\xspace}
\newcommand{\hellingerDistanceMixedFullXiVM}{0.24\xspace}
\newcommand{\hellingerDistanceIsoMax}{0.067\xspace}
\newcommand{\hellingerDistanceIsoMedian}{0.037\xspace}
\newcommand{\hellingerDistanceGWTCTwoUnifMedian}{0.028\xspace}
\newcommand{\hellingerDistanceGWTCTwoJeffMedian}{0.026\xspace}
\newcommand{\hellingerDistanceGWTCTwoSingleMedian}{0.070\xspace}

\newcommand{\hellingerDistanceGWTCTwoSingleMax}{0.10\xspace}
\newcommand{\hellingerDistanceVarma}{0.098\xspace}

\acrodef{O4}[O4]{fourth observing run}
\acrodef{BBH}[BBH]{binary black hole}
\acrodef{LVK}[LVK]{LIGO-Virgo-KAGRA}
\acrodef{PN}[PN]{post-Newtonian}
\acrodef{SOR}[SOR]{spin-orbit resonance}
\acrodef{RMR}[RMR]{reversed mass ratio}
\acrodef{SMR}[SMR]{standard mass ratio}
\acrodef{SNR}[SNR]{signal-to-noise ratio}
\acrodef{VM}[VM]{von Mises}
\acrodef{MC}[MC]{Monte Carlo}
\acrodef{MCMC}[MCMC]{Markov Chain Monte Carlo}
\acrodef{CDF}[CDF]{cumulative distribution function}
\acrodef{KL}[KL]{Kullback-Leibler}
\acrodef{HD}[HD]{Hellinger distance}

\begin{document}
\title{Probing Spin-Orbit Resonances with the Binary Black Hole Population}
\author[0000-0001-7616-7366]{Sylvia Biscoveanu}\email{sbisco@princeton.edu}\thanks{NASA Einstein Fellow}\Princeton\CIERA

\begin{abstract}
Measurements of the binary black hole spin distribution from the growing catalog of gravitational-wave observations can help elucidate the astrophysical processes shaping the formation and evolution of these systems. Spin-orbit resonances are one process of interest, in which the component spin vectors and the orbital angular momentum align into a common plane and jointly precess about the total angular momentum of the system. These resonances, which occur preferentially in systems formed via isolated binary evolution with strong tidal effects, lead to excesses in the distribution of the azimuthal angle between the projections of the component spin vectors onto the orbital plane at $\phi_{12}=0,\pm\pi$. In this work, we conduct the first hierarchical analysis modeling the population-level distribution of $\phi_{12}$ simultaneously with the other mass and spin parameters for simulated binary black hole populations to determine whether spin-orbit resonances can be reliably constrained. While we are unlikely to find definitive evidence for spin-orbit resonances with a population of the size expected by the end of the ongoing LIGO-Virgo-KAGRA fourth observing run, we correctly recover the various $\phi_{12}$ distributions we simulate within uncertainties. We find that we can place meaningful constraints on the relative excesses at $\phi_{12}=0,\pm\pi$, which encodes information about binary mass transfer. We can also distinguish between fully isotropic spin angle distributions and those with features in the spin azimuth and tilt distributions. Thus, we show that population-level measurements of the $\phi_{12}$ distribution offer a reliable, novel way to probe binary formation channels, dynamics, and mass transfer with gravitational-wave observations.
\end{abstract}

\section{Introduction}
The growing catalog of \ac{BBH} mergers detected in gravitational waves has provided an increasingly detailed view of the properties of these systems along with insight into how they form and evolve. The \ac{LVK} detectors~\citep{TheLIGOScientific:2014jea, TheVirgo:2014hva, Aso:2013eba, Somiya:2011np, KAGRA:2020tym} have now observed nearly three hundred gravitational waves from merging \acp{BBH}, including almost two hundred candidates detected during the ongoing \ac{O4}~\citep{KAGRA:2021vkt, GraceDB}. Analyses of this population of sources as a whole have revealed that the mass distribution includes substructure~\citep{KAGRA:2021duu, Tiwari:2020otp, Edelman:2021zkw, Rinaldi:2021bhm} beyond a simple power law~\citep{Fishbach:2017zga} plus Gaussian peak~\citep{Talbot:2018cva}, that \acp{BBH} typically have small spin magnitudes~\citep{Biscoveanu:2020are, Callister:2022qwb, Tong:2022iws, Mould:2022xeu, Galaudage:2021rkt, Roulet:2021hcu}, and that the merger rate distribution evolves with redshift~\citep{KAGRA:2021duu, Fishbach:2021yvy, Edelman:2022ydv, Payne:2022xan, Ray:2023upk, Callister:2023tgi}. Evidence has also been reported for correlations between the \ac{BBH} spin, mass, and redshift distributions~\citep{Callister:2021fpo, KAGRA:2021duu, Tiwari:2021yvr, Adamcewicz:2022hce, Adamcewicz:2023mov, Safarzadeh:2020mlb, Franciolini:2022iaa, Biscoveanu:2022qac, Heinzel:2023hlb}, hinting at the possibility of multiple sub-populations~\citep{Rinaldi:2023bbd, Ray:2024hos, Godfrey:2023oxb, Pierra:2024fbl, Baibhav:2022qxm, Wang:2022gnx, Li:2023yyt, Li:2024jzi}.

The spin orientations of \acp{BBH} have traditionally been regarded as a promising discriminator between binary formation channels~\cite[e.g.,][]{Vitale:2015tea, Stevenson:2017dlk, Talbot:2017yur}, as the primary theoretical models predict different distributions. Those systems that form in field environments via isolated binary evolution are expected to have spins aligned to the orbital angular momentum~\citep{Tutukov:1993, Kalogera:1999tq, Grandclement:2003ck, Postnov:2014tza, Belczynski:2016obo, Mandel:2015qlu, Marchant:2016wow, Rodriguez:2016vmx, Stevenson:2017tfq}, while systems formed dynamically in dense stellar environments should have random spin orientations~\citep{Sigurdsson:1993zrm, Miller:2008yw, Zwart:2010kx, Benacquista:2011kv, Rodriguez:2016vmx}. This basic picture relies on several key astrophysical assumptions; tidal effects in the stellar progenitor of the second-formed black hole must efficiently realign its spin axis with the orbital angular momentum following the first supernova in the binary, and black hole natal kicks must be small to avoid significant misalignment on formation~\citep{Fragos:2010tm, Dominik:2012kk, Fryer:2011cx, OShaughnessy:2017eks, Gerosa:2018wbw, Belczynski:2017gds, Giacobbo:2019fmo, Mandel:2020qwb}. However, this simple model has been recently challenged by the possibility of spin-axis tossing upon black hole birth analogous to what is observed in binary pulsars~\citep{Tauris:2022ggv} or the possibility that the black hole does not inherit its spin direction from its progenitor star~\citep{Baibhav:2024rkn}.

Based on this theoretical picture, most \ac{BBH} population studies---including flagship \ac{LVK} analyses~\citep{KAGRA:2021duu, LIGOScientific:2018jsj, Abbott:2020gyp}---have fit the distribution of spin tilts with a mixture model consisting of an isotropic component representing binaries formed dynamically and a component favoring aligned spins representing binaries formed in the field~\citep{Talbot:2017yur}. Results using the last \ac{LVK} catalog rule out an excess of systems with anti-aligned tilts but are consistent with isotropy~\citep{KAGRA:2021duu}. However, recent work utilizing more flexible models, both phenomenological~\citep{Vitale:2022dpa} and non-parametric~\citep{Golomb:2022bon, Edelman:2022ydv, Callister:2023tgi}, has identified tentative evidence for an excess of spin tilts around $\cos\theta \sim 0.3$, which is not currently explained by theory. 

To leading \ac{PN} order, the effect of spin on the gravitational waveform is captured by the effective aligned spin, \chieff---the mass-weighted spin aligned to the orbital angular momentum~\citep{Damour:2001, Ajith:2009bn, Ajith:2011ec, Santamaria:2010yb, Purrer:2013ojf}. Hence, this parameter is better constrained than the component spin tilts both for individual events~\cite[e.g.,][]{Vitale:2016avz, Shaik:2019dym} and on the population level~\citep{Miller:2020zox, Roulet:2021hcu}. The inferred population-level distribution for \chieff favors positive values, corresponding to spin tilt values $\theta < \pi/2$~\cite[e.g.,][]{Roulet:2021hcu, KAGRA:2021duu, Adamcewicz:2022hce, Banagiri:2025dxo}. This suggests a non-negligible contribution from field binaries, as the assumption of small natal kicks makes it difficult to produce systems with large misalignments ($\theta > \pi/2$) and negative values of \chieff.

In addition to the effective spin and component tilts, the azimuthal spin angle between the projections of the individual black hole spins onto the orbital plane, $\phi_{12}$, can also be used as a tracer of binary formation channel. Binaries with spin tilts misaligned to the orbital angular momentum will undergo general-relativistic spin-induced precession, whereby the orbital plane and the component spin vectors precess about the total angular momentum, changing direction as the binary inspirals~\citep{Apostolatos:1994mx, Kidder:1995zr}. While initially-isotropic distributions of the spin angles remain isotropic throughout this process~\citep{Bogdanovic:2007hp}, binaries formed in the field following the standard astrophysical assumptions presented above are more likely to be caught in a \ac{SOR} as a consequence of \ac{PN} orbital dynamics~\citep{Schnittman:2004vq, Kesden:2010ji, Kesden:2010yp, Berti:2012zp}. Instead of freely precessing, the component spins and orbital angular momentum align into a common plane and jointly precess, such that the azimuthal spin angle librates around a fixed value of either $\phi_{12}=0$ or $\phi_{12}=\pm \pi$, and the spin tilts asymptotically approach specific values that can be predicted in closed form depending on the binary parameters~\citep{Mould:2020cgc}.

Whether the projections of the spins onto the orbital plane are aligned or anti-aligned depends on the relative values of the component tilt angles, which in turn depend on the black hole birth order~\citep{Gerosa:2013laa, Gerosa:2018wbw}. Under the assumption of efficient tides described above, the tilt angle of the second-born black hole will be smaller than that of the first-born black hole, since it effectively only experiences one supernova kick. In most cases, the more massive black hole (the primary) will form first, evolving from the initially more massive star such that $\theta_{1} > \theta_{2}$. However, in the case of efficient mass transfer onto the initially less massive star before the first supernova, the system can undergo mass ratio reversal~\citep{Gerosa:2013laa, Gerosa:2018wbw, Stevenson:2017tfq, Zevin:2022wrw, Broekgaarden:2022nst}. This implies that the primary black hole forms second, evolving from the initially less massive star such that $\theta_{2} > \theta_{1}$. In the \ac{SMR} scenario, resonant binaries librate around $\phi_{12} = \pm\pi$, while \ac{RMR} systems librate around $\phi_{12} = 0$~\citep{Schnittman:2004vq}\footnote{Although not the focus of this work, mass ratio reversal also affects the spin magnitudes. Under the assumption of efficient angular momentum transport in the progenitor stars of stellar-mass black holes~\citep{Fuller:2019sxi}, only the second-born black hole in a binary may acquire considerable spin by the time of merger due to tidal synchronization between its progenitor and the first-born black hole~\cite[e.g.,][]{Qin:2018vaa, Bavera:2020inc}. This means that under the \ac{SMR} scenario, the less massive black hole will be the more rapidly spinning ($\chi_{2}>\chi_{1}$) and vice versa for the \ac{RMR} scenario~\citep{Mould:2022xeu, Broekgaarden:2022nst}.}.

\acp{SOR} occur preferentially for large spin magnitudes, unequal but small spin tilt angles, unequal but comparable mass ratios, and small orbital separations~\citep{Schnittman:2004vq, Gerosa:2013laa, Gerosa:2018wbw}. This is because \ac{PN} spin-orbit couplings are weak at large separations, meaning that the fraction of resonant systems increases closer to merger. In the alternative case of inefficient tidal realignment of the initially less massive stellar progenitor, the tilt angles of the two black holes will be approximately equal, suppressing the probability of getting caught in a \ac{SOR}. Because the dynamical evolution of $\phi_{12}$ is slower when the in-plane spin components are perpendicular to each other, binaries with initial values of $\phi_{12}\approx\pi/2$ get stuck in this regime, leading to a pile-up in the $\phi_{12}$ distribution~\citep{Gerosa:2013laa, Gerosa:2018wbw}. 

The rich physics of \acp{SOR} makes them a key observational target to probe orbital dynamics, binary formation channels, and the strength of tides in massive stars. Identification of this effect in individual binaries is difficult, however, as $\phi_{12}$ is generally poorly constrained. Previous studies using waveform mismatch calculations suggest that the waveforms corresponding to systems with different resonant configurations may be distinguishable between themselves and from freely precessing systems~\citep{Gupta:2013mea, Gerosa:2014kta, Afle:2018slw}. However, investigations using full parameter estimation to characterize the $\phi_{12}$ posterior find this parameter is only well-constrained in certain parts of the binary parameter space, even for high-\ac{SNR} systems~\citep{Trifiro:2015zda, Biscoveanu:2021nvg, Johnson-McDaniel:2023oea}. The constraints on $\phi_{12}$ and the individual azimuthal spin angles $\phi_{1,2}$ are improved when measured at a reference frequency closer to merger, as the waveform is more sensitive to variations in these parameters in this regime~\citep{Varma:2021csh}.

In this work, we study the distinguishability of \acp{SOR} using hierarchical Bayesian inference considering the \ac{BBH} population as a whole rather than seeking to identify this effect in individual binaries. A previous population-level analysis of the azimuthal spin angle of the \ac{BBH} events in the second \ac{LVK} catalog identified hints of substructure in the $\phi_{12}$ distribution that could be interpreted as evidence for a sub-population undergoing \acp{SOR}~\citep{Varma:2021xbh}. We develop an astrophysically-motivated phenomenological population model based on the $\phi_{12}$ distribution considered in \citet{Varma:2021xbh}, allowing for direct constraints on a proxy parameter for the fraction of binaries that have undergone mass ratio reversal. While previous work has constrained this fraction using models that enforce mass-spin correlations~\citep{Mould:2022xeu}, our model includes the effect of the azimuthal spin angle, whose inference we find to drive the \ac{RMR} constraint. By performing full Bayesian parameter estimation on a range of astrophysical populations, we show that the presence of \acp{SOR} could be distinguishable by the end of \ac{O4} depending on the strength of the resonant features. We also find that the the previously-identified weak evidence for \ac{SOR} based on analysis of the GWTC-2 data is consistent with statistical fluctuations in a simulated isotropic population of that size. This work represents the first end-to-end \ac{BBH} simulation study including full parameter estimation and simultaneous hierarchical inference on the masses, redshift, spin magnitudes, tilts, \textit{and} azimuthal angle, accounting for selection effects. While we find a bias in the recovery of the spin magnitude distributions, the recovery of both the spin tilt and azimuth distributions is robust for a range of different simulated populations and population models.

The rest of this work is structured as follows. In Section~\ref{sec:methods}, we describe our four simulated populations---one with strong resonant features, one consisting of a mixture model between an isotropic component and a strongly resonant component, a fully isotropic population, and one with weak resonances. We also describe our choices for the individual-event parameter estimation step and the population models we employ. In Section~\ref{sec:results} we present the inferred spin angle distributions for each of our four simulated populations, compare our results against \citet{Varma:2021xbh}, and apply our population model to constrain the presence of resonant features in the GWTC-3 population. We conclude in Section~\ref{sec:conclusion} and discuss additional analysis details and results, including inference for the mass and spin magnitude distributions and our method for accounting for selection effects, in the Appendix. The individual-event and hierarchical inference posterior samples for all analyses are available on Zenodo~\citep{biscoveanu_2025_17428250}.

\section{Methods}
\label{sec:methods}
\subsection{Simulated population}
To probe whether \acp{SOR} are distinguishable with current detectors, we simulate four different populations of quasi-circular \acp{BBH}. Each population includes 200 events observed by a detector network consisting of LIGO Hanford and Livingston at their predicted optimistic \ac{O4} sensitivities~\citep{O4_psds} with network matched filter \ac{SNR} $\geq 9$ calculated using simulated Gaussian noise. The black hole primary mass ($m_{1} > m_{2}$) is drawn from a \textsc{Power-Law + Peak} mass distribution, and the mass ratio ($q = m_{2}/m_{1}$) is drawn from a power-law distribution conditional on the primary mass~\citep{Talbot:2018cva}. The sources are distributed following a power-law distribution in redshift out to a maximum redshift of $z=1.9$~\citep{Fishbach:2021yvy}. The spin magnitudes are drawn from a Beta distribution~\citep{Wysocki:2018mpo}. The values of the hyper-parameters governing these distributions correspond to the maximum-likelihood values inferred using the observed population published in the previous LVK catalog, GWTC-3~\citep{KAGRA:2021duu}. The definitions of each of the hyper-parameters and their chosen values are given in Table~\ref{tab:true_pars}.

\begin{deluxetable*}{lccr}
\tablecaption{Definitions, true values, and priors on the parameters of the mass, redshift, and spin magnitude distributions used for all simulations\label{tab:true_pars}}
\tablehead{
  \colhead{Parameter} & \colhead{Definition} & \colhead{True value} & \colhead{Prior}
}
\startdata
\multicolumn{4}{c}{\textsc{Power-Law + Peak} Mass Distribution} \\
\hline
$\alpha$ & $m_{1}$ power-law index & 3.51 & U(-4, 12) \\
$m_{\max}$ & maximum $m_{1}$ & 87.7 & U(30, 100) \\
$m_{\min}$ & minimum $m_{1}$ & 5.06 & U(4,10) \\
$\lambda_{\mathrm{peak}}$ & Gaussian sub-population fraction & 0.038 & U(0,1) \\
$\mu_{m}$ & mean of the Gaussian & 33.6 & U(20,50) \\
$\sigma_{m}$ & width of the Gaussian & 4.61 & U(2,10) \\
$\delta_{m}$ & low-mass smoothing parameter & 4.95 & (0,10) \\
$\beta_{q}$ & mass ratio power-law index & 1.09 & U(-2, 7) \\
\hline
\multicolumn{4}{c}{\textsc{Beta} Spin Magnitude Distribution} \\
\hline
$\mu_{\chi}$ & Beta distribution mean & 0.280 & U(0,1) \\
$\sigma^{2}_{\chi}$ & Beta distribution variance & 0.033 & U(0.005, 0.25) \\
\hline
\multicolumn{4}{c}{\textsc{Power-law} Redshift Distribution} \\
\hline
$\lambda_{z}$ & redshift power-law index & 2.86 & U(-2, 10) \\
\enddata
\end{deluxetable*}

\begin{deluxetable*}{lccccc r}
\tablewidth{\textwidth}
\tablecaption{True values of the spin angle parameters used for each of the four simulations along with the fraction of librating systems that librate around $\phi_{12}=0$ corresponding to the \ac{RMR} scenario ($\hat{f}_{\mathrm{RMR}}$) and the fraction of circulating systems ($\hat{f}_{c}$) calculated based on the 200 binaries simulated for each population\label{tab:sims}}
\tablehead{
  \colhead{} & 
  \colhead{$f_{\mathrm{RMR}}$} & 
  \colhead{$\kappa$} & 
  \colhead{$\sigma_{t}$} & 
  \colhead{$\xi$} & 
  \colhead{$\hat{f}_{\mathrm{RMR}}$} & 
  \colhead{$\hat{f}_{c}$}
}
\startdata
Strong resonances & 0.3 & 4 & 0.5 & 1 & 0.361 & 0.515 \\
Strong resonances + isotropic & 0.3 & 4 & 0.5 & 0.644 & 0.494 & 0.585 \\
Isotropic & -- & -- & -- & 0 & 0.638 & 0.710 \\
Weak resonances & 0.3 & 1 & 1.18 & 1 & 0.468 & 0.615 \\
\enddata
\end{deluxetable*}

Each precessing \ac{BBH} system can be classified into one of three spin morphologies~\citep{Kesden:2014sla, Gerosa:2015tea}: circulating (freely precessing), librating around $\phi_{12}=0$, or librating around $\phi_{12}=\pm \pi$. Given that the value of $\phi_{12}$ to which librating systems are drawn depends on the ordering of the spin tilt angles, we use a joint distribution on $\theta_{1}, \theta_{2}$ based on the \textsc{Default} LVK spin tilt model~\citep{Talbot:2017yur, LIGOScientific:2018jsj, KAGRA:2021duu}. The \textsc{Default} LVK tilt model assumes the component spin tilts are identically distributed from a mixture model consisting of an isotropically-oriented component and a preferentially aligned-spin component represented by a truncated Gaussian distribution peaked at $\mu=1$ with variable width and mixture fraction $\xi$. For this work, we simulate a two-component population with a fraction $\xi$ of sources formed via isolated binary evolution with preferentially aligned tilts, while the rest of the population forms dynamically with isotropically distributed spin angles (both $\theta_{1,2}$ and $\phi_{12}$). We fix the fraction of systems that have undergone mass ratio reversal to $f_{\mathrm{RMR}}=0.3$, consistent with predictions from population synthesis simulations for \ac{BBH}~\citep{Gerosa:2013laa, Gerosa:2018wbw, Broekgaarden:2022nst}. This implies that for 30\% of the field binaries, $\theta_{2} > \theta_{1}$. We enforce this ordering using a conditional truncated Gaussian ($\mathcal{N}_{t}$) mixture model,
\begin{widetext}
\begin{align}
\label{eq:spin_tilt}
\pi(\cos\theta_{1}, \cos\theta_{2} &| \xi, \sigma_{t}, f_{\mathrm{RMR}}) = \frac{(1-\xi)}{4}\\ \nonumber &+ \xi\bigg[(1 - f_{\mathrm{RMR}}) \mathcal{N}_{t}(\cos\theta_{1} | \mu=1, \sigma_{t}, \min=-1, \max=1)\mathcal{N}_{t}(\cos\theta_{2} | \mu=1, \sigma_{t}, \min=\cos\theta_{1}, \max=1)\\ \nonumber &+ f_{\mathrm{RMR}}\mathcal{N}_{t}(\cos\theta_{2} | \mu=1, \sigma_{t}, \min=-1, \max=1)\mathcal{N}_{t}(\cos\theta_{1} | \mu=1, \sigma_{t}, \min=\cos\theta_{2}, \max=1)\bigg].
\end{align}
\end{widetext}
For the azimuthal angle distribution, we follow \citet{Varma:2021xbh} and use a \ac{VM} mixture model:
\begin{widetext}
\begin{align}
\label{eq:azimuth}
\pi(\phi_{12} | \xi, f_{\mathrm{RMR}}, \kappa) = \frac{(1-\xi)}{2\pi} + \xi \bigg[ (1 - f_{\mathrm{RMR}})\mathrm{VM}(\phi_{12} | \mu=\pi, \kappa) + f_{\mathrm{RMR}}\mathrm{VM}(\phi_{12} | \mu=0, \kappa)\bigg].
\end{align}
\end{widetext}
The \ac{VM} distribution is akin to a periodic Gaussian with concentration parameter $\kappa = 1/\sigma^{2}$~\citep{Mardia_Jupp_vonMises}. When generating simulated sources, we sample from the joint $p(\cos\theta_1, \cos\theta_2, \phi_{12})$ distribution given by the product of Eqs.~\ref{eq:spin_tilt}-\ref{eq:azimuth} with no cross terms, i.e., all the spin angles are either drawn from isotropic distributions or from the peaked distributions. However, in our hierarchical inference recovery we allow for the more flexible spin angle population model including cross terms, $p(\cos\theta_1, \cos\theta_2, \phi_{12}) = p(\cos\theta_1, \cos\theta_2)p(\phi_{12})$. 

For each of our four simulated \ac{BBH} populations, we choose different \red{true} values of the spin angle hyper-parameters to probe different regimes, summarized in Table~\ref{tab:sims}. We begin with a population with strong resonant features due to strong tidal effects that efficiently realign the spin of the secondary star with the orbital angular momentum. This population is chosen to most closely mimic the expectations from population synthesis from \citet{Gerosa:2013laa, Gerosa:2018wbw}. 
We then modify this population to include an isotropic component, with the mixture fraction $\xi$ corresponding to the maximum-likelihood value inferred with GWTC-3 for the \textsc{Default} spin tilt model~\citep{KAGRA:2021duu}. We also consider a fully isotropic population with no \ac{SOR} features ($\xi=0$) and a final population with weaker resonant features. This last population is chosen to be more realistic given current inferences of the spin tilt distribution, which does not exhibit a strong preference for a narrow aligned-spin population~\cite[e.g.,][]{KAGRA:2021duu, Vitale:2022dpa}.

Given that we are using phenomenological distributions as a proxy for the outputs of population synthesis simulations, not all systems whose $\phi_{12}$ values we draw from $\ac{VM}$ distributions peaked around $\phi_{12} = 0, \pm \pi$ are classified as having a librating morphology~\citep{Gerosa:2015tea}. We use the \texttt{bbh\_spin\_morphology\_prior} package~\citep{spin_morph, Johnson-McDaniel:2023oea} to evaluate the spin morphology of each binary in our simulated populations using the 2.5~\ac{PN} order expression for the orbital angular momentum. In Table~\ref{tab:sims}, we also report the fraction of systems within each population that are circulating and the fraction of librating systems that librate around $\phi_{12}=0$, corresponding to the \ac{RMR} scenario.

\begin{figure}
  \centering
  \includegraphics[width=\columnwidth]{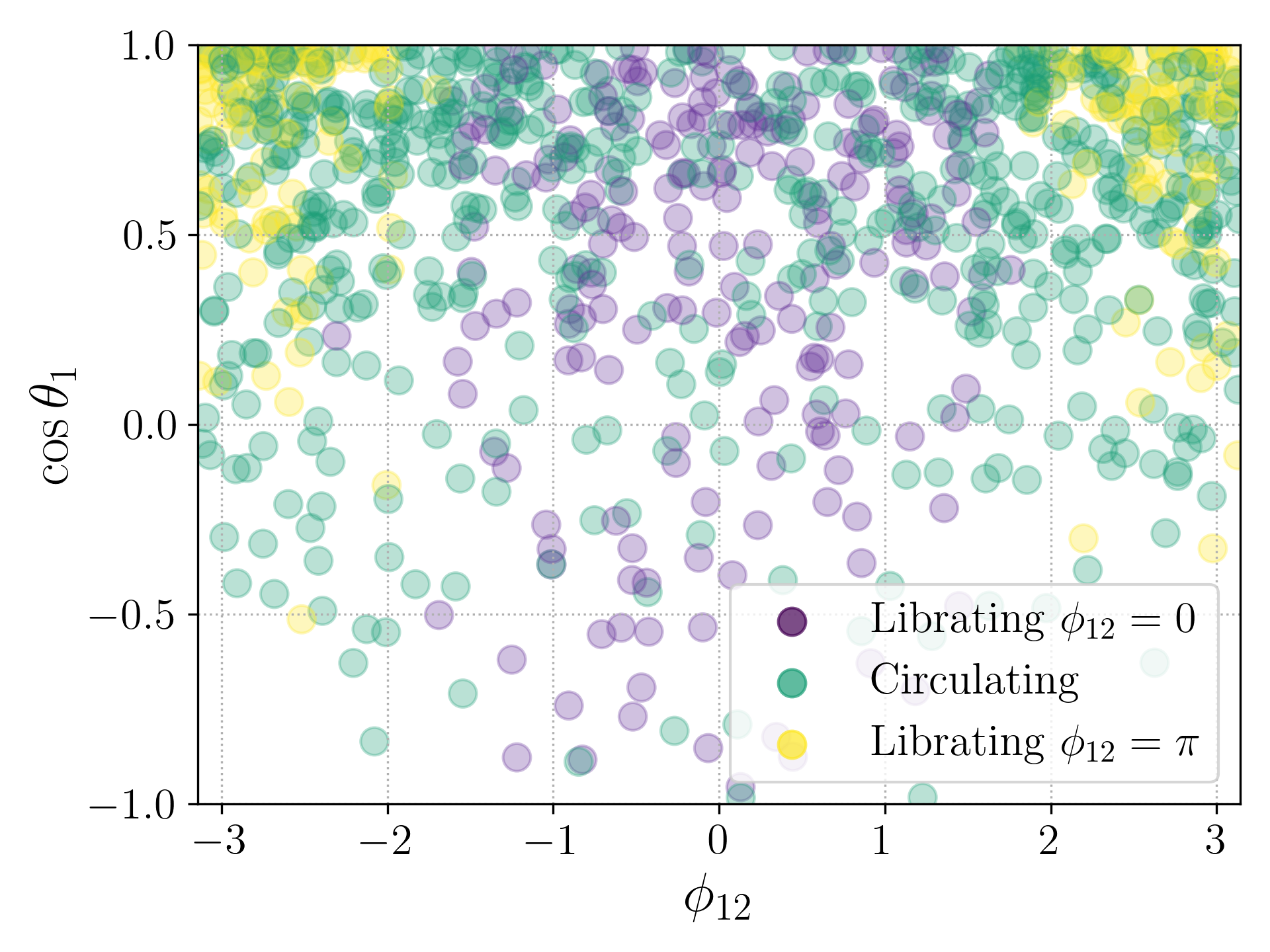}
  \caption{Scatter plot of $\phi_{12}$ vs $\cos{\theta_{1}}$ for the binaries in the weak resonances population colored by the calculated spin morphology of each system: librating around $\phi_{12}=0$ (purple), circulating (teal), and librating around $\phi_{12}=\pi$ (yellow).}
  \label{fig:morphology}
\end{figure}

While the fraction of systems librating around $\phi_{12}=0$ is close to our chosen value of $f_{\mathrm{RMR}}=0.3$ for the all-field population with strong resonant features, this fraction deviates from $f_{\mathrm{RMR}}$ for the other populations. This is because binaries drawn from the isotropic spin angle distributions can randomly correspond to librating morphologies\footnote{For the fully isotropic population, 29\% of systems are in resonant configurations.}, and binaries drawn from the \ac{VM} distribution centered on $\phi_{12}=\pi$ can actually librate around $\phi_{12}=0$ if the distribution is broad enough, as in the weak resonances population shown in Fig.~\ref{fig:morphology}. This means that the parameter $f_{\mathrm{RMR}}$ is just a proxy for the fraction of systems that have undergone mass ratio reversal; in actuality, it corresponds to the fraction of systems whose $\phi_{12}$ values are drawn from a \ac{VM} distribution peaked at $\phi_{12}=0$, even if they are not librating in this configuration once their spin morphology is calculated. This trade-off between simplicity and astrophysical interpretation is a common feature of phenomenological population models.

\subsection{Parameter estimation}
\label{sec:methods_pe}
We perform Bayesian parameter estimation to recover posterior distributions for the binary parameters of each simulated system passing our detection threshold of $\ac{SNR}_{\mathrm{mf, net}} \geq 9$ using the \textsc{Bilby} package~\citep{Ashton:2018jfp, Romero-Shaw:2020owr, colm_talbot_2024_14025522} and the \textsc{Dynesty} nested sampler~\citep{Speagle:2019dynesty}. We use priors that are uniform in the redshifted component masses, spin magnitudes, azimuthal spin angles, and $\cos{\theta}_{1,2}$. The mass prior bounds are chosen to be wide enough to avoid posterior railing depending on the parameters of each binary. For the luminosity distance, we use a prior that is uniform in comoving volume and source-frame time over the range $d_{L} \in [10, 15000]~\mathrm{Mpc}$. We use standard priors~\cite[e.g.,][]{Romero-Shaw:2020owr} for the other extrinsic parameters.

Given the broad mass distribution used for our simulated population, we calculate the analysis segment duration and sampling frequency based on the mass of each system, while the minimum frequency is fixed to $20~\mathrm{Hz}$ for all systems. For the most massive systems, we use the numerical relativity surrogate model NRSur7dq4~\citep{Field:2013cfa, Varma:2019csw} to generate the simulated signal and perform parameter estimation. However, NRSur7dq4 is limited to only ~20 cycles before merger, which means that it cannot produce waveforms down to a starting frequency of $20~\mathrm{Hz}$ for lower-mass signals. In this case, we instead use the phenomenological frequency-domain model IMRPhenomXPHM~\citep{Pratten:2020ceb, Pratten:2020fqn, Garcia-Quiros:2020qpx}.

This choice is motivated by previous work showing that NRSur7dq4 provides significantly more reliable measurements of the azimuthal spin angles compared to other state-of-the-art waveform models, including IMRPhenomXPHM~\citep{Varma:2021csh}. The same analysis also identified that the azimuthal spin angles are better measured close to merger rather than at a fixed reference frequency, which is typically chosen to coincide with the starting frequency of the analysis. As such, we specify the \red{true} binary parameters for each simulated system at a fixed \textit{dimensionless} reference frequency $Mf_{\mathrm{ISCO}} = 6^{-3/2}/\pi$, where $M$ is the redshifted total mass of the system and $f_{\mathrm{ISCO}}$ is the frequency of the innermost stable circular orbit of the Schwarzschild black hole of the same total mass. The true distributions governing the spin tilt angles detailed in Table~\ref{tab:sims} thus represent the \ac{BBH} population at $f_{\mathrm{ISCO}}$.

\begin{figure}
  \centering
  \includegraphics[width=\columnwidth]{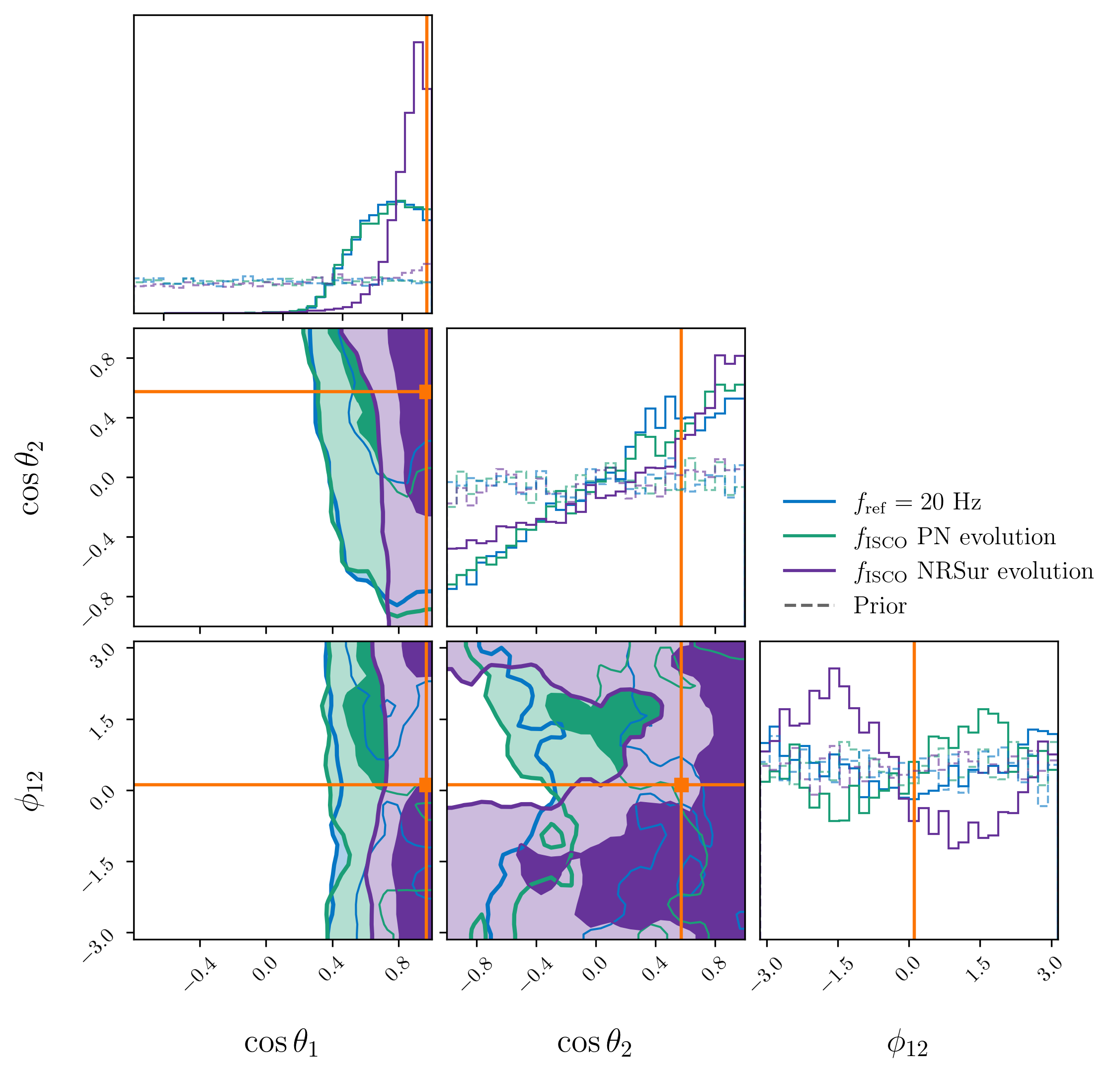}
  \caption{Posterior distributions of the spin tilt angles $\cos\theta_{1,2}$ and azimuthal angle $\phi_{12}$ for one binary system simulated and recovered with NRSur7dq4 at a reference frequency of $f_{\mathrm{ref}}=20~\mathrm{Hz}$ (blue), evolved forward to $f_{\mathrm{ISCO}}$ using the SpinTaylorT5 orbital dynamics (green) and using the surrogate dynamics (purple). The orange lines show the true parameter values at ISCO, and the dashed lines in each color show samples from the prior passed through the same evolution method.}
  \label{fig:compare_evolve}
\end{figure}

For simplicity, we use a fixed reference frequency of $20~\mathrm{Hz}$ during the Bayesian inference step but evolve the posterior samples on the spin angles obtained at this frequency to the ISCO frequency in post-processing. For the systems analyzed with NRSur7dq4, we use the surrogate dynamics implemented in the \textsc{GWSurrogate} package~\citep{Field:2013cfa, Varma:2019csw, gwsurrogate}. For the IMRPhenomXPHM systems, we use the \ac{PN}-based SpinTaylorT5~\citep{Ajith:2011ec} forward evolution implemented in the \textsc{PESummary} package~\citep{Hoy:2020vys}. Given that the latter is an approximation, there are certain spin configurations where the \ac{PN}-based evolved spin posteriors differ significantly from those obtained with the surrogate dynamics for the same system. The posteriors on the spin angles for one such system obtained at $f_{\mathrm{ref}}=20~\mathrm{Hz}$ and at $f_{\mathrm{ISCO}}$ using the two different evolution methods are shown in Fig.~\ref{fig:compare_evolve}. While the primary tilt angle is well-measured, the secondary tilt and azimuthal angle are poorly constrained. However, the posterior using the surrogate evolution is more informative relative to the isotropic prior at $f_{\mathrm{ISCO}}$ compared to $f_{\mathrm{ref}}=20~\mathrm{Hz}$. As we will show in Section~\ref{sec:results}, the approximate nature of the \ac{PN}-based spin evolution does not lead to significant biases in the inferred spin angle distributions.

\subsection{Hierarchical inference}
For each of our simulated populations, we perform hierarchical Bayesian inference~\cite[e.g.,][]{Thrane:2019pe} to obtain posteriors on the hyper-parameters governing the population distributions using the \textsc{GWPopulation} package~\citep{2019PhRvD.100d3030T}. In addition to the mass, redshift, and spin models presented in Table~\ref{tab:true_pars}, we employ a variety of spin angle models that each provide different insights into the ability to probe spin-orbit resonances, whose parameters are summarized in Table~\ref{tab:spin_angle_models}. We refer to the correlated truncated Gaussian + isotropic spin tilt model in Eq.~\ref{eq:spin_tilt} used to generate the simulated systems as the \textsc{Full} tilt model, characterized by the parameters $\sigma_{t}$, the width of the Gaussian components, $\xi$, the fraction of binaries in the Gaussian components, and $f_{\mathrm{RMR}}$, the fraction of those binaries with $\theta_{2} > \theta_{1}$. 

\begin{deluxetable*}{lcr}
\tablewidth{\textwidth}
\tablecaption{Definitions and priors on the parameters of the various spin angle distributions used during hierarchical inference\label{tab:spin_angle_models}}
\tablehead{
  \colhead{Parameter} & \colhead{Definition} & \colhead{Prior}
}
\startdata
\multicolumn{3}{c}{\textsc{von Mises} $\phi_{12}$ Distribution} \\
\hline
$f_{\mathrm{RMR}}$ & mass-ratio-reversed fraction & U(0, 1) \\
$\kappa$ & concentration parameter & U(0, 8) \\
$\mu_{\mathrm{RMR}}$ & peak of the mass-ratio-reversed \ac{VM} sub-population & $\delta(0)$ or $\mathrm{U}(-\pi,\pi)$ \\
$\mu_{\mathrm{SMR}}$ & peak of the standard mass ratio \ac{VM} sub-population & $\delta(\pi)$ or $\mathrm{U}(-\pi,\pi)$ \\
$\xi_{\mathrm{VM}}$ & field binary fraction in the $\phi_{12}$ distribution & $\delta(\xi)$ or U(0,1) \\
\hline
\multicolumn{3}{c}{\textsc{Full} Tilt Distribution} \\
\hline
$f_{\mathrm{RMR}}$ & mass-ratio-reversed fraction & U(0, 1) \\
$\sigma_{t}$ & width of the aligned field binary component & U(0.1,4) \\
$\xi$ & field binary fraction & U(0,1) \\
\hline
\multicolumn{3}{c}{\textsc{Simple} Tilt Distribution} \\
\hline
$\mu_{1}$ & peak of the $\cos{\theta_{1}}$ distribution & U(1,10) or U(5,30) \\
$\mu_{2}$ & peak of the $\cos{\theta_{2}}$ distribution & U(1,10) or U(5,30)\\
$\sigma_{1}$ & width of the $\cos{\theta_{1}}$ distribution & U(0.5,2) or U(1,5) \\
$\sigma_{2}$ & width of the $\cos{\theta_{2}}$ distribution & U(1,2) or U(1,5) \\
\enddata
\end{deluxetable*}

This tilt model introduces correlations between the two spin tilt angles and a sharp discontinuity when $f_{\mathrm{RMR}}\neq 0.5$. Such sharp model features can be difficult to probe and often increase the uncertainty in the \ac{MC} integrals utilized when performing hierarchical inference by ``recycling'' the individual-event binary parameter posteriors (see Appendix~\ref{ap:injections}). This tilt model also depends on the $f_{\mathrm{RMR}}$ parameter, so the contribution of the $\phi_{12}$ inference to the $f_{\mathrm{RMR}}$ posterior cannot be distinguished from the contribution of the tilts, obfuscating the ability to probe spin-orbit resonances independently of the tilt distribution. To avoid these issues, we introduce another \textsc{Simple} tilt model, which assumes that the spin tilt angles for the field binary sub-population are distributed following independent truncated Gaussians:
\begin{widetext}
\begin{align}
\label{eq:simple_tilt}
\pi(\cos\theta_{1}, \cos\theta_{2} | \xi, \mu_{1}, \sigma_{1}, &\mu_{2}, \sigma_{2}) = \frac{(1-\xi)}{4} \nonumber \\
&+ \xi\bigg[\mathcal{N}_{t}(\cos\theta_{1} | \mu_{1}, \sigma_{1}, \min=-1, \max=1)\mathcal{N}_{t}(\cos\theta_{2} | \mu_{2}, \sigma_{2}, \min=-1, \max=1)\bigg].
\end{align}
\end{widetext}
By ignoring the correlation, our hierarchical analyses with this \textsc{Simple} tilt distribution intentionally mismodel the tilts. However, this distribution is sufficiently flexible to fit the true marginalized 1D tilt distributions without introducing biases in the recovery of the population distributions for the other binary parameters. We find that this mismodeling can introduce a bias in the inferred $\xi$ posterior for certain astrophysical populations while still qualitatively recovering the correct marginal tilt angle distributions, as discussed in more detail in Section~\ref{sec:wide_results}.

To probe whether the $\phi_{12}$ distribution contains information about the mixture fraction between dynamically-formed (isotropic) and field binaries (\ac{VM}-distributed), we consider a modification to the \textsc{von Mises} $\phi_{12}$ model where the parameter $\xi_{\mathrm{VM}}$ is independent of the $\xi$ parameter with the same definition in the tilt distribution model. Finally, we consider another modification to the $\phi_{12}$ model where the locations of the \ac{VM} peaks, $\mu_{\mathrm{RMR}}$ and $\mu_{\mathrm{SMR}}$ are free parameters of the model rather than being fixed to the values expected theoretically. This alternative allows us to determine if there is sufficient information in the $\phi_{12}$ distribution to distinguish between the \acp{SOR} expected for strong tides and the weaker peaks at $\phi_{12} = \pm \pi/2$ expected in the case of weak tides.

Because we apply a detection threshold to the simulated systems included in the populations we hierarchically analyze, we must account for selection biases~\cite[e.g.,][]{Loredo:2004nn, Mandel:2018mve, Thrane:2019pe, Vitale:2020aaz}. This is done using a \ac{MC} integral over sensitivity injections meeting our detection criterion of $\mathrm{SNR_{net, mf}} \geq 9$ generated for our chosen detector network of LIGO Hanford and Livingston at their predicted \ac{O4} sensitivities~\cite[e.g.,][]{Essick:2023toz}. To reduce the \ac{MC} integral uncertainty~\citep{Farr_2019, Essick:2022ojx, Talbot:2023pex}, we generate over ten million found injections---two orders of magnitude more than the number used in \ac{LVK} analyses of the \ac{BBH} population in GWTC-3~\citep{KAGRA:2021duu}; further details of our semi-analytic injection campaign are given in Appendix~\ref{ap:injections}.

\section{Results}
\label{sec:results}

\red{We report the results of 42 total hierarchical inference analyses on the four different simulated populations described above and the real GWTC-3 data. All five populations are analyzed with the same base \textsc{Full} and \textsc{Simple} tilt models given in Eqns.~\ref{eq:spin_tilt}-\ref{eq:simple_tilt}; we also analyze the subset of events in each simulated population recovered with NRSur7dq4 separately, in addition to the full population. In Tables~\ref{tab:gwtc3_results}-\ref{tab:results} in Appendix~\ref{ap:result_tables}, we summarize each of the hierarchical inference runs performed and report summary statistics (maximum posterior value and 95\% credible interval width and bounds) for the posteriors on the parameters governing the spin angle distributions: the proxy parameter for the fraction of field binaries that have undergone mass ratio reversal, $f_{\mathrm{RMR}}$, the concentration parameter of the \ac{VM} distribution, $\kappa$, and the mixture fraction between the isotropic and aligned spin tilt components (a proxy for the field binary fraction), $\xi$.}

\red{In the discussion below, we focus on the inferences of the spin angle distributions. However, we fit the mass, spin magnitude, and redshift population-level distributions simultaneously in all cases. In general, we recover the true values of all hyper-parameters within the $3\sigma$ posterior credible intervals. The true redshift power-law slope hyper-parameter is recovered at high credibility for all our hierarchical inference analyses. However, we find a bias in the power-law parameters governing the mass distribution when analyzing the full simulated populations that can be explained by the inconsistency in the choice of waveform between the source simulation and individual-event parameter estimation introduced by our simulation pipeline (see Appendix~\ref{ap:masses} for more details). This bias is ameliorated when analyzing the NRSur7dq4-subset of each simulated population individually. We also find a bias in the recovered spin magnitude distributions, as we generally recover distributions that are too narrowly peaked. The spin magnitude bias is explored in more detail in Appendix~\ref{ap:spin_mag}. As we will demonstrate below, neither the mass nor spin magnitude biases affect the spin angle inference highlighted in the rest of this work.}

\subsection{Strong resonances population}
\label{sec:strong_resonances}
\begin{figure}
  \centering
  \includegraphics[width=\columnwidth]{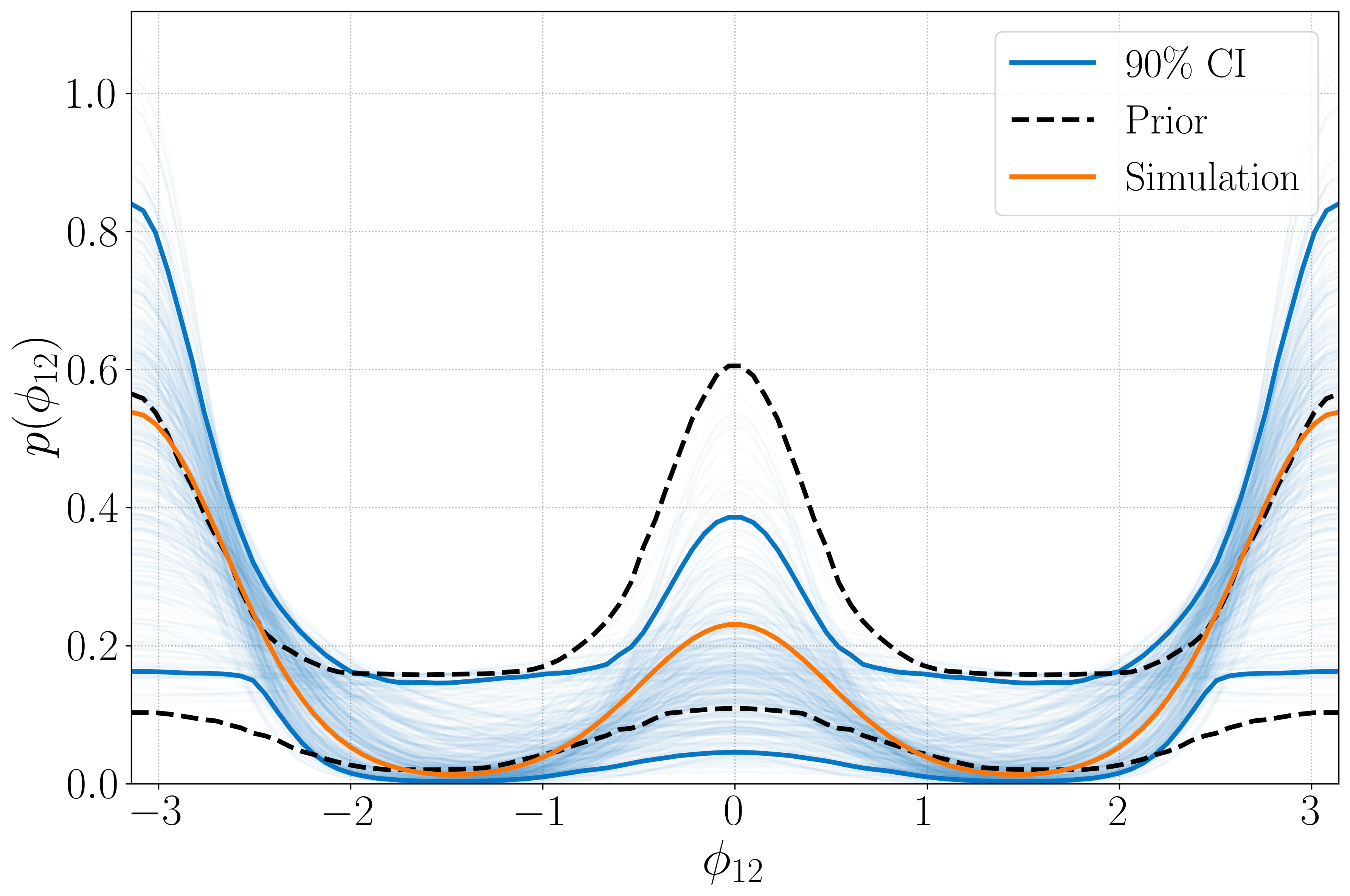}
  \caption{Inferred $\phi_{12}$ distribution for the strong resonances population under the \textsc{Full} tilt model. Individual light blue traces show the distributions corresponding to individual hyper-parameter posterior samples, the dark blue lines bound the 90\% posterior credible interval, and the dashed black lines bound the 90\% prior credible interval. The true simulated distribution is shown in orange.}
  \label{fig:all_field_iso_mix_full_new_lightning_phi12}
\end{figure}

\begin{figure}
  \centering
  \includegraphics[width=\columnwidth]{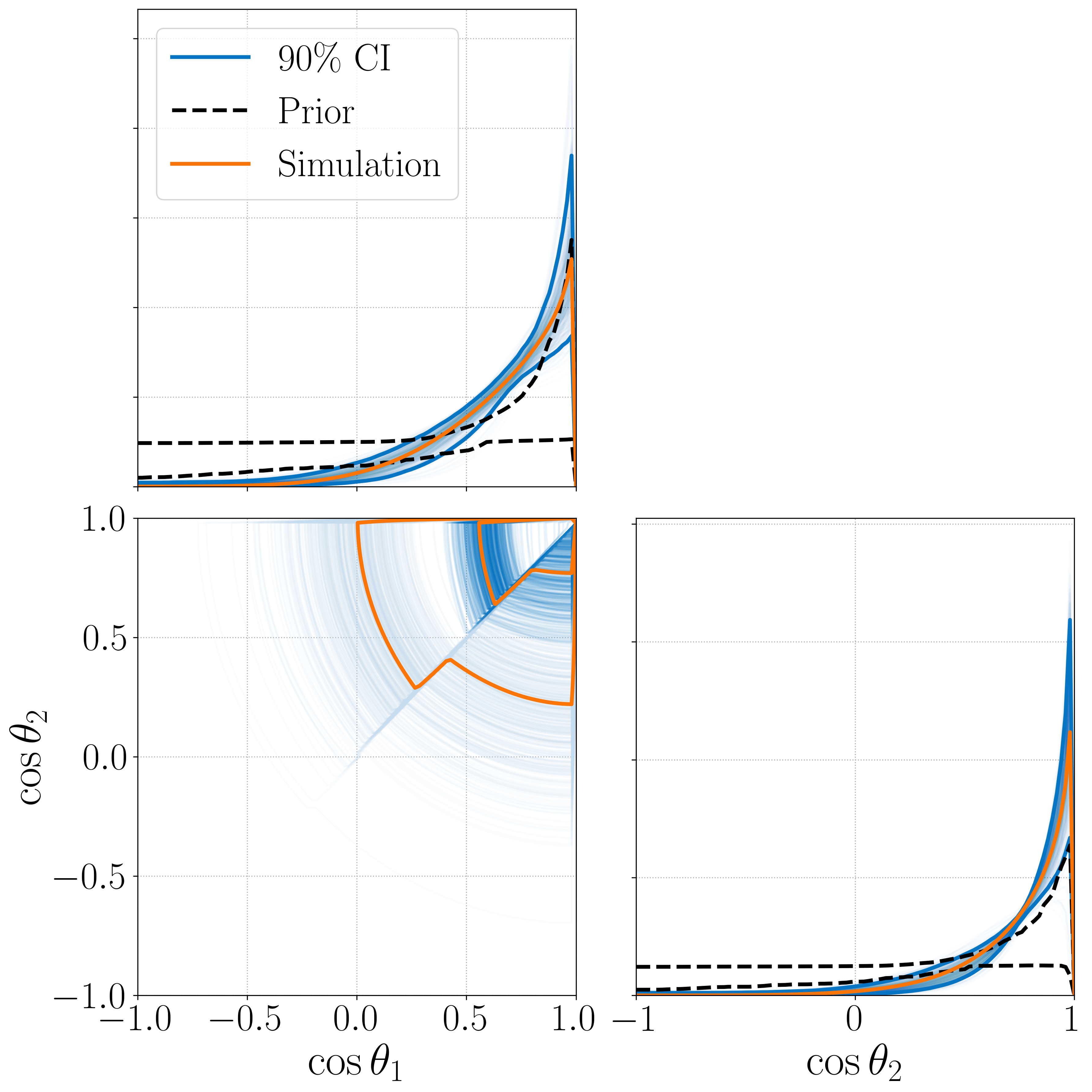}
  \caption{Inferred $p(\cos\theta_1, \cos\theta_2)$ distribution for the strong resonances population under the \textsc{Full} tilt model. Individual light blue traces show the distributions corresponding to individual hyper-parameter posterior samples, the dark blue lines bound the marginalized 1D 90\% posterior credible intervals for the individual tilts, and the dashed black lines bound the 90\% prior credible intervals. The true simulated distribution is shown in orange.}
  \label{fig:all_field_iso_mix_full_new_lightning_tilt}
\end{figure}

For the strong resonances population, we generally recover the true hyper-parameter values within the $3\sigma$ credible intervals for the spin angle distributions\footnote{For the \textsc{Simple} tilt model, we find the best-fitting values of $\mu_{1}, \sigma_{1}, \mu_{2}, \sigma_{2}$ using the non-linear least squares fitting implemented in \texttt{scipy} as a proxy for the ``true'' parameter values.}. In Figs.~\ref{fig:all_field_iso_mix_full_new_lightning_phi12}-\ref{fig:all_field_iso_mix_full_new_lightning_tilt}, we show the inferred population-level distributions for the $\cos\theta_1, \cos\theta_2$, and $\phi_{12}$ parameters recovered under the \textsc{Full} tilt model when we analyze the full population of 200 events. The true simulated distribution is recovered within the 90\% credible interval of the posterior for both the spin tilt and azimuthal angles. 

The recovery of the true hyper-parameter values for the spin angles is more marginal for the \textsc{Simple} tilt model, though still within the $3\sigma$ credible intervals. In order to verify if this is a robust conclusion of the analysis or instead sensitive to the particular realization of binary parameters for the 200 simulated events in the analyzed population, we simulate another independent population of 200 events drawn from the same true distributions, repeating both the individual-event parameter estimation and hierarchical inference steps. While for one population, the true values are recovered at lower credibility with the \textsc{Full} tilt model, the inverse is true for the other simulated population; further comparison of the results obtained for these two distinct realizations of the strong resonances population is included in Appendix~\ref{ap:strong_resonances}. 

However, for both population realizations we are able to obtain a meaningful constraint on the $f_{\mathrm{RMR}}$ parameter. We obtain 95\% credible interval posterior widths $\lesssim 0.7$ for both the \textsc{Simple} and \textsc{Full} tilt models, compared to the prior width of 0.95 (see full results summary in Table~\ref{tab:results}). Given that the posterior constraints on $f_{\mathrm{RMR}}$ are similar for both tilt models, this means that the $\phi_{12}$ parameter drives the inference of the mass ratio reversal fraction. While the spin tilt ordering imposed by the \textsc{Full} tilt model may add information to the $f_{\mathrm{RMR}}$ posterior in some cases depending on the observed population, the spin tilt distribution is not the dominant source of information for the mass ratio reversal fraction parameter.

For the strong resonances population, we generally find that
\begin{itemize}
\item The $\phi_{12}$ distribution is successfully recovered \red{within the 90\% posterior credible interval} for a population with resonant features that are this narrowly peaked.
\item The proxy parameter for the fraction of sources undergoing mass ratio reversal in the population can be constrained \red{(95\% posterior credible interval widths $\leq 0.7$)} using only information from the $\phi_{12}$ distribution (\textsc{Simple} tilt model).
\item The posterior on the mixture fraction is narrowly constrained to $\xi > 0.9$ for both populations and tilt models at $>95\%$ credibility, meaning that the isolated binary sub-population can be confidently identified for such a strongly peaked tilt distribution.
\end{itemize}

\subsection{Strong resonances + isotropic population}
We now turn to the more realistic population that includes an isotropic component consistent with the mixture fraction inferred under the \textsc{Default} spin model in the \ac{LVK} GWTC-3 analysis~\citep{KAGRA:2021duu}. Again, we generally recover the true values of all hyper-parameters within the $3\sigma$ posterior credible interval. In addition to the \textsc{Simple} and \textsc{Full} tilt models used to analyze the population consisting only of field binaries with strong resonances presented above, we decouple the mixture fraction parameter on which the $\phi_{12}$ \ac{VM} distribution depends from the mixture fraction parameter in the tilt distribution. 

\begin{figure}
  \centering
  \includegraphics[width=\columnwidth]{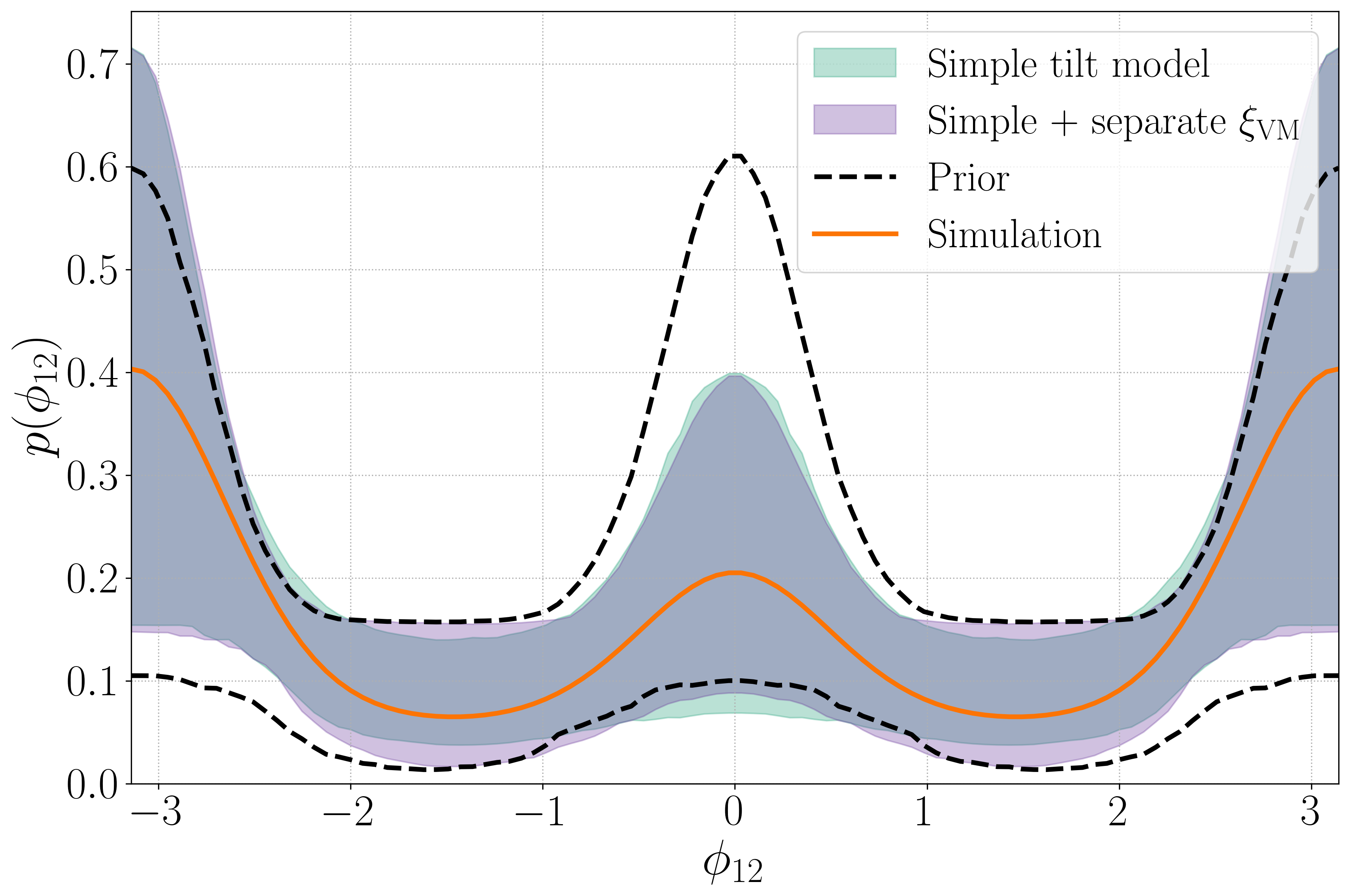}
  \caption{Inferred $\phi_{12}$ distribution for the strong resonances + isotropic population under the \textsc{Simple} tilt model when the \ac{VM} mixture fraction tracks the tilt mixture fraction (green) and when it is an independent parameter (purple). The shaded region bounds the 90\% posterior credible interval, and the dashed black lines bound the 90\% prior credible interval. The true simulated distribution is shown in orange.}
  \label{fig:field_frac_0.64_mix_simple_separate_phi12_comp}
\end{figure}

\begin{figure}
  \centering
  \includegraphics[width=\columnwidth]{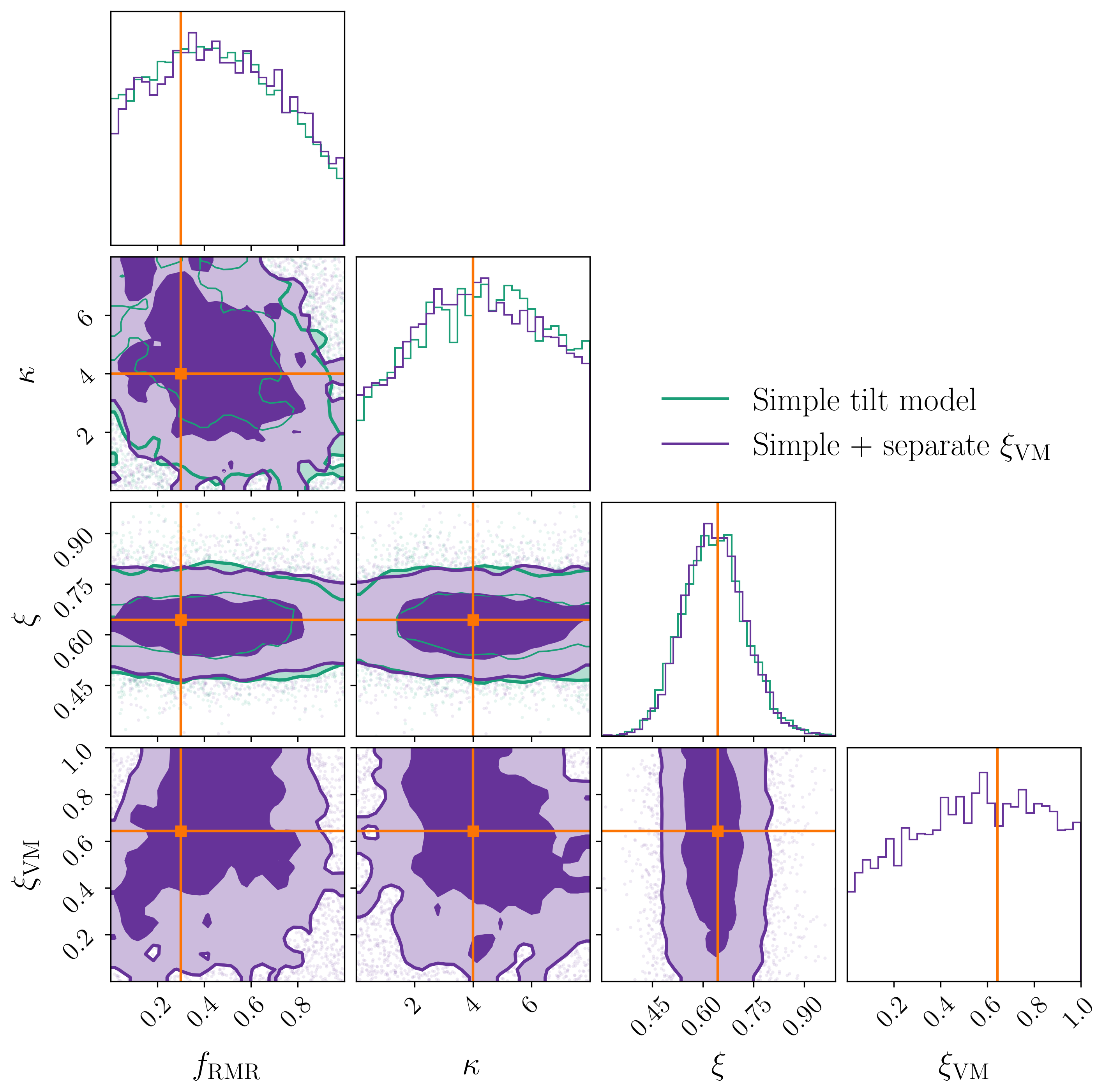}
  \caption{Corner plot of the strong resonances + isotropic population analyzed with the \textsc{Simple} (green) and \textsc{Simple} + separate $\xi_{\mathrm{VM}}$ (purple) tilt models showing the posteriors on the $\phi_{12}$ hyper-parameters. The orange lines indicate the true hyper-parameter values.}
   \label{fig:field_frac_0.64_mix_simple_separate_phi12_corner_comp}
\end{figure}

In Fig.~\ref{fig:field_frac_0.64_mix_simple_separate_phi12_comp}, we show the $\phi_{12}$ distribution inferred for this population under the original $\textsc{Simple}$ tilt model and this modification, dubbed \textsc{Simple} + separate $\xi_{\mathrm{VM}}$. The corresponding posteriors on the $\phi_{12}$ hyper-parameters are shown in Fig.~\ref{fig:field_frac_0.64_mix_simple_separate_phi12_corner_comp}. The posteriors on $f_{\mathrm{RMR}}, \kappa$ are generally less informative than for the original Strong resonances population, which is expected given that only $64\%$ of the 200 events contribute information to these parameters. 

We accurately measure the mixture fraction between the isotropic and aligned spin components; fully-isotropic and fully-aligned populations are excluded from the $\xi$ posterior, which has a 95\% credible interval width $< 0.5$ (see Table~\ref{tab:results}). While the posterior on $\xi$ is dominated by the information from the spin tilt distribution, the $\xi_{\mathrm{VM}}$ posterior differs from the flat prior and weakly favors the true value. To quantify the amount of information contributed by the $\phi_{12}$ distribution to the constraint of $\xi$, we calculate the \ac{HD}~\citep{Hellinger1909} between the posteriors on $\xi,\ \xi_{\mathrm{VM}}$ and the prior for both the \textsc{Simple} and \textsc{Full} tilt models; the \ac{HD} is maximized when computed between completely disjoint distributions, $H_{D}(p, q) = 1$ and $H_{D}(p, p) = 0$ for identical distributions. We find 
\begin{align*}
&H_{D}(\pi(\xi), p(\xi)) = \hellingerDistanceMixedSimpleXi ,\\ &H_{D}(\pi(\xi_{\mathrm{VM}}), p(\xi_{\mathrm{VM}})) = \hellingerDistanceMixedSimpleXiVM 
\end{align*} 
for the \textsc{Simple} tilt model and 
\begin{align*}
&H_{D}(\pi(\xi), p(\xi)) = \hellingerDistanceMixedFullXi ,\\ &H_{D}(\pi(\xi_{\mathrm{VM}}), p(\xi_{\mathrm{VM}})) = \hellingerDistanceMixedFullXiVM 
\end{align*}
for the \textsc{Full} model. Despite the additional uncertainty in the recovered $\phi_{12}$ distribution under the model variations with a separate $\xi_{\mathrm{VM}}$ parameter, the true distribution is recovered within the 90\% credible interval of the posterior. 

We thus conclude that
\begin{itemize}
\item The $\phi_{12}$ distribution is successfully recovered \red{within the 90\% posterior credible interval} for a population with strong resonances mixed with a significant isotropic component.
\item The azimuthal spin angle inference carries some information ($H_{D, \mathrm{max}} =  \hellingerDistanceMixedFullXiVM$ for this simulated population) on the mixture fraction between the aligned and isotropic components.
\item For a population of the size expected by the end of \ac{O4} where the aligned-spin component is this narrowly peaked, the mixture fraction is well-measured with a 95\% credible interval width of $\sim 0.4$.
\end{itemize}

\subsection{Isotropic population}
We next analyze a fully isotropic population to verify whether evidence for \acp{SOR} can spuriously appear in populations without a resonant component. In addition to the spin angle models previously explored, we allow the peaks in the $\phi_{12}$ distributions to be free parameters. A deviation from the prior in the inference of these parameters, $\mu_{\mathrm{RMR}}, \mu_{\mathrm{SMR}}$, would indicate a false preference for substructure in the $\phi_{12}$ distribution. 

\begin{figure}
  \centering
  \includegraphics[width=\columnwidth]{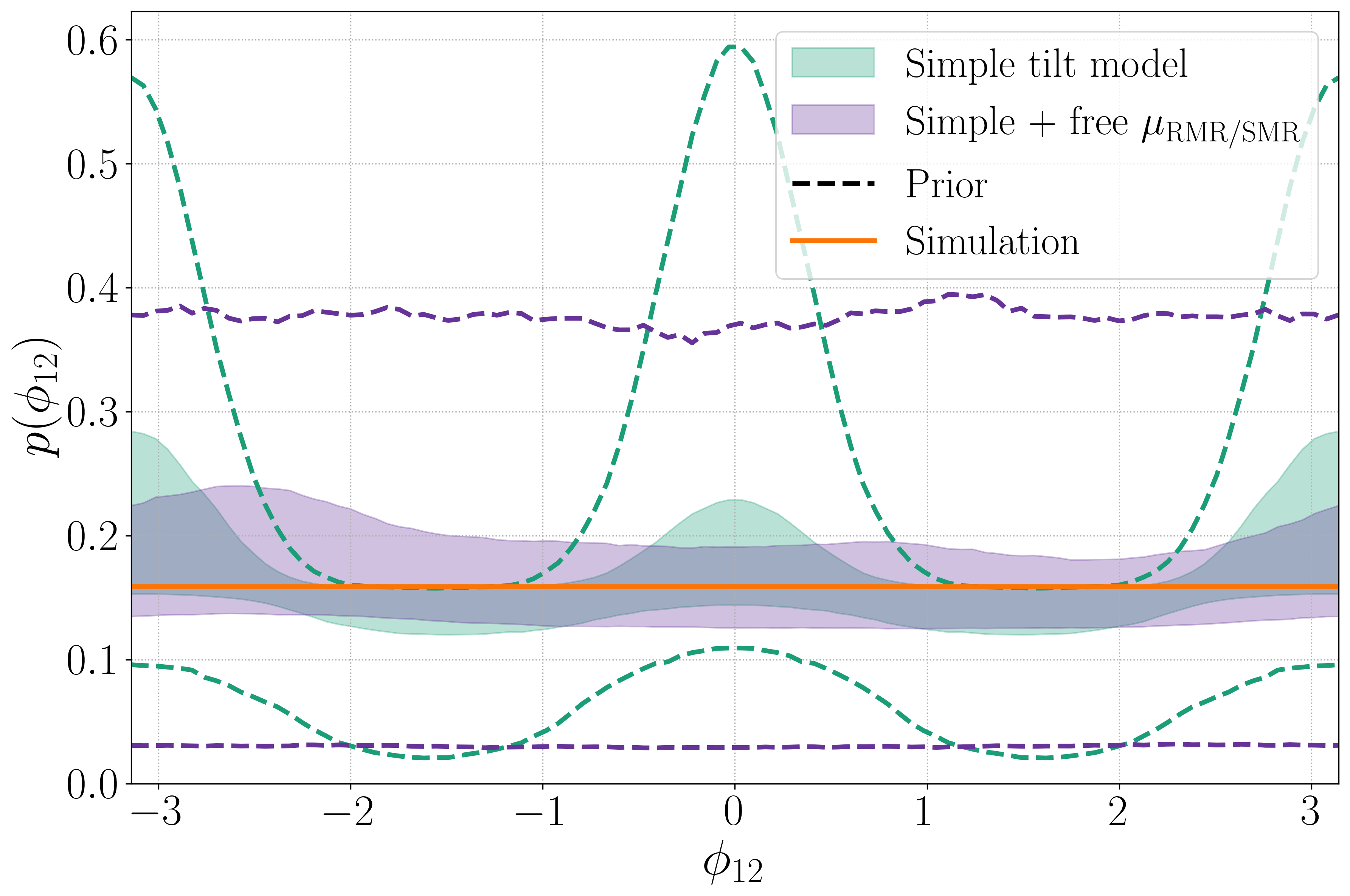}
  \caption{Inferred $\phi_{12}$ distribution for the isotropic population under the \textsc{Simple} tilt model with the peaks of the resonant component fixed (green) and inferred as free parameters (purple). The shading bounds the 90\% posterior credible intervals, the dashed lines bound the 90\% prior credible interval, and the true simulated distribution is shown in orange.}
  \label{fig:iso_mix_simple_free_mu_lightning_phi12}
\end{figure}

In Fig~\ref{fig:iso_mix_simple_free_mu_lightning_phi12}, we show the $\phi_{12}$ distribution inferred under the \textsc{Simple} tilt model for the full population of 200 simulated events. When  $\mu_{\mathrm{RMR}}, \mu_{\mathrm{SMR}}$ are treated as free parameters (purple), we recover a strong preference for an isotropic distribution \red{($\xi < 0.29$)}, with no evidence for additional features in the distribution. To quantify this, the maximum \ac{HD} we find between the posteriors on $\mu_{\mathrm{RMR}},\ \mu_{\mathrm{SMR}}$ and their priors among all the analyses of the isotropic population where these are free parameters in Table~\ref{tab:results} is \hellingerDistanceIsoMax, and the median is \hellingerDistanceIsoMedian.
As expected for an isotropic distribution with no resonant component, the posteriors on $f_{\mathrm{RMR}}, \kappa$ are uninformative. In the case where $\mu_{\mathrm{RMR}}, \mu_{\mathrm{SMR}}$ are fixed (green), the prior still leads to excesses in the $\phi_{12}$ distribution at their theoretically expected locations, but we find a much stronger preference for isotropic distributions \red{($\xi < 0.33$)} compared to the simulated populations with a resonant component. We find no significant differences in the recovered spin angle distributions depending on the tilt model considered or whether we analyze the full population or just the NRSur7dq4 subset.

\subsubsection{Spin tilt distribution}
\begin{figure}
  \centering
  \includegraphics[width=\columnwidth]{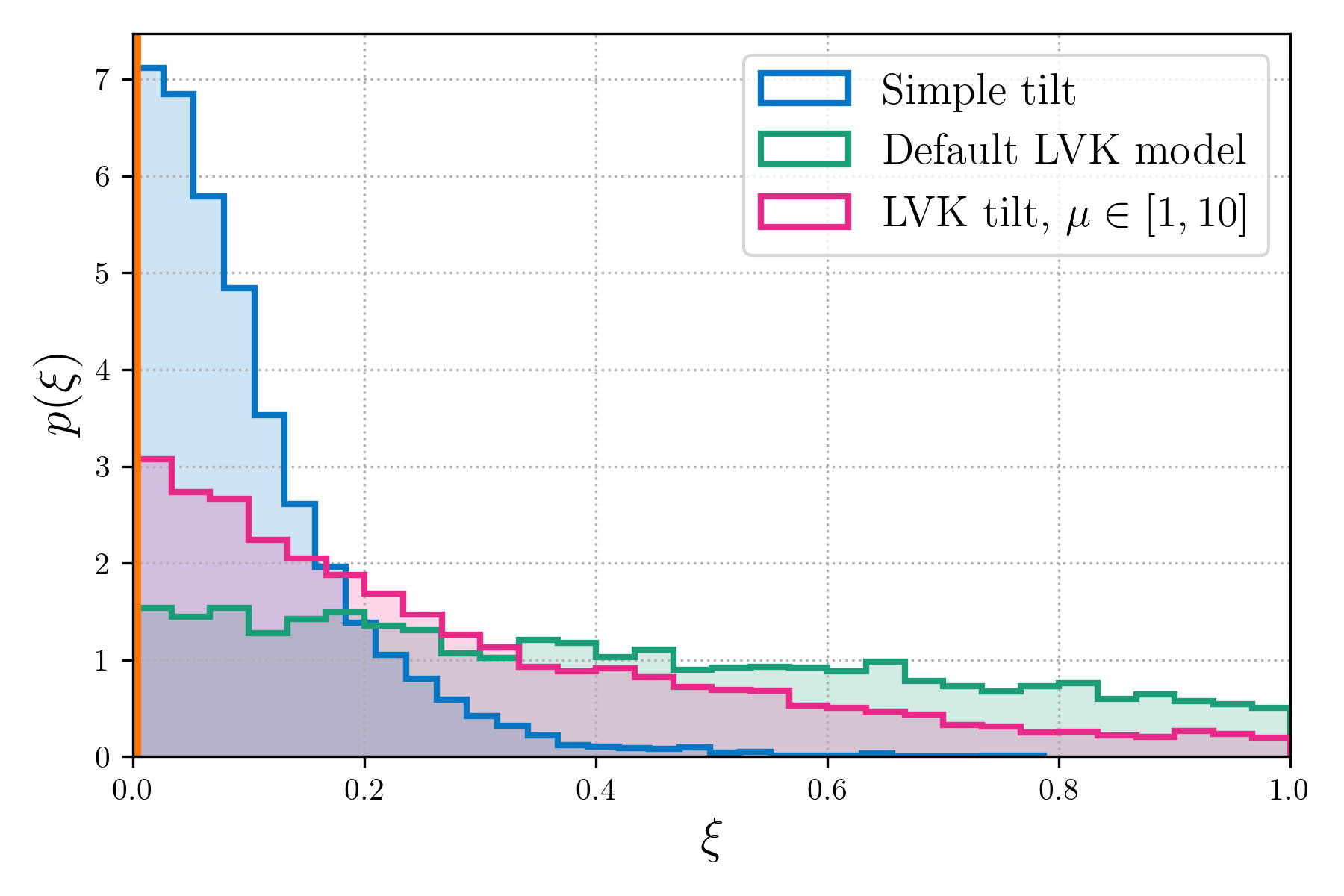}
  \caption{Posteriors on the mixture fraction between the isotropic and aligned spin tilt components under our \textsc{Simple} tilt model in blue, the \textsc{Default} LVK model in green, and the LVK model with a flexible mean of the aligned-spin component in pink. Under the \textsc{Default} LVK model, the mean of the distribution of the aligned-spin component is fixed to $\mu=1$; the prior for the flexible mean model is $\mathrm{U}[1,10]$, based on the priors chosen for $\mu_{1,2}$ for our \textsc{Simple} tilt model.}
  \label{fig:iso_xi_comp}
\end{figure}

Similar to the $\phi_{12}$ distribution, the spin tilt distributions are well-measured to be roughly flat. The posterior on $\xi$ is narrowly constrained to $\lesssim 0.3$ at 95\% credibility for both the \textsc{Simple} and \textsc{Full} tilt models considered in this work. However, we find that these constraints obtained on $\xi$  are much stronger than those obtained when we reanalyze the isotropic population with the \textsc{Default} LVK tilt model, $\xi \leq 0.90$. 
In Fig.~\ref{fig:iso_xi_comp}, we show the posteriors on $\xi$ for the full isotropic population under the \textsc{Simple} tilt model, the \textsc{Default} LVK model, and the \textsc{Default} LVK model with a more flexible prior on the mean of the aligned-spin component. The \textsc{Default} LVK model assumes the two component tilt distributions are identically distributed with fixed $\mu_{1}=\mu_{2}=1$, while our more flexible modification uses a uniform prior on $\mu_{1}=\mu_{2} \in [1,10]$. Under this modified LVK model, we obtain $\xi \leq 0.78$. The prior on $\sigma_{1}=\sigma_{2}$ for the LVK models is $\mathrm{U}(0.01, 4)$. We note that both of these models are subsets of our \textsc{Simple} tilt model, which uses the same functional form for the distribution but does not assume that the two component tilts are identically distributed.

\begin{figure*}
  \centering
  \includegraphics[width=\textwidth]{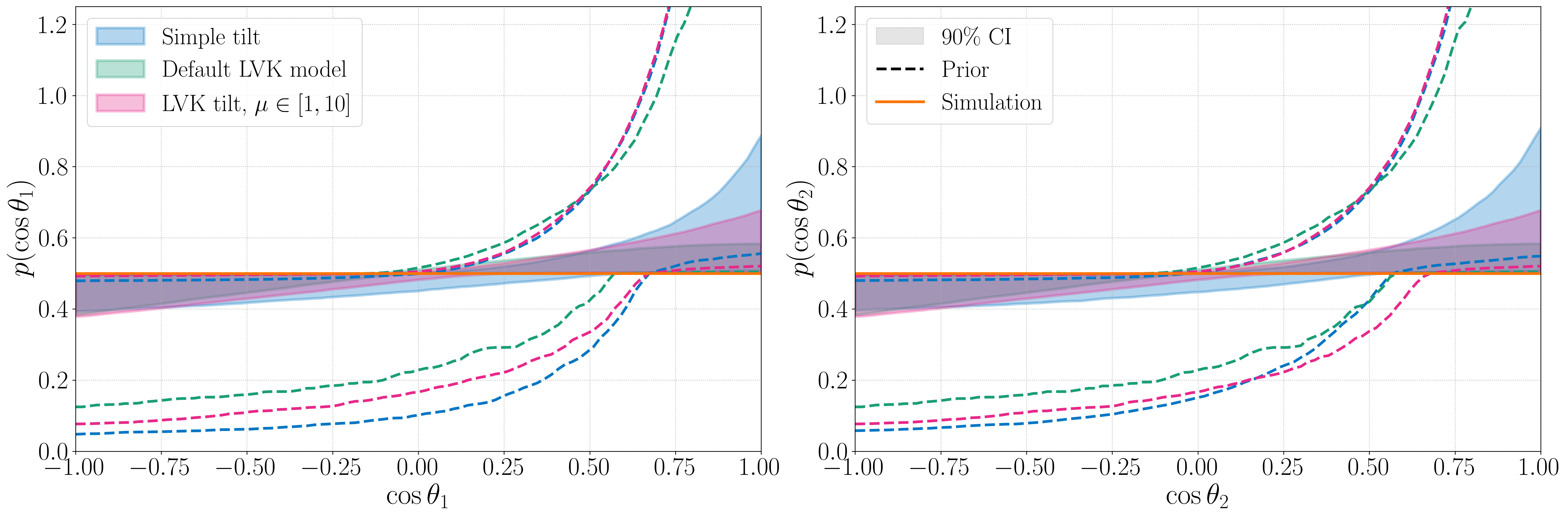}
  \caption{Inferred $p(\cos\theta_1, \cos\theta_2)$ distribution for the strong resonances population under the \textsc{Simple} tilt model in blue, the \textsc{Default} LVK model in green, and the LVK model with a flexible mean of the aligned-spin component in pink. The shading bounds the 90\% posterior credible intervals, the dashed lines bound the 90\% prior credible interval, and the true simulated distribution is shown in orange.}
  \label{fig:iso_tilt_comp}
\end{figure*}

In Fig.~\ref{fig:iso_tilt_comp}, we show the population-level distributions inferred for $\cos\theta_{1,2}$ under these three models. Despite the significant differences in the $\xi$ posteriors, the shapes of the resulting tilt distributions and their uncertainties are very similar. In fact, the tilt distributions appear to be measured with the \textit{least} uncertainty under the \textsc{Default} LVK model, despite the fact that the posterior on $\xi$ obtained under this model is the \textit{most} uncertain. This apparent contradiction can be explained in terms of a prior volume effect. The models where $\mu_{1,2} \geq 1$ allow for more strongly peaked distributions than the \textsc{Default} LVK model, which has much more relative prior volume that supports flat distributions. Even when $\xi=0.8$ under this model, meaning that only 20\% of sources are drawn from an isotropic spin tilt distribution, a sufficiently flat distribution can be obtained if the width of the aligned-spin component distribution is large enough. Indeed, the part of parameter space where $\sigma_{1}=\sigma_{2}$ is small and $\xi$ is large is ruled out by the posterior (narrow, preferentially aligned distributions). The value of the $\xi$ parameter does not have as significant an effect on the shape of the resulting tilt distribution under the \textsc{Default} LVK model as under the models where $\mu_{1,2} \geq 1$. 

The significant differences obtained in the $\xi$ posteriors depending on the prior assumed on the other spin tilt hyper-parameters suggest caution when interpreting this parameter astrophysically. While all three posteriors are consistent with the true value of $\xi=0$ in the case of a fully isotropic distribution, such large variations indicate that even with a population of 200 events at \ac{O4} sensitivity, robust inferences cannot be made for this parameter. Rather than interpreting $\xi$ strictly as the mixture fraction between \acp{BBH} formed dynamically vs via isolated binary evolution, it is prudent to remember that phenomenological models such as those employed here and by the LVK are only weak proxies for the underlying astrophysics. A more model-independent probe of the degree of isotropy or asymmetry in the tilt distribution could be calculated by comparing the probability in narrow regions around aligned and anti-aligned spins~\citep{Vitale:2022dpa}.

\subsubsection{GWTC-2-sized catalogs}
\label{sec:gwtc2_test}
Weak evidence for peaks in the $\phi_{12}$ distribution at $\pm \pi$ was previously identified using GWTC-2 data in \citet{Varma:2021xbh}, which could be interpreted as hints of \acp{SOR} in the \ac{BBH} population. To provide statistical context for this previous result, we randomly down-sample the full isotropic population of 200 events to eleven different GWTC-2-sized catalogs consisting of 46 events each. We independently analyze the NRSur7dq4 subset of each of these catalogs using the \textsc{Simple} tilt model with the peaks of the $\phi_{12}$ distribution as free parameters to determine whether we can recover as significant prior deviations in the posteriors on these parameters as in the real data. 

In order to more closely match the \citet{Varma:2021xbh} hierarchical analysis, instead of using a uniform prior on $\kappa$, we explore two different priors. The first is uniform in $\sigma_{\phi_{12}} = 1/\sqrt{\kappa}$ and the second is the Jeffreys prior for both $\sigma_{\phi_{12}}$ and $\kappa$, $\pi(\sigma_{\phi_{12}}) \propto \sigma_{\phi_{12}}^{-1},\ \pi(\kappa) \propto \kappa^{-1}$. The Jeffreys prior is an uninformative prior which is invariant under a change in coordinates for the parameter vector ($\sigma_{\phi_{12}} \leftrightarrow \kappa$), making it an appealing choice for scale parameters. We note that the spin tilt and azimuth angle distributions used in this analysis are nonetheless different from those used in \citet{Varma:2021xbh} (the Default LVK tilt model and a single-component \ac{VM} distribution for $\phi_{12}$ with peak $\mu_{\phi_{12}}$). 

\begin{figure*}
  \centering
  \includegraphics[width=\textwidth]{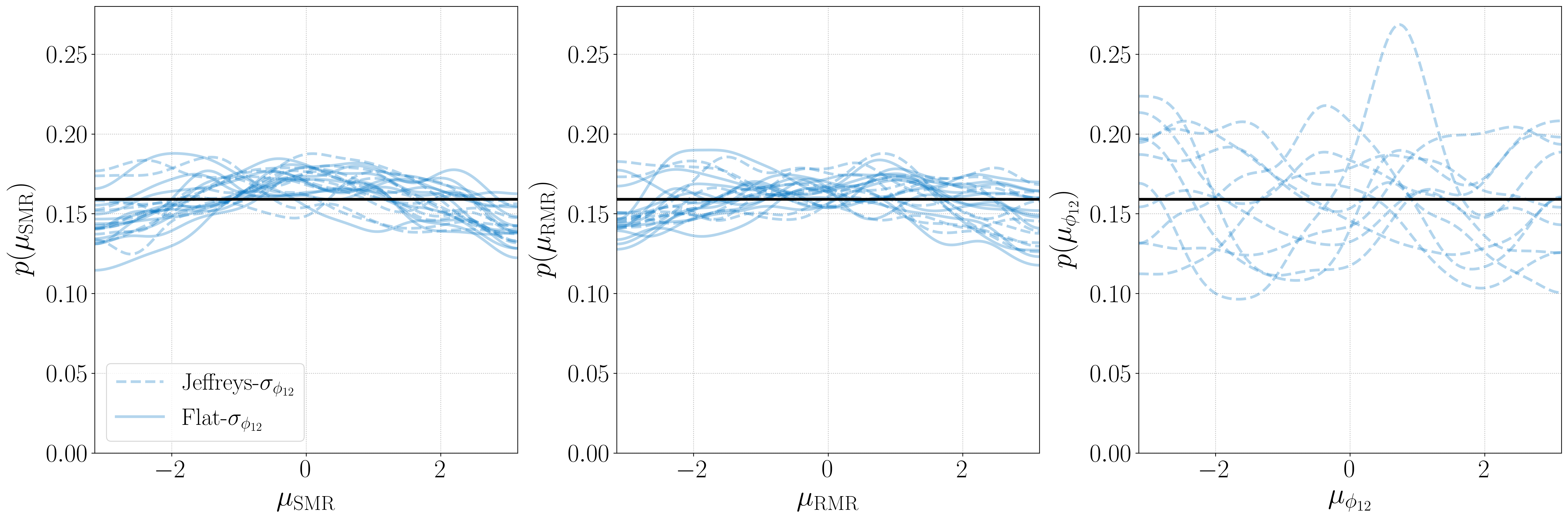}
  \caption{Kernel density estimates of the posterior distributions on the azimuthal spin angle hyper-parameters $\mu_{\mathrm{SMR}}$ (left) and $\mu_{\mathrm{RMR}}$ (middle) under the \textsc{Simple} tilt model with our two-component \ac{VM} model for the $\phi_{12}$ distribution for eleven different GWTC-2-sized sub-catalogs of the full isotropic population. The right panel shows the posteriors on $\mu_{\phi_{12}}$ when we reanalyze the same eleven populations using the single-component \ac{VM} model of \citet{Varma:2021xbh}. The solid blue lines represent the posteriors obtained under a flat prior on $\sigma_{\phi_{12}}$, and the dashed lines show the Jeffreys prior results. The priors themselves are shown in the black lines.}
  \label{fig:gwtc2_phi12_kde_comp}
\end{figure*}

The hyper-parameter posteriors we infer on $\mu_{\mathrm{SMR}}, \mu_{\mathrm{RMR}}$ are shown in Fig.~\ref{fig:gwtc2_phi12_kde_comp}. While \citet{Varma:2021xbh} find that the Jeffreys prior yields more informative posteriors on $\mu$, we do not find any significant differences between the $\mu$ posteriors obtained under the two different priors. In both cases, the posteriors are uninformative, as expected for an isotropic distribution. We find median \acp{HD} between the priors and posteriors of \hellingerDistanceGWTCTwoUnifMedian versus \hellingerDistanceGWTCTwoJeffMedian for the results obtained with uniform and Jeffrey's priors, respectively. None of our GWTC-2-sized catalogs yield as strong constraints on $\mu_{\mathrm{SMR}}, \mu_{\mathrm{RMR}}$ as found on $\mu_{\phi_{12}}$ in \citet{Varma:2021xbh} using the Jeffreys prior, with $H_{D} = \hellingerDistanceVarma$.

However, when we reanalyze the same downsampled catalogs using the single-component \ac{VM} distribution used in \citet{Varma:2021xbh} (still using our different, \textsc{Simple} tilt model), we find much stronger deviations from the prior in the posteriors on the $\mu_{\phi_{12}}$ parameter, shown in the right panel of Fig.~\ref{fig:gwtc2_phi12_kde_comp}. In this case, we find a median \ac{HD} of \hellingerDistanceGWTCTwoSingleMedian and a maximum of \hellingerDistanceGWTCTwoSingleMax. The additional complexity of the two-component \ac{VM} distribution used in the rest of this work protects against inferring evidence for spurious features in the $\phi_{12}$ distribution, which are more prone to appear using the single-component \ac{VM} distribution. The informative posterior on $\mu_{\phi_{12}}$ obtained in \citet{Varma:2021xbh} is consistent with the random fluctuations we obtain in the posteriors on this parameter for these simulated isotropic populations, meaning that the \citet{Varma:2021xbh} result is unlikely to be a significant measurement of evidence for \acp{SOR}.

To summarize,
\begin{itemize}
\item We recover strong evidence for isotropically distributed spin tilt and azimuthal angles in a \ac{BBH} population of the size expected by the end of O4 simulated with isotropic spin orientations, \red{constraining the fraction of sources drawn from a preferentially aligned spin tilt distribution (field binaries) to $\xi \leq 0.33$.}
\item Care should be taken when astrophysically interpreting the mixture fraction parameter $\xi$ between the aligned and isotropically-distributed spin angle components, as the posteriors on this parameter are strongly model-dependent even for models that produce qualitatively similar constraints on the resulting tilt distributions.
\item We do not find spurious evidence for deviations from isotropy in this population when using our two-component \ac{VM} model for the $\phi_{12}$ distribution, even when considering smaller GWTC-2-sized catalogs for comparison with the evidence for \acp{SOR} in \citet{Varma:2021xbh}.
\item The single-component \ac{VM} model for the $\phi_{12}$ distribution used in \citet{Varma:2021xbh} is more prone to inferences of spurious features in the $\phi_{12}$ distribution than our two-component model, quantified by larger \acp{HD} between the priors and posteriors on the hyper-parameters governing the peaks in the $\phi_{12}$ distribution. 
\end{itemize}

\subsection{Weak resonances population}
\label{sec:wide_results}
Next, we analyze the population with weaker resonant features drawn from a broader distribution of spin tilts for the preferentially-aligned component representing binaries formed in the field. This population is more consistent with the GWTC-3 inference than the strong resonances population discussed above, as there is no significant evidence in the current data for a narrow, aligned component in the inferred tilt distribution. In order to accommodate the broader tilt distribution shape of this simulated population, we use slightly different priors on the hyper-parameters of the \textsc{Simple} tilt model for this analysis, shown at the bottom of Table~\ref{tab:spin_angle_models}.

While we generally find slightly weaker constraints on the $\phi_{12}$ distribution for this simulated population compared to the strong resonances populations (see credible intervals in Table~\ref{tab:results}), we are still able to recover the true distribution shape \red{within the 90\% credible interval of the posterior. Comparing the median 95\% posterior credible interval widths averaged over all the analyses in Table~\ref{tab:results}, we obtain $$\mathrm{CI}_{95}(f_{\mathrm{RMR}})=0.828,\ \mathrm{CI}_{95}(\kappa)=7.38$$ for the weak resonances population compared to $$\mathrm{CI}_{95}(f_{\mathrm{RMR}})=0.680,\ \mathrm{CI}_{95}(\kappa)=7.07$$ for the strong resonances population.}
We find weaker constraints on $f_{\mathrm{RMR}}, \kappa$ when using the \textsc{Simple} tilt model compared to the \textsc{Full} model and when $\xi_{\mathrm{VM}}$ is introduced as an independent parameter. This is consistent with the picture that the spin tilts contribute information to the inference of these hyper-parameters, which we exclude when using these model variations.

\begin{figure*}
  \centering
  \includegraphics[width=\textwidth]{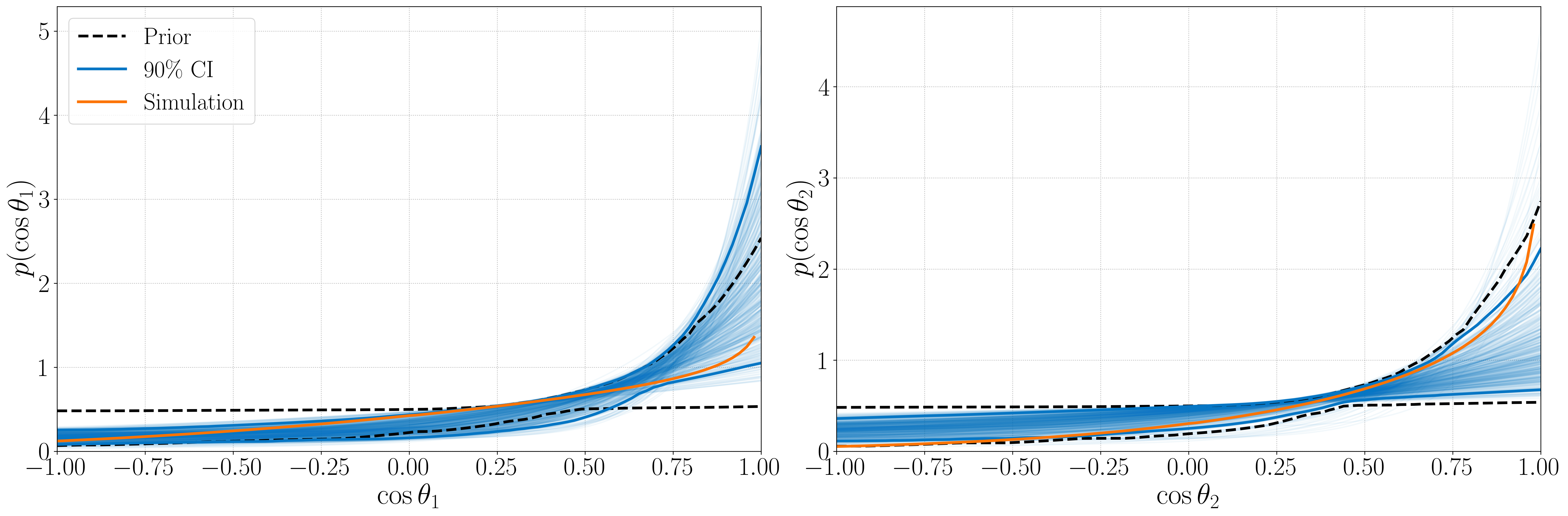}
  \caption{Inferred $p(\cos\theta_1, \cos\theta_2)$ distributions for the weak resonances population under the \textsc{Simple} tilt model. Individual light blue traces show the distributions corresponding to individual hyper-parameter posterior samples, the dark blue lines bound the marginalized 1D 90\% posterior credible intervals for the individual tilts, and the dashed black lines bound the 90\% prior credible intervals. The true simulated distribution is shown in orange.}
  \label{fig:all_field_weak_mix_simple_new_lightning_tilt}
\end{figure*}

\begin{figure}
  \centering
  \includegraphics[width=\columnwidth]{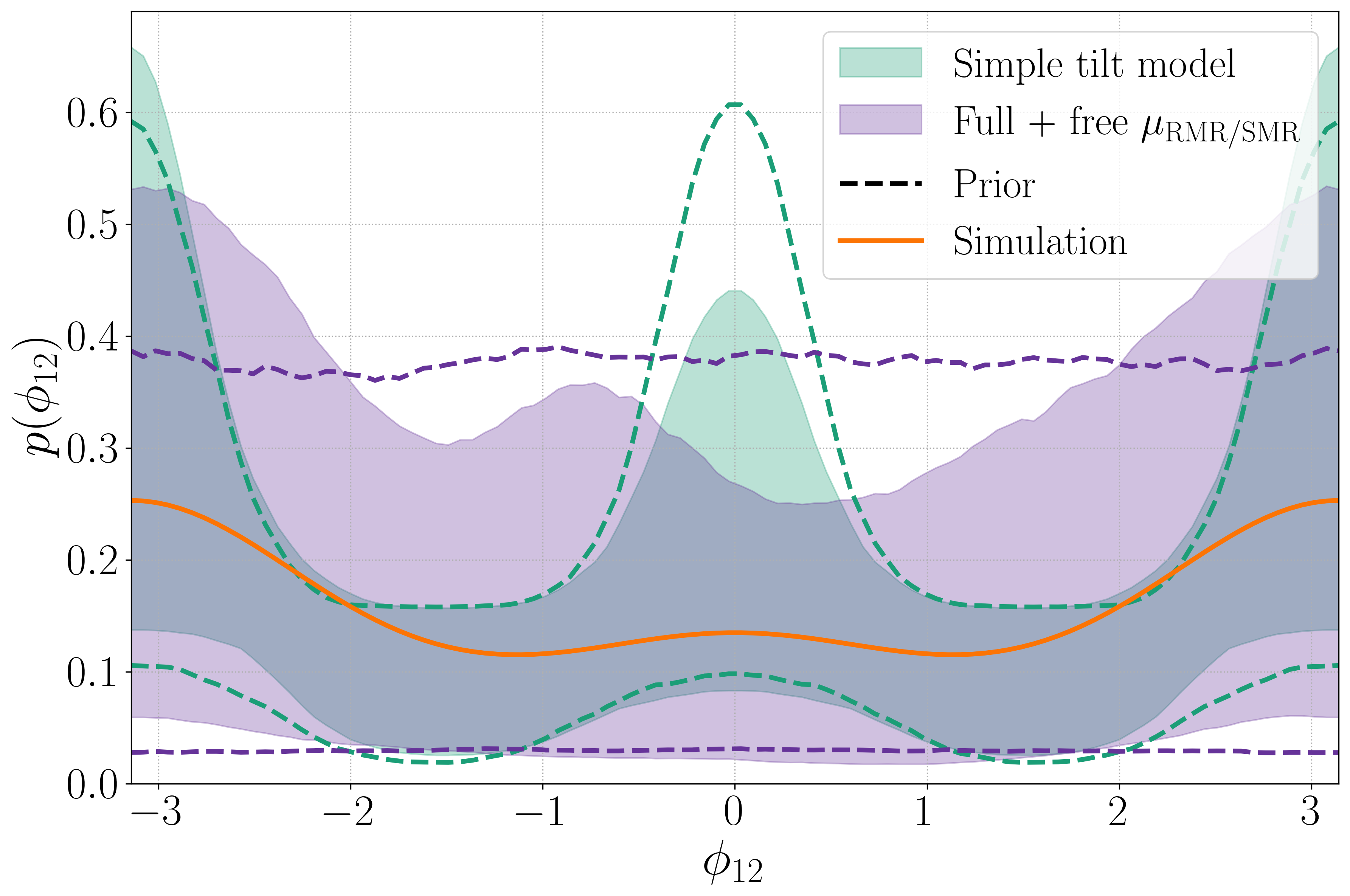}
  \caption{Inferred $\phi_{12}$ distribution for the weak resonances population under the \textsc{Simple} tilt model (green) and under the \textsc{Full} tilt model with the peaks of the resonant component as free parameters (purple). The shading bounds the 90\% posterior credible intervals, the dashed lines bound the 90\% prior credible interval, and the true simulated distribution is shown in orange.}
  \label{fig:all_field_weak_mix_shaded_phi12_comp}
\end{figure}

Interestingly, we find a pervasive bias in the posterior on the mixture fraction $\xi$ when using the \textsc{Simple} tilt model for this simulated population. This bias is not due to issues with the individual-event parameter estimation, as it is still present when we remove the statistical uncertainty in the measurement of the individual-event parameters (i.e., perform the hierarchical inference using delta function posteriors at the true binary parameter values). For this choice of true hyper-parameter values, neglecting to model the correlation between $\theta_{1}, \theta_{2}$ forces the inference to prefer lower values of $\xi \approx 0.72$ in order to qualitatively reproduce the shape of the marginal tilt angle distributions. In this case, $\xi$ loses its interpretation as the fraction of systems of that formed via isolated binary evolution with preferentially aligned spins, but the true marginal tilt and azimuthal angle distributions are recovered within the bulk of the posterior support, as shown in Figs.~\ref{fig:all_field_weak_mix_simple_new_lightning_tilt}-\ref{fig:all_field_weak_mix_shaded_phi12_comp} (purple). The bias in $\xi$ does not lead to a qualitative misestimation of the $\phi_{12}$ distribution because the shape of the distribution changes very little between $\xi=1$ and $\xi=0.72$ given the other hyper-parameter values. This further indicates that caution should be taken when interpreting the posteriors on $\xi$ astrophysically.

We also check whether the locations of the peaks in the $\phi_{12}$ distribution can be independently inferred even with weak resonant features by allowing $\mu_{\mathrm{RMR}}, \mu_{\mathrm{SMR}}$ to be free parameters. This would potentially allow us to distinguish between a population with strong tides but weak resonant features and a population with weak tides, where we expect excesses in the $\phi_{12}$ distribution at $\pm \pi/2$. Under the \textsc{Full} tilt model, we correctly infer that $\sim 70\%$ of sources are librating around $\phi_{12}=\pm \pi$ and recover the true $\phi_{12}$ distribution within the 90\% posterior credible interval, shown in green in Fig.~\ref{fig:all_field_weak_mix_shaded_phi12_comp}. However, we \textit{incorrectly} infer that this corresponds to the \ac{RMR} subpopulation rather than the \ac{SMR} subpopulation, as we find $f_{\mathrm{RMR}} = 0.76^{+0.24}_{-0.49}$, and the $\mu_{\mathrm{RMR}}$ posterior peaks strongly at the edges of the prior at $\pm \pi$. This implies a bias in our recovered tilt distribution, as we incorrectly infer that $\theta_{2} > \theta_{1}$ ($\cos\theta_{2} < \cos\theta_{1}$) more frequently than the other way around. 

This bias indicates that the constraint on $f_{\mathrm{RMR}}$ is driven by the $\phi_{12}$ inference; we are able to qualitatively recover the features in the $\phi_{12}$ distribution at the expense of the tilt distribution. In this case, the differences between the true $\theta_{1}$ and $\theta_{2}$ distributions are not significant enough (due to the wider truncated Gaussians) to overcome the degeneracy in our overlapping priors for $\mu_{\mathrm{RMR}}, \mu_{\mathrm{SMR}}$. Even in the presence of this bias in the $f_{\mathrm{RMR}}$ hyper-parameter affecting the tilt distribution, the true marginal $\cos\theta_{1}, \cos\theta_{2}$ distributions are still recovered at or within the edges of the 90\% posterior credible intervals shown in Fig.~\ref{fig:all_field_weak_mix_simple_new_lightning_tilt}. With the \textsc{Simple} + free $\mu_{\mathrm{RMR/SMR}}$ tilt model, the hyper-parameters governing the $\phi_{12}$ distribution are largely unconstrained. The model is too flexible, and the features are not pronounced enough to obtain a significant constraint.

For the weak resonances population, our main take-aways are that
\begin{itemize}
\item We are still able to gain information about the $f_{\mathrm{RMR}}, \kappa$ parameters governing the $\phi_{12}$ distribution, particularly when using the \textsc{Full} tilt model \red{($\mathrm{CI}_{95}(f_{\mathrm{RMR}})=0.654,\ \mathrm{CI}_{95}(\kappa)=7.39$)}.
\item Mismodeling the correlation in the tilt distribution using the \textsc{Simple} tilt model leads to a bias in the inference of $\xi$.
\item While this means that this parameter can no longer be correctly astrophysically interpreted as the isotropic vs aligned-spin mixture fraction, we still recover the \red{true marginal tilt and azimuthal spin angle distributions within the 90\% posterior credible interval}.
\item We can correctly infer the fraction of the population with azimuthal spin angles librating around $\phi_{12}=\pi,0$ \red{($f_{\mathrm{RMR}} = 0.76^{+0.24}_{-0.50}$)}, but we misidentify whether that fraction corresponds to the \ac{RMR} ($\theta_{2} > \theta_{1}$) vs \ac{SMR} ($\theta_{1} > \theta_{2}$) scenario. This is because the differences between these broader $\cos\theta_{1,2}$ distributions are not significant enough to break the prior degeneracy.
\end{itemize}

\subsection{GWTC-3 results}
Finally, we analyze the real \ac{LVK} data from GWTC-3 to search for evidence of \acp{SOR}. To match the analysis setup used for the simulated populations in the rest of this work, we use the publicly-released individual-event parameter estimation posteriors obtained using NRSur7dq4 from \citet{Islam:2023zzj} where available; otherwise, we use the IMRPhenomXPHM results released by the LVK~\citep{ligo_scientific_collaboration_and_virgo_2021_5117703, ligo_scientific_collaboration_and_virgo_2021_5546663}. We evolve the spin angles forward to $f_{\mathrm{ISCO}}$ using the same methodology described in Section~\ref{sec:methods_pe}. Following \ac{LVK} convention, we impose a detection threshold of FAR (false alarm rate) $<  1/\mathrm{yr}$~\citep{KAGRA:2021duu}, leaving a population of 69 events, 41 of which have available NRSur7dq4 posteriors. To account for selection effects, we use the combined O1-O3 \ac{BBH} sensitivity injection set released by the LVK~\citep{ligo_scientific_collaboration_and_virgo_2021_5636816}. We impose a cut on the variance in the hierarchical likelihood, $\sigma_{\mathrm{tot}}^{2}<5$, to obtain hyper-parameter posteriors consistent with the official \ac{LVK} results by ensuring convergence of the \ac{MC} integral. We use the same fiducial mass, redshift, and spin magnitude distributions employed by the \ac{LVK} and in the rest of this work. 

We analyze the GWTC-3 population using both the \textsc{Simple} and \textsc{Full} tilt models, keeping $\mu_{\mathrm{RMR/SMR}}$ fixed to their theoretically-predicted values and allowing them to be free parameters. For the \textsc{Simple} tilt model, we show results obtained using the prior on the tilt hyper-parameters for the strong resonances and isotropic populations by default but also repeat the analysis using the prior choices for the weak resonances population. We recover posteriors on the mass, spin magnitude, and redshift distribution hyper-parameters that are in excellent agreement with the published \ac{LVK} results. 

\begin{figure}
  \centering
  \includegraphics[width=\columnwidth]{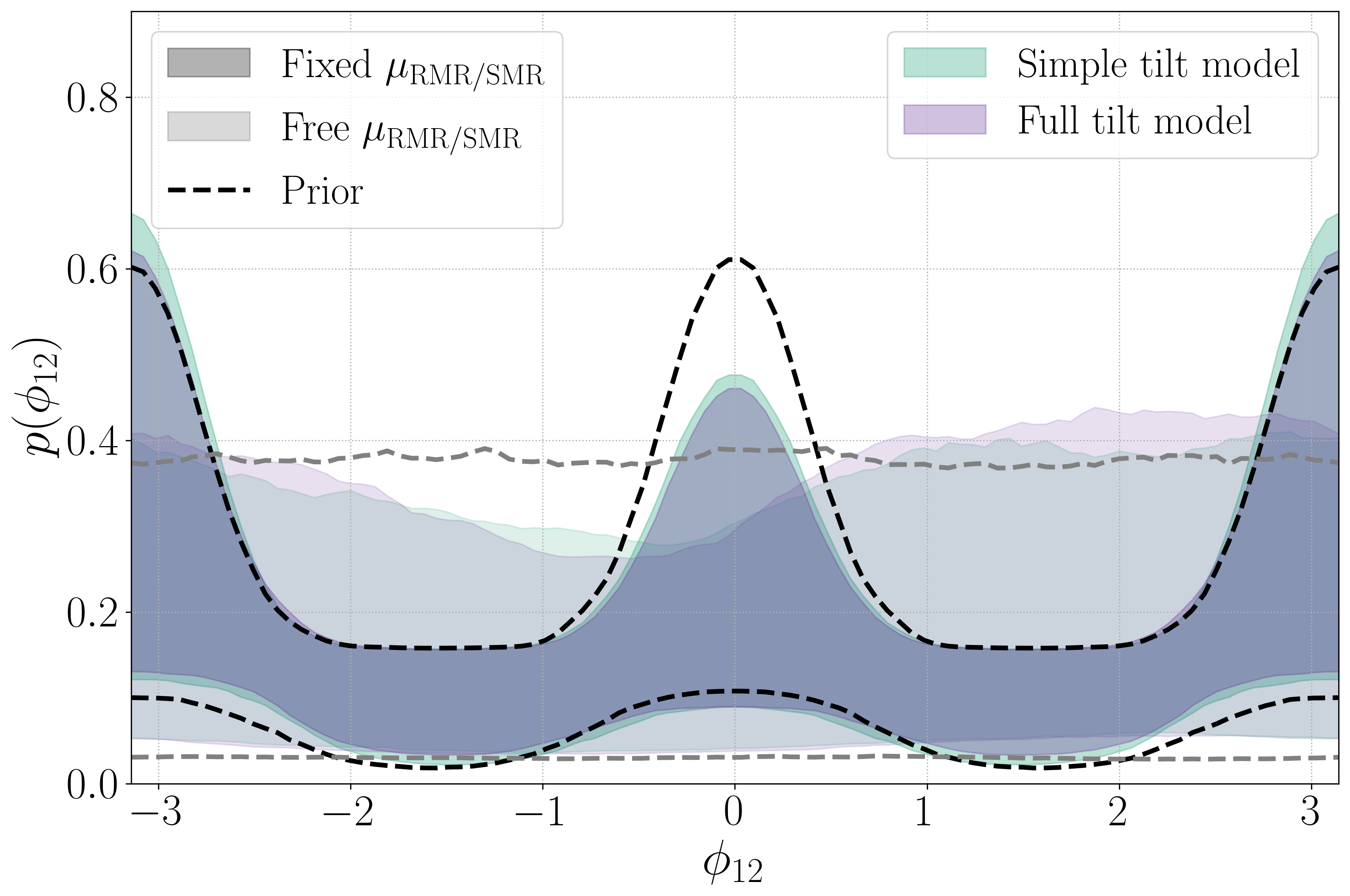}
  \caption{Inferred $\phi_{12}$ distribution for GWTC-3 under the \textsc{Simple} (green) and \textsc{Full} (purple) tilt models, with the peaks of the resonant component fixed (dark) and as free parameters (light). The shading bounds the 90\% posterior credible intervals and the dashed lines bound the 90\% prior credible interval.}
  \label{fig:gwtc3_phi12_comp}
\end{figure}

\begin{figure}
  \centering
  \includegraphics[width=\columnwidth]{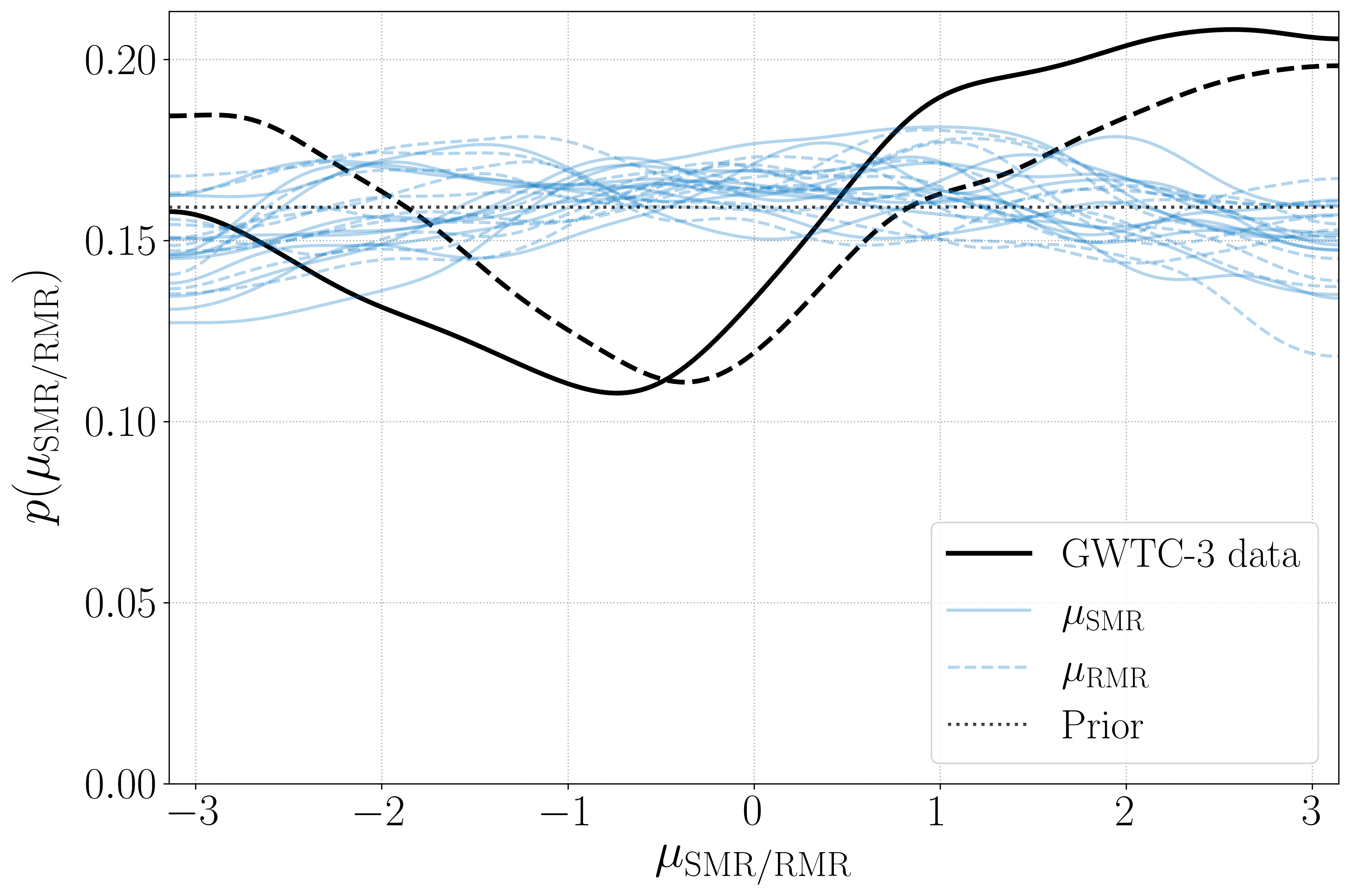}
  \caption{Kernel density estimates of the posterior distributions on the azimuthal spin angle hyper-parameters $\mu_{\mathrm{SMR}}$ (solid) and $\mu_{\mathrm{RMR}}$ (dashed) under the \textsc{Full} tilt model with our two-component \ac{VM} model for the $\phi_{12}$ distribution. The posteriors for eleven different GWTC-3-sized sub-catalogs of the full isotropic population are shown in blue, while the results obtained with the real GWTC-3 data are shown in black. The prior is denoted by the gray dotted line.}
  \label{fig:gwtc3_mus}
\end{figure}

In Fig.~\ref{fig:gwtc3_phi12_comp}, we show the inferred $\phi_{12}$ distribution under the four different model variations described above. Neither the choice of prior on the tilt hyper-parameters nor the choice of tilt model has a significant effect on the inferred $\phi_{12}$ distribution.%
The $f_{\mathrm{RMR}}$ posteriors under the fixed-$\mu_{\mathrm{RMR/SMR}}$ models favor small values, while for the free-$\mu_{\mathrm{RMR/SMR}}$ models, they are uninformative. The $\kappa$ posteriors are uninformative regardless of the choice of $\mu_{\mathrm{RMR/SMR}}$ prior. However, for the free-$\mu_{\mathrm{RMR/SMR}}$ models, the posteriors on both $\mu_{\mathrm{RMR/SMR}}$ parameters peak weakly at $\mu_{\mathrm{RMR/SMR}}\approx \pm\pi$ at the edges of the prior. To quantify the statistical significance of these peaks in the $\mu_{\mathrm{RMR/SMR}}$ posteriors using the real GWTC-3 data, we repeat the isotropic catalog analysis of Section~\ref{sec:gwtc2_test} with 11 GWTC-3-sized catalogs (69 events) analyzed with the \textsc{Full} tilt model with $\mu_{\mathrm{RMR/SMR}}$ as free parameters. This is the model for which we obtain the most informative posteriors on $\mu_{\mathrm{RMR/SMR}}$ using the real data. We find that the \acp{HD} between the prior and the posterior obtained for these parameters (\hellingerDistanceGWTCThreeSMR and \hellingerDistanceGWTCThreeRMR for $\mu_{\mathrm{SMR}}$ and $\mu_{\mathrm{RMR}}$, respectively) are larger than the largest \ac{HD} obtained for the isotropic GWTC-3-sized catalogs analyzed with the same population model (\hellingerDistanceGWTCThreeISO). The posteriors on these parameters for each of the isotropic catalogs and the real GWTC-3 data are shown in Fig.~\ref{fig:gwtc3_mus}.
Thus, the inferred $\phi_{12}$ distributions obtained using both $\mu_{\mathrm{RMR/SMR}}$ prior choices demonstrate a weak preference for an excess of systems with $\phi_{12}=\pm \pi$, which could be interpreted as evidence for \ac{SOR} under the \ac{SMR} scenario.

\begin{figure}
  \centering
  \includegraphics[width=\columnwidth]{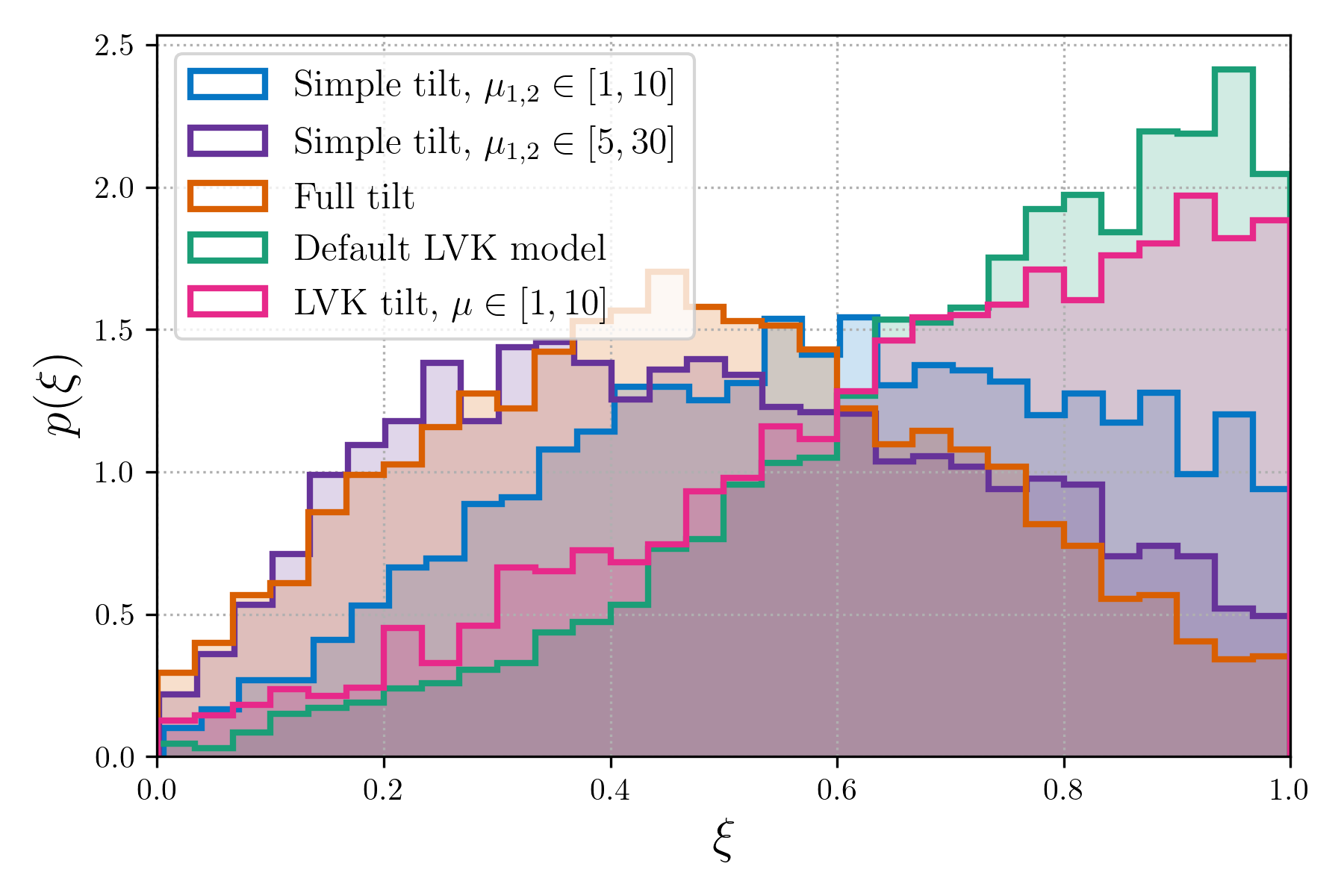}
  \caption{Posteriors on the mixture fraction between the isotropic and aligned spin tilt components under our \textsc{Simple} tilt model with the hyper-parameter priors used for the strong resonances population in blue, with the hyper-parameters priors used for the weak resonances population in purple, the \textsc{Full} tilt model in orange, the \textsc{Default} LVK model in green, and the LVK model with a flexible mean of the aligned-spin component in pink.}
  \label{fig:gwtc3_xi_comp}
\end{figure}

The tilt distributions inferred under all model variations are in qualitative agreement, despite using different hyper-parameter priors. Consistent with previous studies~\citep{KAGRA:2021duu, Vitale:2022dpa, Golomb:2022bon, Edelman:2022ydv, Callister:2023tgi}, we find that the GWTC-3 data do not rule out an isotropic distribution, but prefer a distribution with more aligned than anti-aligned spins. Note that all of our model variations require a peak at $\cos\theta_{1,2}=1$, if there is a peak. However, as for the simulated isotropic distribution, we find considerable variation in the posterior on the mixture fraction between the isotropic and aligned components, shown in Fig.~\ref{fig:gwtc3_xi_comp}. In addition to the four model variations discussed above, we include the $\xi$ posteriors obtained under the \textsc{Default} LVK tilt model and the LVK model with a flexible peak of the aligned-spin component in the comparison. While the posteriors for the two LVK model variations that require that the component tilts are identically distributed (orange and pink) both peak at $\xi=1$ (all aligned), those obtained with the tilt models used in the rest of this work that allow for independent $\cos\theta_{1,2}$ distributions are less informative and instead peak weakly at $\xi \sim 0.4-0.6$. The variation in these posteriors depending on the tilt model again suggests that care should be taken in the astrophysical interpretation of this parameter.

In summary, upon analysis of the GWTC-3 data using our phenomenological models targetting \acp{SOR}, we find that 
\begin{itemize}
\item There is no statistically significant evidence for \acp{SOR} based on the observed current population of \ac{BBH}.
\item Nonetheless, regardless of the population model and prior choice used for the spin angles, there is a weak preference for an excess of events with $\phi_{12}=\pm\pi$, which could be explained by \ac{SOR} under the \ac{SMR} scenario.
\item The posterior on the mixture fraction between the aligned-spin and isotropic components varies considerably depending on the choice of spin tilt population model and hyper-parameter priors.
\end{itemize}

\section{Summary and Conclusion}
\label{sec:conclusion}
In this work, we have conducted the first end-to-end analysis of simulated \ac{BBH} populations including simultaneous modeling of the mass, redshift, spin magnitude, tilt, and azimuthal angle population-level distributions with the goal of determining whether we can probe \acfp{SOR} with gravitational-wave observations. This astrophysical process occurs preferentially in systems formed via isolated binary evolution whereby the component spin vectors and orbital angular momentum align into a common plane and jointly precess about the total angular momentum vector. This leads to features in the resulting distribution of the azimuthal spin angle between the projections of the component spins onto the orbital plane at $\phi_{12}=0, \pm \pi$. Which feature is more pronounced depends on the fraction of systems in the population that have undergone mass ratio reversal, such that the initially more massive star ends up as the less massive black hole. The presence and strength of these features encode information about dynamics, binary formation channels, and tides, making them an important measurement target for gravitational-wave observatories.

We develop an astrophysically-informed phenomenological mixture model for both the spin tilt and azimuthal angle distributions. This model is designed to capture any underlying population that consists of some fraction of systems formed via isolated binary evolution with preferentially aligned spin tilts and features in the $\phi_{12}$ distribution due to \acp{SOR} depending on the fraction of this sub-population that has undergone mass ratio reversal. The remaining fraction of the population is assumed to have formed dynamically with isotropically-oriented spins. 

We simulate four different populations of \ac{BBH} that could be detected by the end of the ongoing fourth LVK observing run (O4) to determine how the measurability of this effect depends on the properties of the astrophysical \ac{BBH} population: 1) a population with strong resonant features represented by spin tilt distributions narrowly peaked around aligned spin ($\cos\theta_{1,2}=1$) and with narrow peaks in the $\phi_{12}$ distribution with a fraction $f_{\mathrm{RMR}}=0.3$ of sources in the librating spin morphology corresponding to mass ratio reversal, 2) a two-component population with fraction $\xi=0.64$ of sources drawn from the previously-described population and the rest drawn from an isotropic spin distribution, 3) a fully isotropic distribution, and 4) a population with weak resonant features represented by spin angle distributions with broader peaks with the same fraction of mass ratio reversed systems.

We find that we are able to correctly qualitatively recover the simulated spin tilt and azimuthal angle distributions for all four of our simulated populations. The inference of the $f_{\mathrm{RMR}}$ parameter is driven by the information in the individual-event $\phi_{12}$ posteriors rather than the spin tilts, while the inference of the mixture fraction between the aligned and isotropic components of the population is driven by the tilts. Our simulations suggest that we will not recover smoking-gun evidence for \acp{SOR} with a population of the size expected by the end of O4, but we can obtain informative measurements of the $\phi_{12}$ distribution and start building evidence for this effect, even in the case where resonant features are weak. We can also potentially distinguish between a population with weak resonances (features at $\phi_{12}=0,\pm\pi$) from one with weak tides (features at $\phi_{12}=\pm\pi/2$). The uncertainties in our inferred spin tilt and azimuth distributions are much smaller for the simulated isotropic population, suggesting that such a population is easier to confidently identify compared to the populations including a significant aligned-spin component.

In addition to the four simulated populations discussed above, we analyze the GWTC-3 catalog of \acp{BBH} observed by the LVK using multiple variations of our population model designed to capture \acp{SOR}. We find no compelling evidence for such features, although we do find a weak preference for an excess of events with $\phi_{12}=\pm\pi$, which could be interpreted to come from \acp{SOR} under the \acf{SMR} scenario. 

We emphasize that care should be taken when interpreting the posteriors on the hyper-parameters of our phenomenological model astrophysically. The mixture fraction parameters $\xi$ and $f_{\mathrm{RMR}}$ are just weak proxies for the astrophysical fraction of systems formed via isolated binary evolution and the fraction of this sub-population that has undergone mass ratio reversal, respectively. The posteriors on $\xi$ can change significantly depending on the prior choice for the other tilt model hyper-parameters, and just because a particular set of binary parameters was drawn from a distribution peaked at $\phi_{12}=0$ (\ac{RMR}), does not mean its spin morphology is necessarily librating around $\phi_{12}=0$. Conversely, the assumption that $\theta_{1} > \theta_{2}$ for the \ac{SMR} scenario under the canonical picture of efficient tides and weak natal kicks is complicated by the potential of spin-axis tossing~\citep{Tauris:2022ggv} and by the idea that black holes may not inherit the spin orientation of their progenitor star at all~\citep{Baibhav:2024rkn}. 
Despite these variations in the interpretation of these parameters, we find that the features identified in the resulting spin tilt and azimuth distributions are robust against model choices. Future work could explore fitting the $\phi_{12}$ distribution with more data-driven models, but phenomenological, parameterized models provide a reasonable trade-off between astrophysical interpretability and flexibility for this initial study.

\vspace{7pt}
S.B. thanks Colm Talbot, Vicky Kalogera, Salvo Vitale, and the anonymous referee for insightful suggestions and discussions and Matt Mould for providing helpful comments during internal LVK document review. 
This material is based upon work supported by NSF's LIGO Laboratory which is a major facility fully funded by the National Science Foundation. LIGO was constructed by the California Institute of Technology and Massachusetts Institute of Technology with funding from the National Science Foundation and operates under cooperative agreement PHY-0757058. S.B. is supported by NASA through the NASA Hubble Fellowship grant HST-HF2-51524.001-A awarded by the Space Telescope Science Institute, which is operated by the Association of Universities for Research in Astronomy, Inc., for NASA, under contract NAS5-26555.
This research was supported in part through the computational resources and staff contributions provided for the Quest high performance computing facility at Northwestern University which is jointly supported by the Office of the Provost, the Office for Research, and Northwestern University Information Technology.
The author is grateful for computational resources provided by CIERA funded by NSF PHY-1726951 and by the Caltech LIGO Laboratory supported by NSF PHY-0757058 and PHY-0823459. This paper carries LIGO document number LIGO-P2500036.
\bibliography{references}

\appendix
\section{Strong resonances population comparison}
\label{ap:strong_resonances}
\begin{figure*}
  \centering
  \includegraphics[width=0.4\columnwidth]{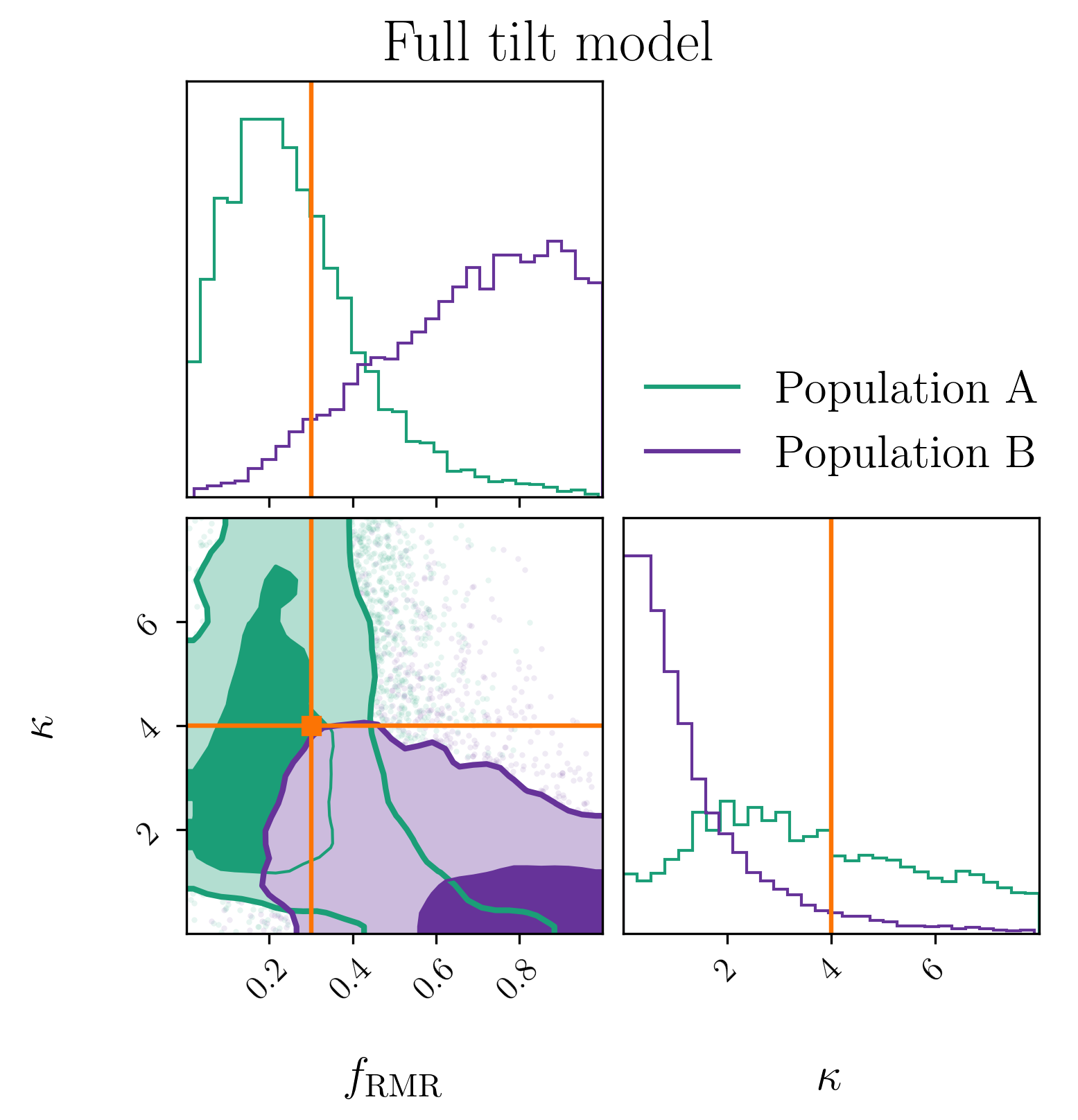}
  \includegraphics[width=0.4\columnwidth]{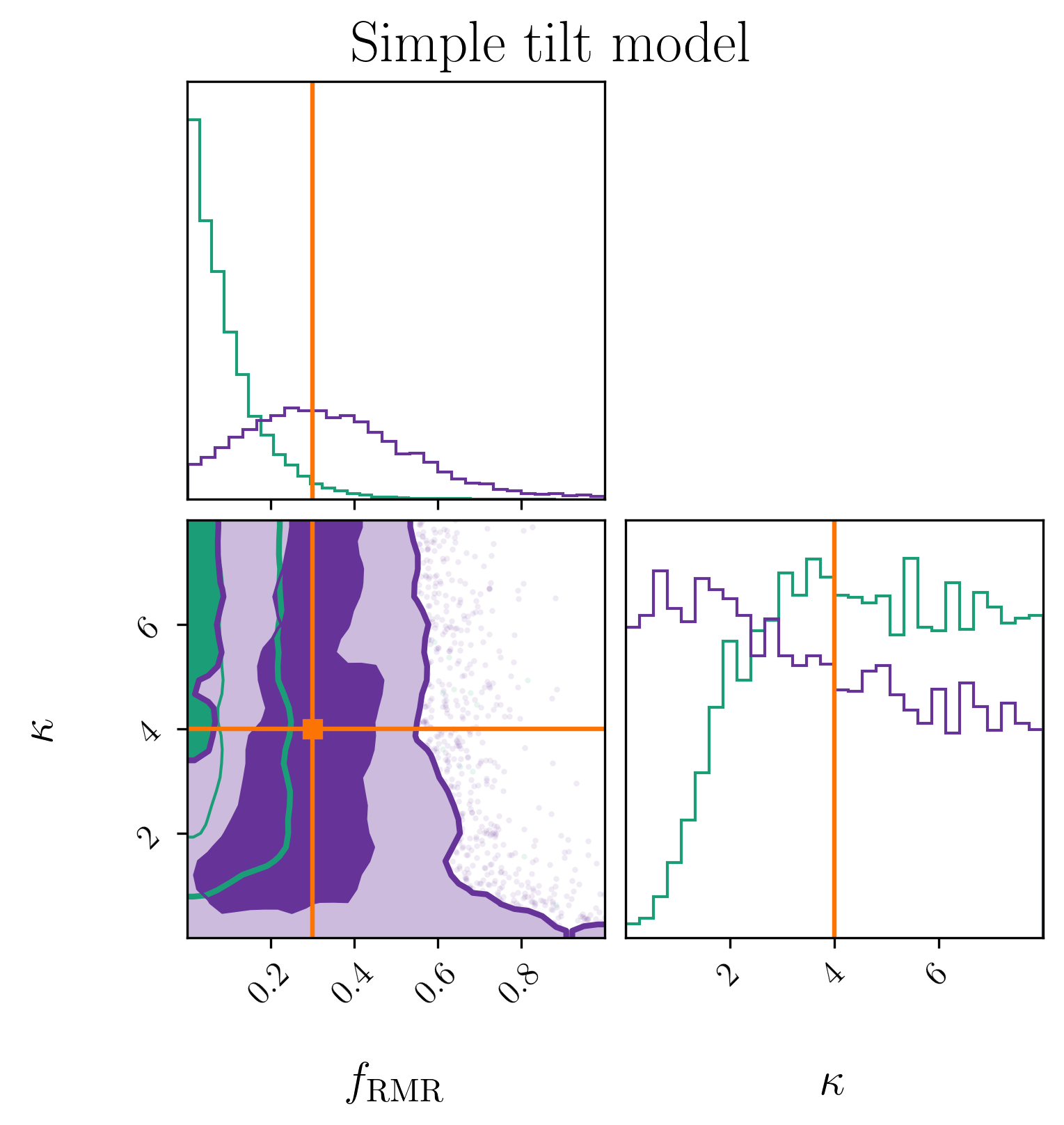}
  \caption{Corner plot of two independent sets of 200 events (green and purple) drawn from the strong resonances population analyzed with the \textsc{Full} (left) and \textsc{Simple} (right) tilt models showing the posteriors on the $\phi_{12}$ hyper-parameters $f_{\mathrm{RMR}}$ and $\kappa$. The orange lines denote the true parameter values chosen for this simulated population.}
  \label{fig:all_field_seed_comp}
\end{figure*}

In our initial analysis of a simulated population with strong resonances, we find that the \textsc{Full} tilt model leads to a more accurate recovery of the hyper-parameters governing the $\phi_{12}$ distribution. To check if this conclusion depends on the exact realization of 200 binary parameter draws from this population model, we simulate another independent population of 200 events drawn from the same distributions.
In Fig.~\ref{fig:all_field_seed_comp}, we show the $f_{\mathrm{RMR}}, \kappa$ hyper-parameter posteriors obtained for both tilt models for both populations. While for one population (dubbed ``Population A'' in Fig.~\ref{fig:all_field_seed_comp}), the true values are recovered at lower credibility with the \textsc{Full} tilt model, the inverse is true for the other simulated population. The preference for values of $f_{\mathrm{RMR}}$ at the edges of the prior interval $f_{\mathrm{RMR}} \in (0,1)$ may be driven by increased uncertainty in the hierarchical likelihood at these points in the parameter space; see Appendix~\ref{ap:injections} for further discussion. Thus, we conclude that the qualitative features of the hyper-parameter posteriors may vary due to the limited sample size of $\sim 200$ events expected to be detected during \ac{O4}. When we analyzed both populations together (400 total events), the posteriors qualitatively split the difference between the features of the two smaller populations, indicating that increasing the sample size will help reduce these fluctuations in the future.

\begin{figure*}
  \centering
  \includegraphics[width=0.45\columnwidth]{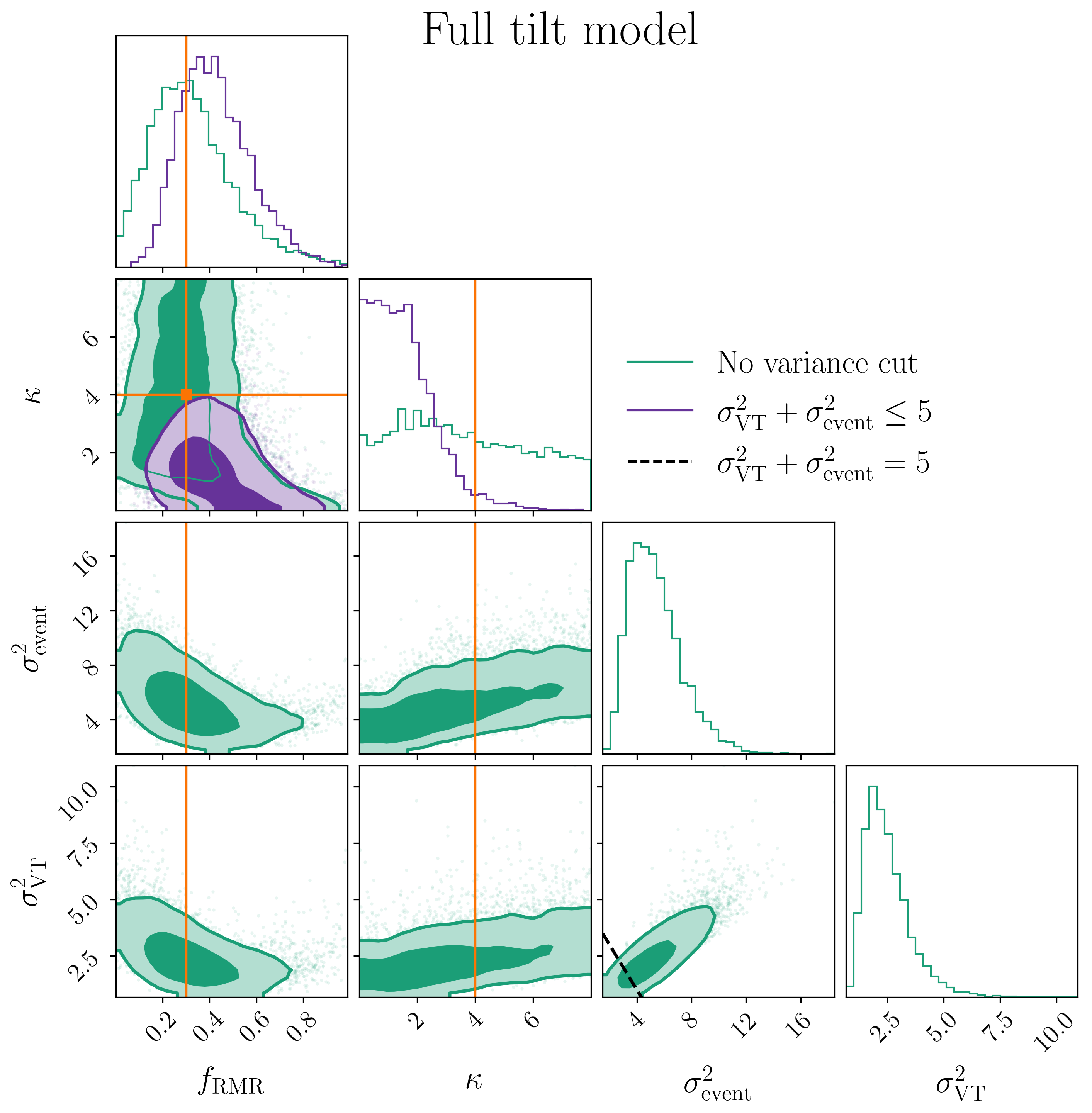}
  \includegraphics[width=0.45\columnwidth]{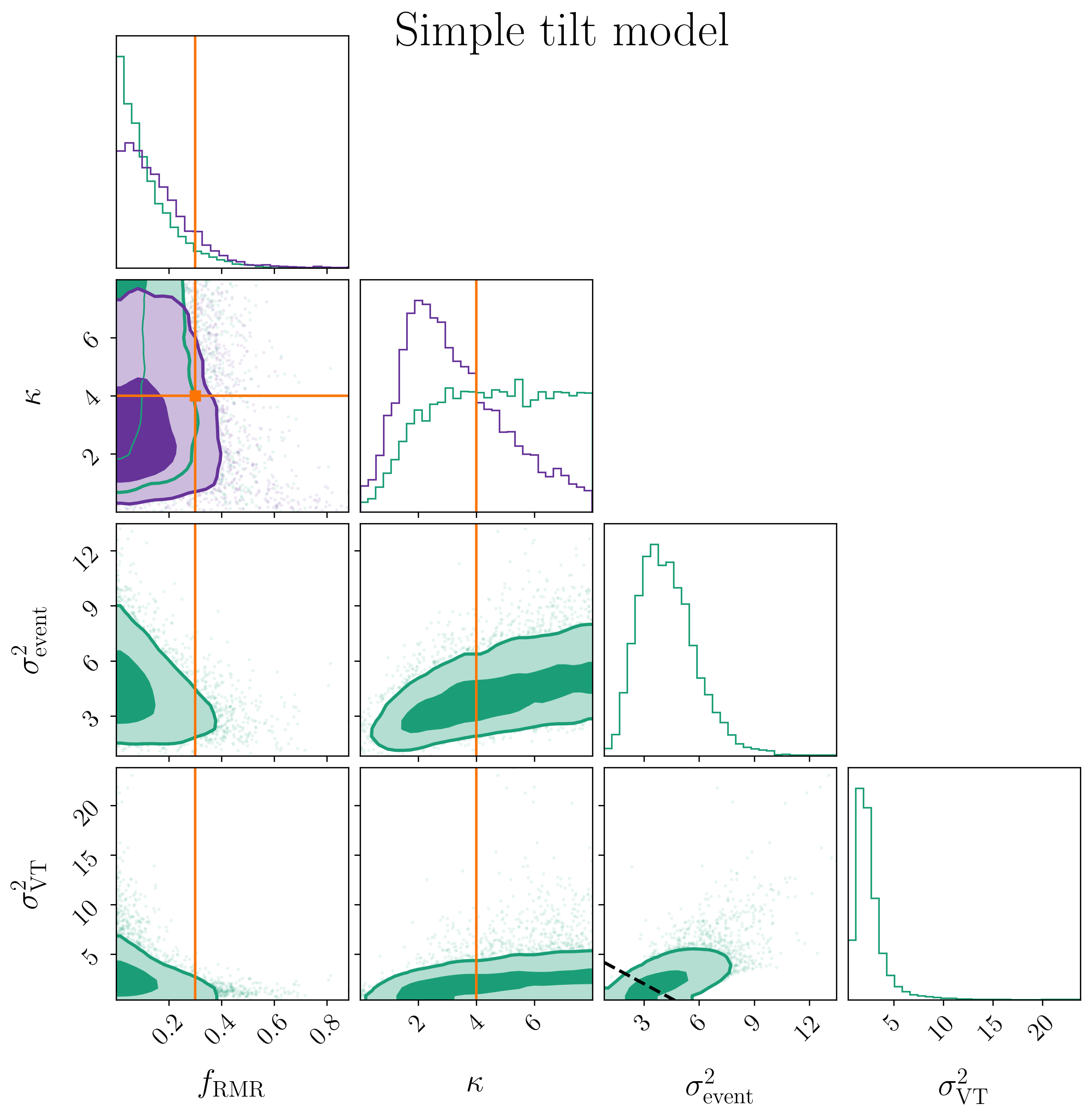}
  \caption{Corner plot of the strong resonances Population A analyzed with the \textsc{Full} (left) and \textsc{Simple} (right) tilt models showing the posteriors on the $\phi_{12}$ hyper-parameters $f_{\mathrm{RMR}}$ and $\kappa$ along with the \ac{MC} integral variances for both the individual-event posteriors and the sensitivity (VT) injection reweighting. The green shows the posteriors without a cut on the maximum total variance in the likelihood and the purple shows the posteriors obtained imposing $\sigma^{2}_{\mathrm{VT}} + \sigma^{2}_{\mathrm{event}} \leq 5$.}
  \label{fig:var_cut_comp}
\end{figure*}

\section{Sensitivity injections}
\label{ap:injections}
To account for selection effects in the recovery of our simulated populations, we perform a semi-analytic injection campaign to determine the sensitivity of our simulated detector network to \acp{BBH} with different properties. The hierarchical inference likelihood including selection effects can be written in terms of \ac{MC} integrals as
\begin{align}
\mathcal{L}(\{d\} | \boldsymbol{\Lambda}, \mathrm{det}) &= \frac{1}{\alpha(\boldsymbol{\Lambda})^{N_{\mathrm{det}}}}\prod_{n}^{N_{\mathrm{det}}}\mathcal{Z}(d_{n} | \mathrm{PE})\sum_{k}^{N_n}\frac{\pi(\boldsymbol{\theta}_{k}|\boldsymbol{\Lambda})}{\pi(\boldsymbol{\theta}_{k}|\mathrm{PE})}, \\
 \alpha(\boldsymbol{\Lambda}) &= \frac{1}{N_{\mathrm{tot}}}\sum_{j}^{N_{\mathrm{found}}}\frac{\pi(\boldsymbol{\theta}_{j}|\boldsymbol{\Lambda})}{p_{\mathrm{true}}(\boldsymbol{\theta}_{j})}.\label{eq:VT_integral}
\end{align}
In this expression, $\pi(\boldsymbol{\theta}_{k} | \boldsymbol{\Lambda})$ is the population model for the binary parameters $\boldsymbol{\theta}$ characterized by hyper-parameters $\boldsymbol{\Lambda}$, $\pi(\boldsymbol{\theta}_{k}|\mathrm{PE})$ is the prior on the binary parameters used during the initial individual-event parameter estimation step, $\mathcal{Z}$ is the Bayesian evidence obtained during this step, and $\alpha(\boldsymbol{\Lambda})$ is the detectable fraction of events drawn from a population characterized by hyper-parameters $\boldsymbol{\Lambda}$. The index $k$ denotes the individual-event posterior sample out of $N_{n}$ total samples for each event, $n$ is the event within the total population of $N_{\mathrm{det}}$ detected events, and $j$ denotes the sensitivity injection out of $N_{\mathrm{found}}$ total injections found in our injection campaign, imposing a detection threshold of $\mathrm{SNR_{net, mf}} \geq 9$ in a Hanford-Livingston network at their projected \ac{O4} sensitivities~\citep{O4_psds}. 

The \ac{BBH} injections are drawn from the following mass and spin distributions:
\begin{widetext}
\begin{align}
p_{\mathrm{true}}(m_{1}, q, \chi_{1}, \chi_{2}, \cos\theta_{1}, \cos\theta_{2}) &= \frac{n_{\mathrm{U}}}{N_{\mathrm{tot}}} \mathrm{U}(m_1; 4~M_{\odot}, 100~M_{\odot})\mathrm{U}^{2}(\cos\theta; -1, 1) \\ \nonumber
&+ \frac{n_{\mathrm{PL}}}{N_{\mathrm{tot}}}\mathrm{PL}(m_1; \alpha=-3.5)\mathrm{U}^{2}(\cos\theta)\\ \nonumber
&+ \frac{n_{\mathrm{PLTG}}}{N_{\mathrm{tot}}}\mathrm{PL}(m_1; \alpha=-3.5)\mathcal{N}_{t}^{2}(\cos\theta | \mu=1, \sigma=0.5)\\ \nonumber
&+ \frac{n_{\mathrm{PLTGM}}}{N_{\mathrm{tot}}}\mathrm{PL}(m_1; \alpha=-3.5)\mathcal{N}_{t}^{2}(\cos\theta | \mu=1, \sigma=0.5)\mathcal{N}_{t}^{2}(\chi | \mu=0, \sigma=0.3, 0, 0.99)\\
N_{\mathrm{tot}} &= n_{\mathrm{U}} + n_{\mathrm{PL}} + n_{\mathrm{PLTG}} + n_{\mathrm{PLTGM}}\\
n_{\mathrm{U}} = 22406657, \quad n_{\mathrm{PL}} &= 564571237, \quad n_{\mathrm{PLTG}} = 504366682, \quad n_{\mathrm{PLTGM}} = 461987020.
\end{align}
\end{widetext}
In the expression above, $\mathrm{U}(x)$ represents a uniform distribution, $\mathrm{PL}(x)$ is a power-law distribution, and $\mathcal{N}_{t}^{2}(x)$ is a truncated Gaussian distribution. In each case, the hyper-parameters of the distribution are written explicitly, and the bounds are passed as the last two values in parentheses. We distribute the sensitivity injections uniformly in source-frame time and comoving volume and use standard priors for the remaining extrinsic parameters~\cite[e.g.,][]{Romero-Shaw:2020owr}.

This distribution was determined empirically through an iterative process where we sought to minimize the spurious effects of large \ac{MC} integral variance on the resulting hyper-parameter posterior, particularly for the \textsc{Full} tilt model with a sharp discontinuity. For the results shown in the main text, we do not impose a limit on the variance of the hierarchical likelihood~\citep{Talbot:2023pex}. Instead, we use the default behavior of \textsc{GWPopulation}, which enforces the criterion proposed in \citet{Farr_2019} by rejecting samples in hyper-parameter space that do not pass the accuracy requirement $N_{\mathrm{eff}} > 4N_{\mathrm{det}}$, where $N_{\mathrm{eff}}$ is the number of effective samples going into the integral and $N_{\mathrm{events}}=200$ four our simulated populations.

The likelihood variances that we obtain using this injection set are roughly of order unity for both tilt models for the strong resonances population; the total variance can reach $\mathcal{O}(10)$, particularly for the \textsc{Full} tilt model. The variance contributed by the \ac{MC} integral over individual-event posterior samples dominates the uncertainty over the variance contributed by the selection effects integral, as the posteriors for each event only include about $\sim 5700$ samples.
In order to determine the effect of the \ac{MC} integral uncertainty on the hyper-parameter posteriors, we repeat the inference on a subset of the simulated populations and tilt models imposing a limit on the total variance, $\sigma^{2}_{\mathrm{tot}} = \sigma^{2}_{\mathrm{VT}} + \sigma^{2}_{\mathrm{event}} \leq 5$. A comparison of the posterior obtained on the $\phi_{12}$ hyper-parameters $f_{\mathrm{RMR}}$ and $\kappa$ is shown in Fig.~\ref{fig:var_cut_comp}. 

From the correlations shown in Fig.~\ref{fig:var_cut_comp}, it is clear that the variance increases for smaller values of $f_{\mathrm{RMR}}$ and larger values of $\kappa$. This can be understood as a decrease in the number of effective samples going into the \ac{MC} integral as the \ac{VM} distribution becomes narrower ($\kappa$ increases) and more strongly peaked at one specific value (the full population is in the standard mass ratio configuration). In the case of the \textsc{Simple} tilt model inference, the variance cut seems to improve the inference of $f_{\mathrm{RMR}}$, pulling the posterior away from zero by reducing the support in this region of the parameter space due to spuriously large likelihood values. However, the variance cut has a negative effect on the inference of $\kappa$ for the $\textsc{Full}$ tilt model, incorrectly excluding the large-$\kappa$ regions of parameter space with narrow $\ac{VM}$ distributions. The variation in the posteriors with the imposition of different \ac{MC} integral convergence criteria indicates that the population models chosen in this work push the limits of the reliability of this method of hierarchical inference. We leave the exploration of alternative approaches (e.g., using density estimators, emulators, or interpolants~\citep{Wysocki:2020myz, Landry:2018prl, DEmilio:2021laf, Golomb:2021tll, Callister:2024qyq}) for future work.

\section{Mass and spin magnitude inference}
\subsection{Masses}
\label{ap:masses}
\begin{figure*}
  \centering
  \includegraphics[width=0.66\columnwidth]{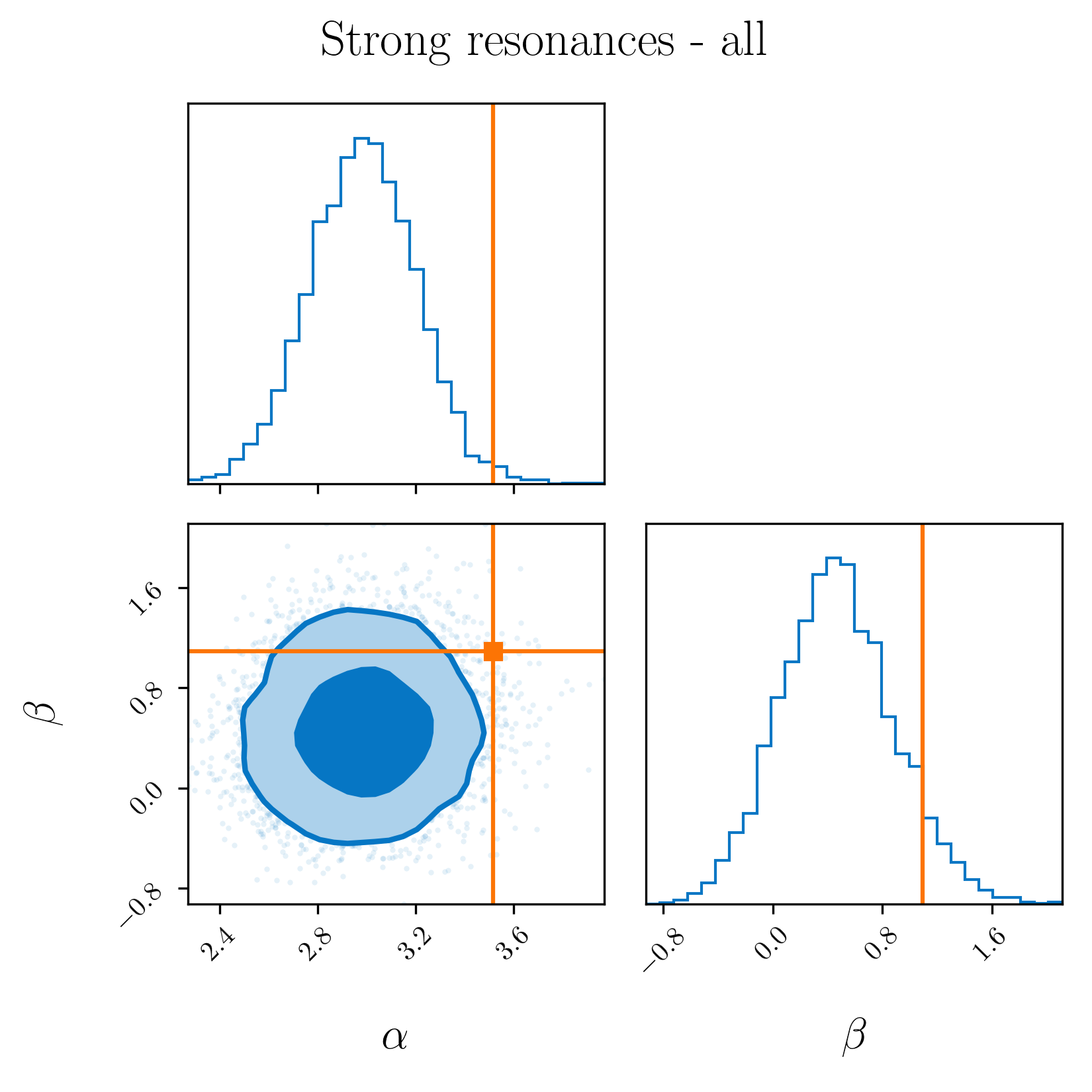}
  \includegraphics[width=0.66\columnwidth]{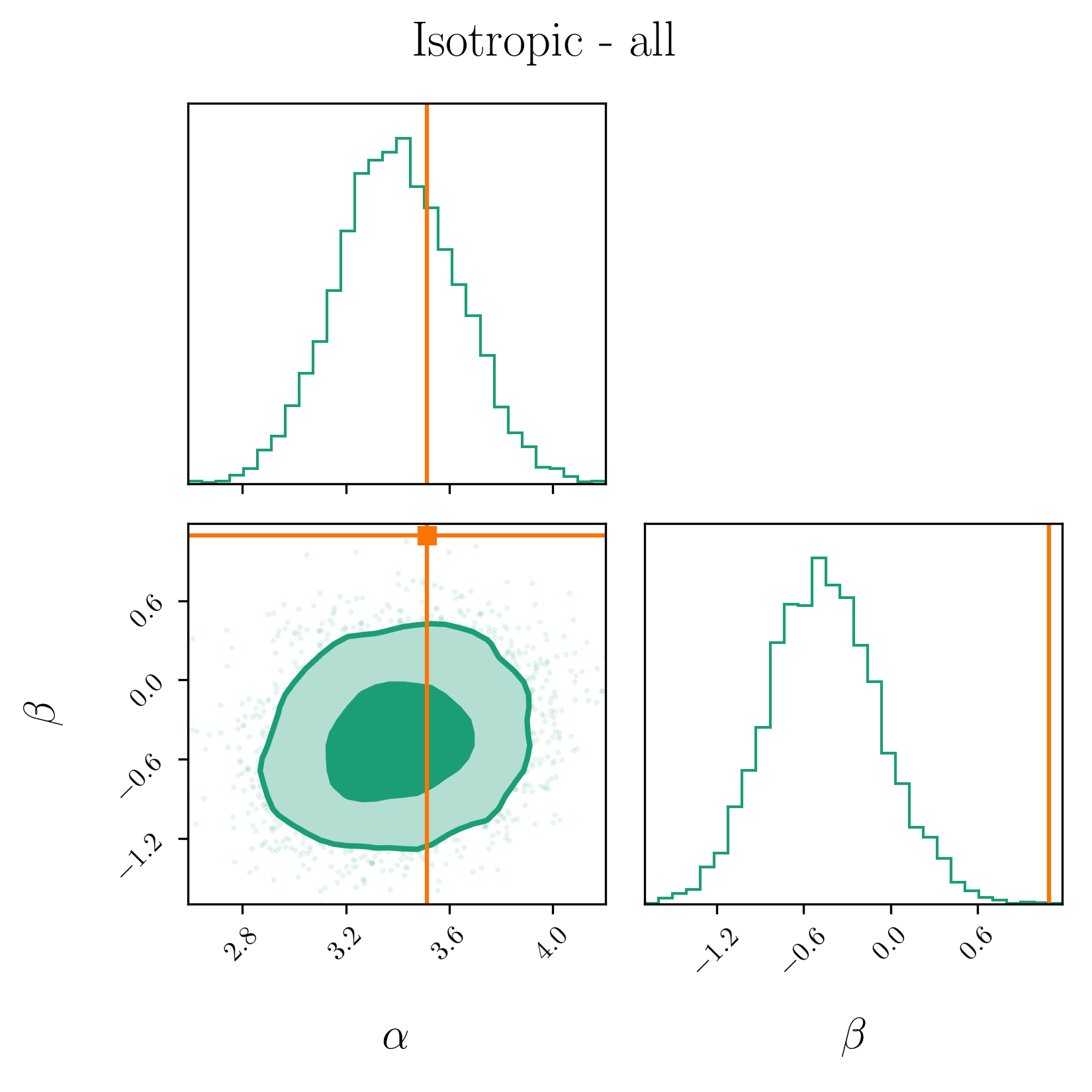}
  \includegraphics[width=0.66\columnwidth]{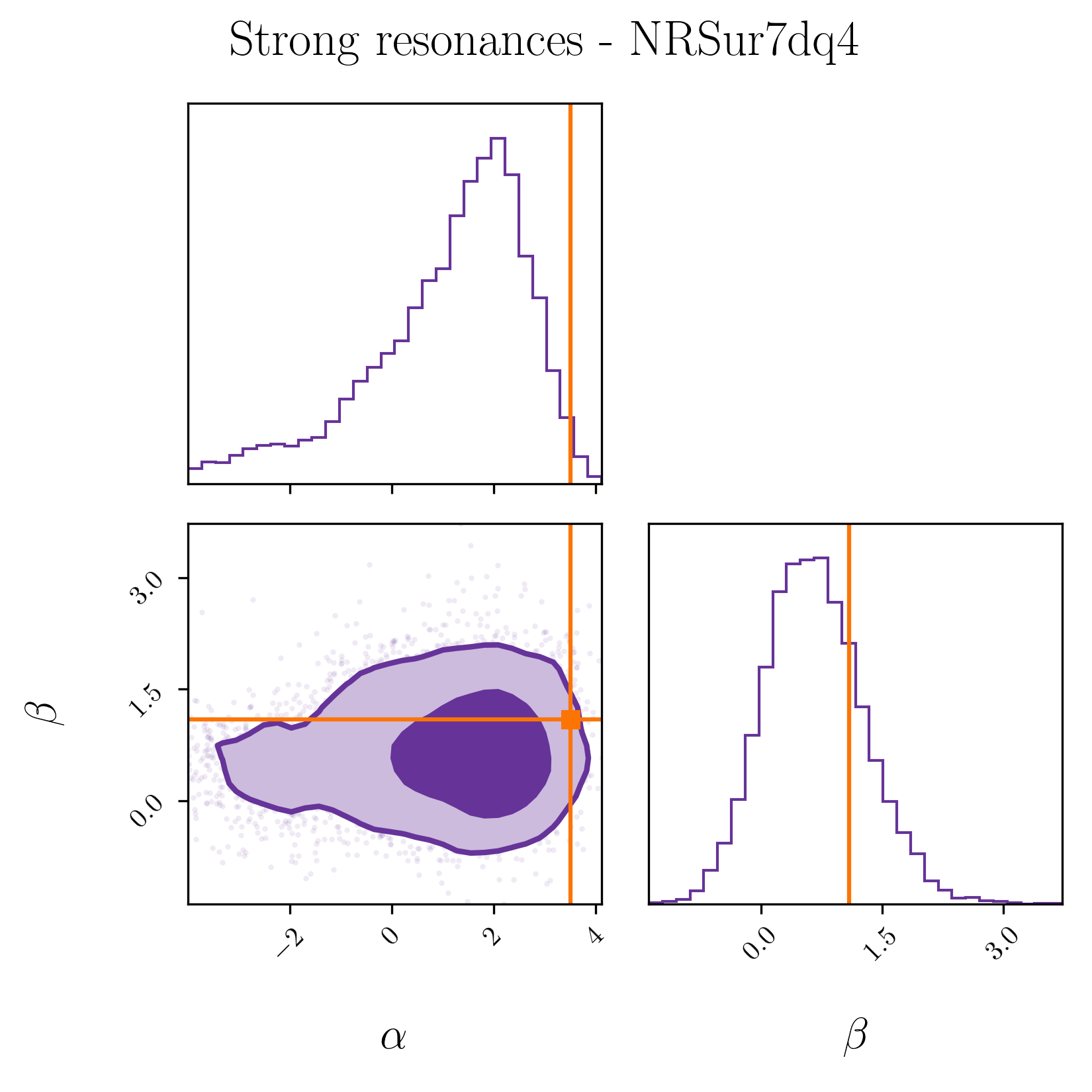}  
  \caption{Corner plot of the mass power-law parameters $\alpha$ ($m_{1}$) and $\beta$ ($q$) for the Strong resonances (all events - left; NRSur7dq4 event only - right) and Isotropic (middle) populations both recovered with the \textsc{Simple} tilt model. While the results of the analyses including all events generated with both NRSur7dq4 and IMRPhenomXPHM demonstrate a biased recovery of the mass power-law indices, the results of analyzing only the events generated with one waveform at a time are unbiased.}
  \label{fig:mass_corners}
\end{figure*}

When analyzing each of the full populations of 200 events including sources generated with both the NRSur7dq4 and IMRPhenomXPHM waveforms, we generally obtain a biased recovery of the power-law hyper-parameters governing the mass distribution. However, the mass distribution recovery is unbiased when restricting the population to only the events generated with one waveform model at a time. The posteriors on the mass power-law parameters $\alpha$ ($m_{1}$) and $\beta$ ($q$) are shown in Fig.~\ref{fig:mass_corners}. While we only compare the mixed-waveform results to the NRSur7dq4 results in the figure, we generally find an unbiased recovery of the mass power-law indices when analyzing only the IMRPhenomXPHM-generated subset of each population as well.

This can likely be explained by the inconsistency in the choice of waveform between the source simulation and individual-event parameter estimation introduced by our simulation pipeline. Our simulation pipeline is designed to analyze an event whose source waveform was generated with NRSur7dq4 only with the NRSur7dq4 waveform at the individual-event parameter estimation stage (the same goes for the events whose source waveforms were generated with IMRPhenomXPHM). However, the posteriors for sources with masses near the edge of the regime of validity of NRSur7dq4 span the region where individual waveforms should be generated with both approximants. We effectively throw away the parts of the prior space where IMRPhenomXPHM would be used when NRSur7dq4 is invalid instead of employing XPHM in those parts of the space. This choice is made to maintain the computational feasibility of the inference, as mixing between waveforms during an individual-event parameter estimation analysis is a technique still under active development~\citep{Ashton:2021cub, Hoy:2024vpc}.

This analysis choice is impossible for real events, because we cannot know a priori if the true source parameters fall in the regime of validity of one waveform versus another, as the true source parameters are unknown. Thus, this inconsistency in our simulation pipeline corresponds to an unphysical directed acyclic graph (DAG)~\citep{Pearl:2009}, which has previously been shown to introduce biases in hierarchical inference~\citep{Essick:2023upv}. For real events, a physical DAG could be maintained by conditioning the choice of waveform during the parameter estimation step on the maximum-likelihood source parameters returned by matched-filter search pipelines. 

In this work, we choose to prioritize minimizing the computational cost associated with running a full matched-filter search on our simulated events at the expense of a physical DAG. The bias in the mass distribution does not affect the spin angle inference, which is our main focus. We do maintain physical consistency in the selection effects step of the pipeline, generating and down-selecting the sensitivity injections described in the previous Appendix~\ref{ap:injections} using the same waveform choice mechanism as the analyzed events in each simulated population. However, an inconsistency is introduced in our choice of waveform models for the GWTC-3 analysis, as the waveform used for the sensitivity injections performed by the LVK (SEOBNRv4PHM~\citep{Ossokine:2020kjp}), which we use to account for selection effects, does not match the models used for the individual-event parameter estimation (IMRPhenomXPHM and NRSur7dq4 when available). The detectability of a given source is unlikely to depend significantly on the waveform model used, so this formal source of bias should not affect our results.

\subsection{Spin magnitudes}
\label{ap:spin_mag}
\begin{figure}
  \centering
  \includegraphics[width=\columnwidth]{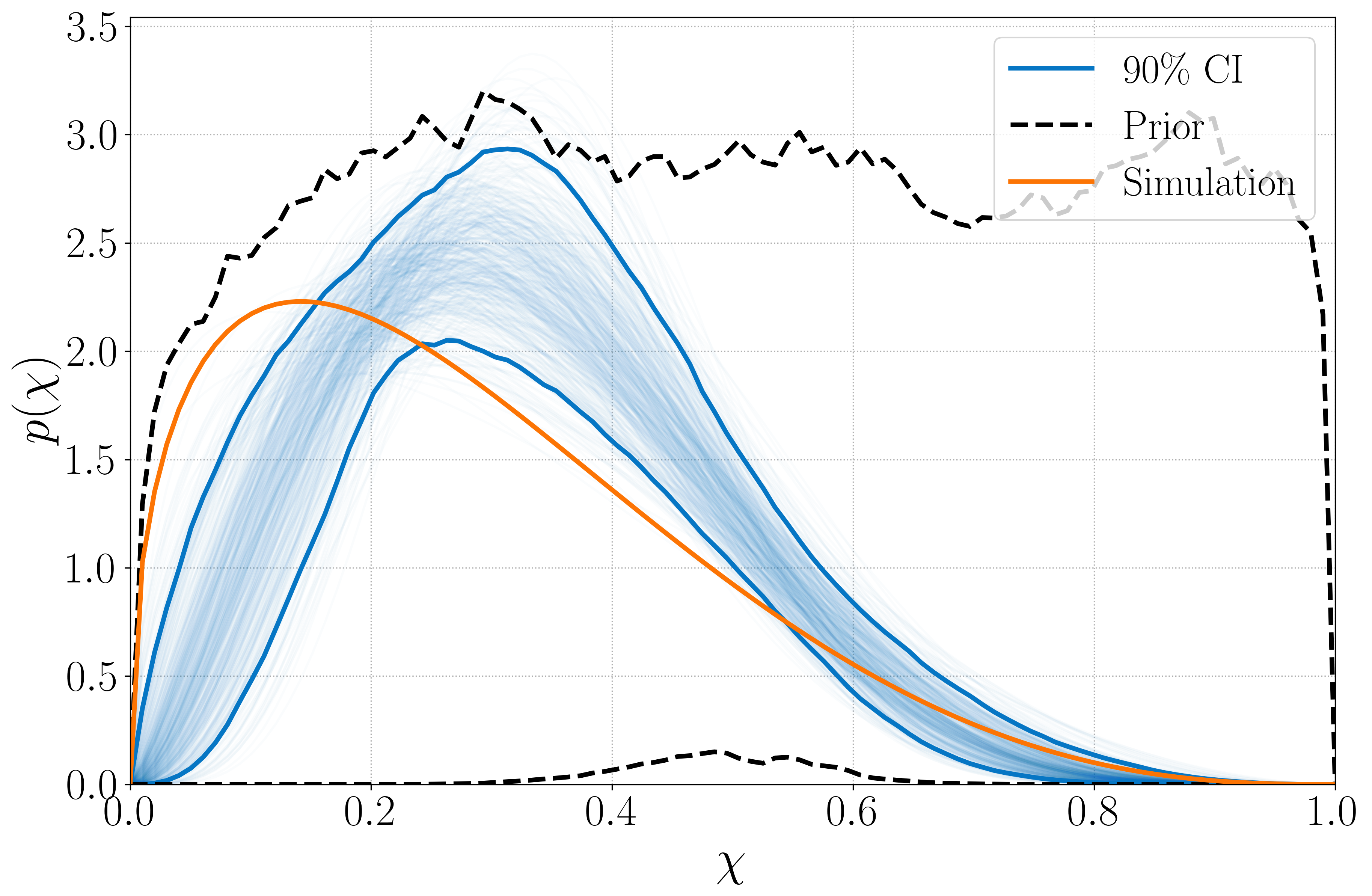}
  \caption{Inferred spin magnitude distribution for the strong resonances population under the \textsc{Simple} tilt model. Individual light blue traces show the distributions corresponding to individual hyper-parameter posterior samples, the dark blue lines bound the 90\% posterior credible interval, and the dashed black lines bound the 90\% prior credible interval. The true simulated distribution is shown in orange.}
  \label{fig:all_field_old_seed_spin_mag}
\end{figure}

\begin{figure*}
  \centering
  \includegraphics[width=\textwidth]{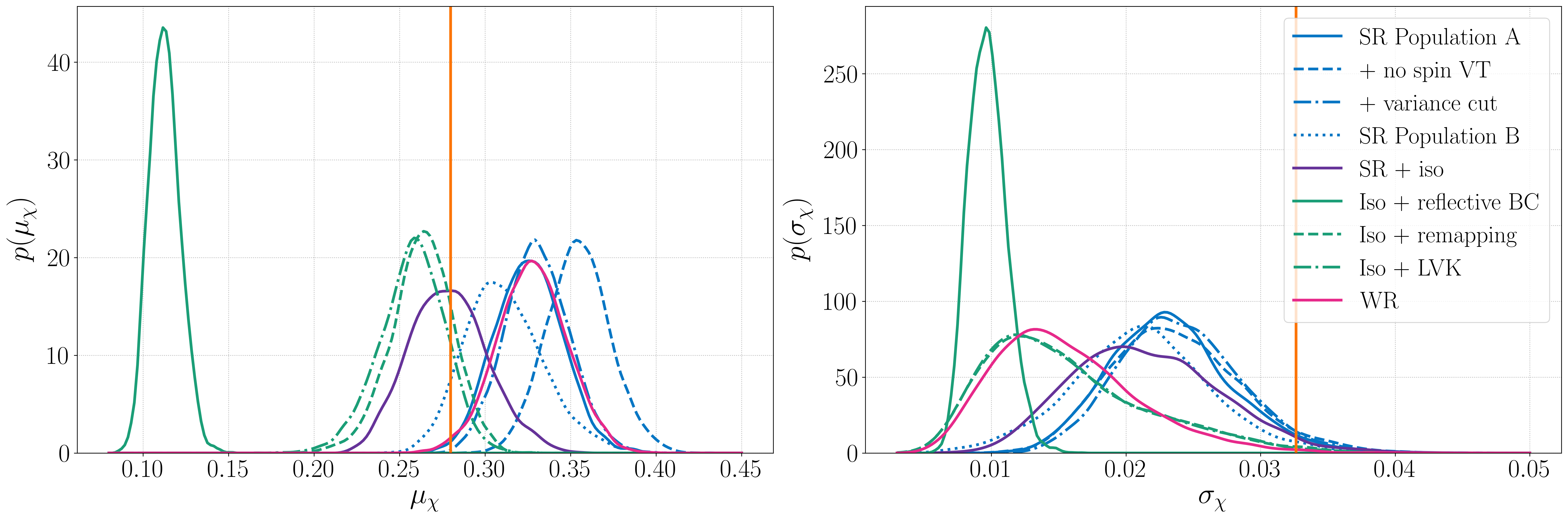}
  \caption{Kernel density estimates of the posterior distributions on the spin magnitude hyper-parameters $\mu_{\chi}, \sigma_{\chi}$ for a variety of different analyses. All results are shown using the full population of 200 events and the \textsc{Simple} tilt model. The strong resonances population is shown in blue, the strong resonances + isotropic population in purple, and the isotropic population in green. The true values are given by the orange lines.}
  \label{fig:spin_mag_hist_comp}
\end{figure*}

The recovery of the spin magnitude hyper-parameters is consistently biased for all of our hierarchical inference analyses. The inferred Beta distribution shown in Fig.~\ref{fig:all_field_old_seed_spin_mag} for the strong resonances Population B is characterized by a width that is too narrow and a mean that is too high (see posteriors in Fig.~\ref{fig:spin_mag_hist_comp}). To verify whether this is a general issue or just dependent on the random seed used to generate the binary parameters for this particular set of 200 simulated events, we change the random seed to generate the independent Population A described in Section~\ref{sec:strong_resonances}. The posterior on $\mu_{\chi}$ peaks closer to the true value than for Population B, but the same bias is present. While restricting the analysis to a single waveform ameliorated the mass hyper-parameter bias discussed above in Appendix~\ref{ap:masses}, the spin magnitude bias persists when we independently analyze the NRSur7dq4 and IMRPhenomXPHM subsets of both strong resonances Populations A and B. The biases are generally less severe for the IMRPhenomXPHM-only analyses, but this is likely because that subset of the population has approximately half as many events as the NRSur7dq4 subset.

One possible explanation for this bias is the uncertainty in the selection effects Monte Carlo integral described in Appendix~\ref{ap:injections}. Previous works have found that for GWTC-3-sized catalogs, the recovery of the spin magnitude distribution is particularly sensitive to using insufficient sensitivity injections in the selection effects integral (Eq.~\ref{eq:VT_integral})~\citep{Talbot:2023pex, Golomb:2022bon}. Conversely, these same studies find that the spin magnitude does not strongly affect the detectability of a particular source; they obtain unbiased hierarchical inference results when omitting the spins from the selection effects integral. When implementing this as a possible solution to the spin magnitude bias, we recover $\mu_{\chi}$ posteriors that peak even higher (stronger bias) than our original results. While the effect of spin magnitude on detectability is weak compared to the mass, systems with larger (aligned) spin magnitudes are generally easier to detect. When this is not accounted for in the analysis, a preference for larger spin magnitudes is recovered, particularly for the larger catalog size considered in this work.

We additionally compare the posteriors on the spin magnitude hyper-parameters when various \ac{MC} integral convergence criteria are used. As explained in Appendix~\ref{ap:injections}, the default analyses in this work do not impose any convergence criteria in the form of a maximum likelihood variance. We find the same bias when a cut on the total variance, $\sigma_{\mathrm{tot}}^{2} < 5$, is imposed. Finally, we repeat the analysis with no statistical uncertainty on the parameters of the individual events; we use only the true values of the binary parameters rather than the full posterior, finding unbiased results in this case. These tests indicate that the bias likely stems from an issue with the individual-event posterior samples rather than the way that selection effects are accounted for in the analysis.

Because the black hole spins for this choice of hyper-parameters are generally small, one potential explanation for this biased recovery is that the sampler cannot sufficiently explore the small-spin region at the edge of the spin magnitude prior space during the individual-event analysis. While the sampler settings used in this work were tested and shown to produce robust results using standard performance metrics, like probability-probability (PP) plots~\cite[e.g.,][]{Romero-Shaw:2020owr}, we have previously found that a stealth bias can appear in hierarchical inference results even when sampler performance is sufficient to produce diagonal PP plots~\citep{Biscoveanu:2021eht}. 

In an attempt to improve the sampler coverage of the small-spin region, we initially analyzed the isotropic population with reflective boundary conditions imposed for the uniform spin magnitude priors used in the individual-event parameter estimation stage. Rather than rejecting a proposed point outside the unit cube to which the prior transform is applied during nested sampling, the prior transform is applied to the reflected value of this point across the prior boundary. This breaks detailed balance~\cite[e.g.,][]{Suwa:2010} but can improve sampling efficiency for posteriors peaking near the prior edge. However, we find that the reflective boundary conditions actually lead to the opposite kind of bias in the $\mu_{\chi}$ posterior from the original results; it peaks at very low values rather than too high. The reflective boundary conditions cause the sampler to overcompensate and spend too much time exploring the small-spin part of the parameter space.

\begin{figure}
  \centering
  \includegraphics[width=\columnwidth]{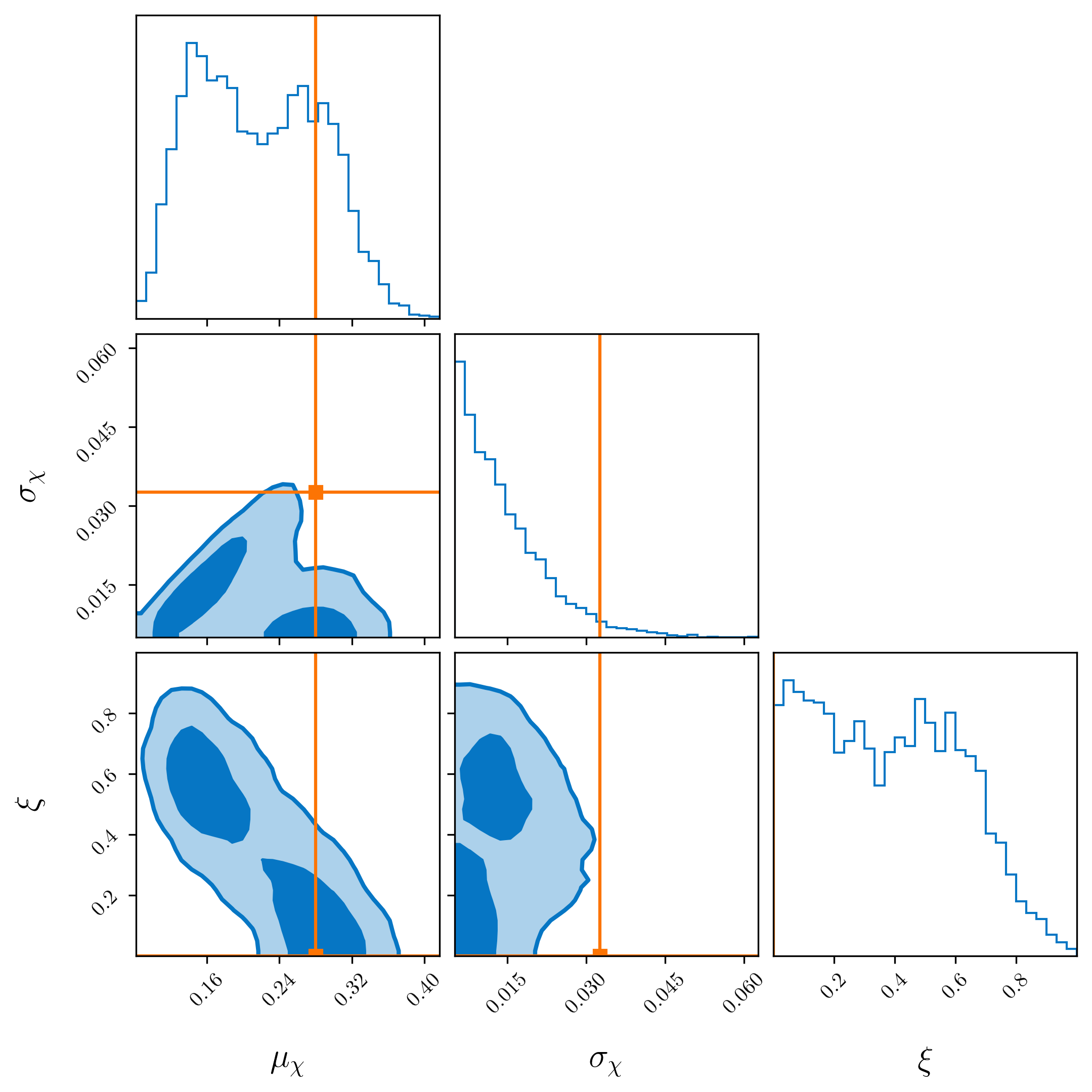}
  \caption{Corner plot of the posteriors on the spin magnitude hyper-parameters, $\mu_{\chi}$ and $\sigma_{\chi}$, and the isotropic vs. aligned tilt mixing fraction, $\xi$, for one of the isotropic downsampled GWTC-2-sized catalogs consisting only of 44 NRSur7dq4 events analyzed individually with reflective boundary conditions on the spin magnitude priors using the \textsc{Simple} tilt model. The shaded regions bound the 50\% and 90\% credible intervals, and the true simulated values are shown in orange.}
  \label{fig:iso_spin_mag_correlation}
\end{figure}

The significant bias towards small $\mu_{\chi}$ values introduced by the reflective boundary conditions reveals a strong correlation between the spin magnitude hyper-parameters and the isotropic vs. aligned tilt mixing fraction, $\xi$. In Fig.~\ref{fig:iso_spin_mag_correlation}, we show the posteriors on these three hyper-parameters for one of the GWTC-2-sized catalogs generated from the isotropic population when only the NRSur7dq4 events are analyzed, described in Section~\ref{sec:gwtc2_test}. The posterior on $\mu_{\chi}$ is multi-modal; one of the peaks corresponds to the low value recovered in the analyses of all the other randomly downsampled catalogs and the full set of 200 events, while the other peak corresponds to the true value of $\mu_{\chi} = 0.280$. Because of the correlation, the smaller values of $\mu_{\chi}$ preferred by the analysis with reflective boundary conditions lead to a spurious preference for larger values of $\xi \sim 0.6$. 

This correlation can be explained in terms of the effect of these hyper-parameters on the resulting distribution of $\chi_{\mathrm{eff}}$. Because $\chi_{\mathrm{eff}}$ is better constrained for individual events than either of the component spin magnitudes and tilts, the distinct probability modes with posterior support conspire to produce similar $\chi_{\mathrm{eff}}$ distributions. Smaller spin magnitude values require more aligned spin tilts (higher $\xi$) to reproduce the same $\chi_{\mathrm{eff}}$ distribution as that implied by a spin magnitude distribution peaked at larger values but paired with more misaligned tilts. The $\xi$ and $\sigma_{\chi}$ posteriors are positively correlated; when the width of the spin magnitude distribution increases, the spin tilt distribution can be more peaked (higher $\xi$) and still produce the same $\chi_{\mathrm{eff}}$ distribution.

Given that the use of reflective boundary conditions introduced a more severe bias in the spin magnitude hyper-parameter recovery in the opposite direction from the original bias, we re-analyzed the isotropic population without the reflective boundary conditions. Nested samplers like \textsc{Dynesty} traditionally generate samples from the prior by drawing points from a unit cube, $x$, and then rescaling by the inverse of the prior \ac{CDF}: $\chi(x) = \mathrm{CDF}^{-1}(x)$. Akin to the remapping proposed in \citet{Biscoveanu:2021eht} for mass ratio, we instead implement a two-step rescaling, moving the lower edge of the spin magnitude prior to the middle of the unit cube,
\begin{align}
u = 2\max(x, 1 - x) - 1,\\
\chi = \mathrm{CDF}^{-1}(u).
\end{align}
Unlike the reflective boundary conditions, this remapping does not break detailed balance during sampling. However, we recover posteriors on $\mu_{\chi}, \sigma_{\chi}$ similar to those obtained for the other simulated populations with the standard uniform prior, so we do not re-analyze any of the other populations with this remapping given the high computational cost of repeating the individual-event parameter estimation. 

\begin{deluxetable*}{lccccccr}
\tablewidth{\textwidth}
\tablecaption{Results summary table for all analyses performed in this work on the GWTC-3 \ac{BBH} population\tablenotemark{a}\label{tab:gwtc3_results}}
\tablehead{
\colhead{Tilt Model} & 
\colhead{$\mu_{\mathrm{RMR/SMR}}$} & 
\colhead{$f_{\mathrm{RMR}}$} & 
\colhead{$\mathrm{CI}_{95, f_{\mathrm{RMR}}}$} & 
\colhead{$\kappa$} & 
\colhead{$\mathrm{CI}_{95, \kappa}$} & 
\colhead{$\xi$} & 
\colhead{$\mathrm{CI}_{95, \xi}$}
}
\startdata
\textsc{simple-wr} & \checkmark & $0.60^{+0.40}_{-0.54}$ & 0.94 & $6.77^{+1.23}_{-6.34}$ & 7.58 & $0.55^{+0.45}_{-0.38}$ & 0.83 \\
\textsc{simple} & \checkmark & $0.60^{+0.40}_{-0.54}$ & 0.94 & $5.93^{+1.59}_{-5.93}$ & 7.52 & $0.43^{+0.49}_{-0.38}$ & 0.87 \\
\textsc{simple-wr} & -- & $0.04^{+0.86}_{-0.04}$ & 0.90 & $6.80^{+1.20}_{-6.34}$ & 7.54 & $0.58^{+0.42}_{-0.40}$ & 0.81 \\
\textsc{simple} & -- & $0.00^{+0.91}_{-0.00}$ & 0.91 & $4.60^{+3.40}_{-4.10}$ & 7.50 & $0.35^{+0.61}_{-0.27}$ & 0.88 \\
\textsc{full} & \checkmark & $0.13^{+0.81}_{-0.13}$ & 0.94 & $4.19^{+3.33}_{-4.19}$ & 7.52 & $0.54^{+0.40}_{-0.47}$ & 0.87 \\
\textsc{full} & -- & $0.06^{+0.82}_{-0.06}$ & 0.88 & $6.17^{+1.83}_{-5.72}$ & 7.54 & $0.45^{+0.47}_{-0.40}$ & 0.87 \\
\textsc{lvk} & -- & $0.00^{+0.84}_{-0.00}$ & 0.84 & $6.74^{+1.26}_{-6.20}$ & 7.46 & $0.97^{+0.03}_{-0.74}$ & 0.77 \\
\textsc{lvk-free} & -- & $0.00^{+0.88}_{-0.00}$ & 0.88 & $6.61^{+1.39}_{-6.17}$ & 7.56 & $0.95^{+0.05}_{-0.72}$ & 0.77 \\
Prior & -- & $0.50^{+0.47}_{-0.47}$ & 0.95 & $4.00^{+3.80}_{-3.80}$ & 7.60 & $0.50^{+0.47}_{-0.47}$ & 0.95 \\
\enddata
\tablenotetext{a}{The \textsc{simple-wr} model corresponds to the \textsc{Simple} tilt model with the priors on the tilt hyper-parameters that we used for the simulated weak resonances population, and the \textsc{lvk-free} model is the flexible variation of the \textsc{Default} LVK model where the peak of the aligned component of the distribution is treated as a free parameter. For the $f_{\mathrm{RMR}}, \kappa, \xi$ hyper-parameters governing the $\phi_{12}$ distribution, we report the maximum posterior value and bounds and width of the 95\% credible interval calculated using the highest posterior density method.}
\end{deluxetable*}

We also verify whether our choice of tilt model could potentially lead to biases in the spin magnitude hyper-parameters by analyzing the isotropic population with the \textsc{Default} \ac{LVK} spin model. The isotropic population we simulate is fully consistent with the \ac{BBH} population properties inferred by the LVK using GWTC-3 data, meaning there is no mismodeling between the simulation and the recovery. However, we recover a similar level of bias to the results obtained with the \textsc{Simple} and \textsc{Full} tilt models explored in the rest of this work. We generally find no significant difference in the spin magnitude hyper-parameter posteriors regardless of the tilt model used. Similar biases are recovered when analyzing only the NRSur7dq4 subset of each population and when analyzing the full population. 

In a final attempt to ameliorate the spin magnitude bias, we launched reruns of the individual-event parameter estimation for the strong resonances + isotropic population but increased the sampling argument \texttt{maxmcmc} from 5000 to 20000. In the sampling mode recommended for parameter estimation of individual compact binary coalescences implemented in \textsc{Bilby} (\texttt{acceptance-walk}), the sampler uses a random walk based on standard \ac{MCMC} techniques~\citep{Metropolis:1953am, Hastings:1970aa} to generate samples from the bounded slices of the posterior distribution used in each iteration of the nested sampling algorithm~\cite[see][for more details on the nested sampling implementation]{Skilling:2004pqw, Speagle:2019dynesty}. Each walk must be long enough such that \texttt{naccept} points would be accepted by the nested sampler---meaning each accepted point has a larger likelihood than the live point with the lowest likelihood---but the number of steps in the walk is capped at \texttt{maxmcmc}. We use \texttt{naccept}=60 in our analyses. If the sampler is routinely hitting \texttt{maxmcmc} without reaching the specified \texttt{naccept}, this can indicate that the start point of the walk (a random live point) and the ending point (that eventually replaces the lowest-likelihood live point) are correlated. This violates one of the fundamental assumptions underpinning nested sampling---that each of the proposed live points is independent---which can potentially lead to biases in the resulting posterior due to under-sampling. When we increased \texttt{maxmcmc} to 20000, the sampler was still hitting this value for many events, and the runtime increased roughly in proportion to the factor of additional steps taken during the random walk at each iteration of the sampler. Because of this increased computational cost with what seemed like minimal gain in the performance of the sampler, we reset \texttt{maxmcmc} to 5000 for the majority of the individual-event analyses in this simulated population and for all subsequent populations analyzed.

While previous works~\cite[e.g.,][]{Vitale:2025lms, Wolfe:2025yxu} have demonstrated a successful recovery of the spin magnitude distribution with a similar number of events, these used a waveform model without higher-order modes, with the relative binning likelihood approximation~\citep{Cornish:2010kf, Zackay:2018qdy, Krishna:2023bug}. %
A stealth bias at the individual-event sampling stage due to sampling difficulties would be ameliorated by using a phenomenologically simpler waveform without higher-order modes and a likelihood approximation that assumes the waveform varies smoothly between neighboring points in parameter space, which helps with sampler convergence.
A similar bias in the spin magnitude hyper-parameter inference was found in \citet{Miller:2024sui} for their \textsc{LowSpinAligned} population using IMRPhenomXPHM, although we do not see the same bias in the spin tilt distribution.
Despite this persistent bias, the true values of $\mu_{\chi}, \sigma_{\chi}$ are generally recovered within the $3\sigma$ credible interval for all analyses (except those employing reflective boundary conditions for the individual-event parameter estimation), and the biases do not affect the recovery of the spin angle distributions. As such, we leave further investigation into ameliorating this bias to future work.

\section{Results summary}
\label{ap:result_tables}
\begin{figure*}
  \centering
  \includegraphics[width=0.9\textwidth]{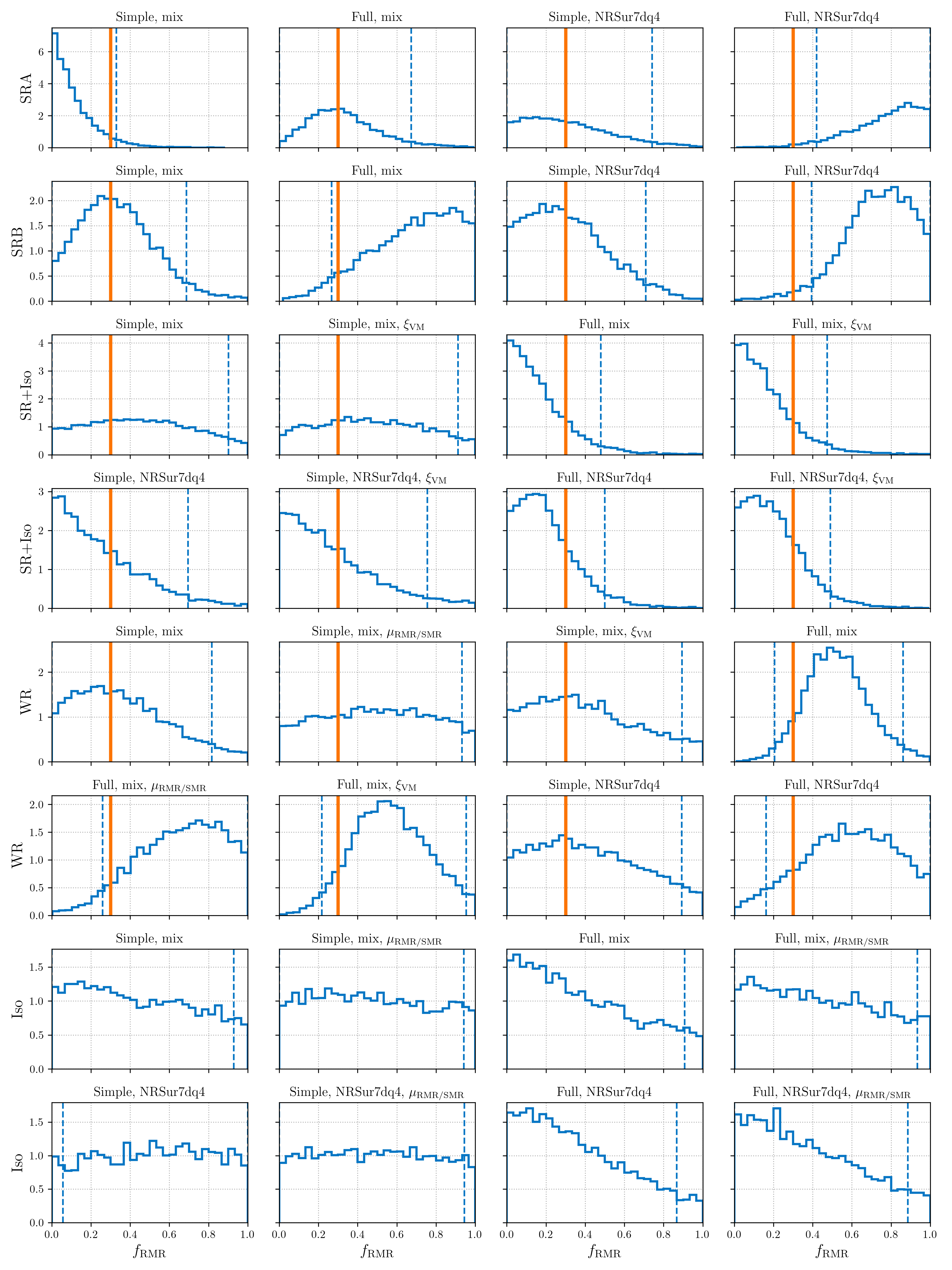}
  \caption{\red{Histograms of the posteriors on the proxy parameter for the fraction of field binaries that have undergone mass ratio reversal, $f_{\mathrm{RMR}}$, for the hierarchical inference runs summarized in Table~\ref{tab:results}. For the sake of space, the two runs on the NRSur7dq4 subset of the isotropic population with a separate $\xi_{\mathrm{VM}}$ parameter are omitted. The vertical orange lines show the true value, and the dashed blue lines bound the 95\% credible interval. The name of the analyzed population is given on the y-axis, and the choice of spin tilt model and waveform is given as the title of each panel. The default is that $\xi_{\mathrm{VM}}=\xi$ and $\mu_{\mathrm{RMR/SMR}}$ are fixed to their true values unless noted in each panel title.}}
  \label{fig:frmr_hist_grid}
\end{figure*}

\begin{figure*}
  \centering
  \includegraphics[width=0.9\textwidth]{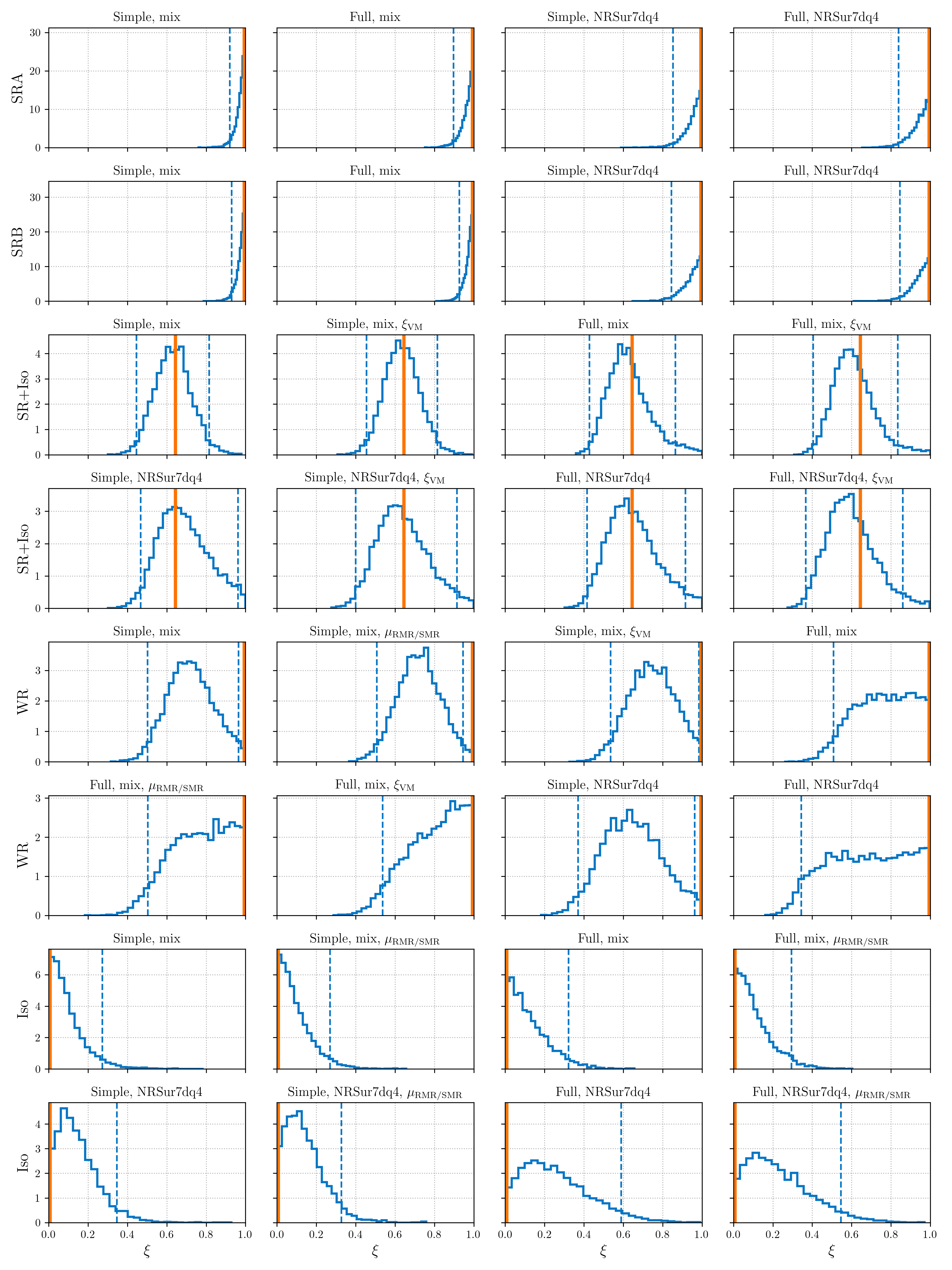}
  \caption{\red{Histograms of the posteriors on the fraction of binaries formed in the field drawn from non-isotropic spin angle distributions, $\xi$, for the hierarchical inference runs summarized in Table~\ref{tab:results}. For the sake of space, the two runs on the NRSur7dq4 subset of the isotropic population with a separate $\xi_{\mathrm{VM}}$ parameter are omitted. The vertical orange lines show the true value, and the dashed blue lines bound the 95\% credible interval. The name of the analyzed population is given on the y-axis, and the choice of spin tilt model and waveform is given as the title of each panel. The default is that $\xi_{\mathrm{VM}}=\xi$ and $\mu_{\mathrm{RMR/SMR}}$ are fixed to their true values unless noted in each panel title.}}
  \label{fig:xi_hist_grid}
\end{figure*}

In Tables~\ref{tab:gwtc3_results}-\ref{tab:results}, we report the maximum posterior values and 95\% credible intervals recovered for the $f_{\mathrm{RMR}}, \kappa, \xi$ hyper-parameters governing the $\phi_{12}$ distributions for all model variations explored in this work, including both GWTC-3 and the four simulated populations. \red{We also include plots of the 1D posteriors on $f_{\mathrm{RMR}}$ and $\xi$ for the hierarchical inference runs summarized in Table~\ref{tab:results} in Figs.~\ref{fig:frmr_hist_grid}-\ref{fig:xi_hist_grid}, respectively.}

\begin{deluxetable*}{lcccccccccr}
\tabletypesize{\scriptsize}
\tablecaption{Results summary table for all analyses performed in this work on simulated populations\label{tab:results}}
\tablehead{
\colhead{Population} &
\colhead{Tilt Model} &
\colhead{Waveform} &
\colhead{$\xi_{\mathrm{VM}}$} &
\colhead{$\mu_{\mathrm{RMR/SMR}}$} &
\colhead{$f_{\mathrm{RMR}}$} &
\colhead{$\mathrm{CI}_{95, f_{\mathrm{RMR}}}$} &
\colhead{$\kappa$} &
\colhead{$\mathrm{CI}_{95, \kappa}$} &
\colhead{$\xi$} &
\colhead{$\mathrm{CI}_{95, \xi}$}
}
\startdata
      Strong resonances A & \textsc{simple} & mix & $-$ & $-$ & $0.00^{+0.33}_{-0.00}$ & $0.33$ & $4.72^{+3.28}_{-3.40}$ & $6.68$ & $1.00^{+0.00}_{-0.08}$ & $0.08$\\
Strong resonances A & \textsc{full} & mix & $-$ & $-$ & $0.29^{+0.39}_{-0.29}$ & $0.67$ & $1.94^{+5.57}_{-1.94}$ & $7.51$ & $1.00^{+0.00}_{-0.10}$ & $0.10$\\
Strong resonances A & \textsc{simple} & NRSur7dq4 & $-$ & $-$ & $0.16^{+0.58}_{-0.16}$ & $0.74$ & $1.07^{+5.81}_{-1.07}$ & $6.88$ & $1.00^{+0.00}_{-0.15}$ & $0.15$\\
Strong resonances A & \textsc{full} & NRSur7dq4 & $-$ & $-$ & $\mathbf{0.89}^{+0.11}_{-0.47}$ & $0.58$ & $0.70^{+6.56}_{-0.70}$ & $7.26$ & $1.00^{+0.00}_{-0.16}$ & $0.16$\\
Strong resonances B & \textsc{simple} & mix & $-$ & $-$ & $0.27^{+0.41}_{-0.27}$ & $0.69$ & $1.56^{+5.87}_{-1.56}$ & $7.43$ & $1.00^{+0.00}_{-0.07}$ & $0.07$\\
Strong resonances B & \textsc{full} & mix & $-$ & $-$ & $0.88^{+0.12}_{-0.61}$ & $0.73$ & $0.00^{+4.12}_{-0.00}$ & $4.12$ & $1.00^{+0.00}_{-0.07}$ & $0.07$\\
Strong resonances B & \textsc{simple} & NRSur7dq4 & $-$ & $-$ & $0.19^{+0.52}_{-0.19}$ & $0.71$ & $1.89^{+5.38}_{-1.89}$ & $7.27$ & $1.00^{+0.00}_{-0.16}$ & $0.16$\\
Strong resonances B & \textsc{full} & NRSur7dq4 & $-$ & $-$ & $\mathbf{0.79}^{+0.21}_{-0.40}$ & $0.61$ & $\mathbf{0.00}^{+3.62}_{-0.00}$ & $3.62$ & $1.00^{+0.00}_{-0.15}$ & $0.15$\\
Strong resonances + iso & \textsc{simple} & mix & \checkmark & $-$ & $0.35^{+0.56}_{-0.35}$ & $0.91$ & $4.16^{+3.84}_{-3.53}$ & $7.38$ & $0.62^{+0.19}_{-0.17}$ & $0.36$\\
Strong resonances + iso & \textsc{simple} & mix & $-$ & $-$ & $0.43^{+0.47}_{-0.43}$ & $0.90$ & $5.26^{+2.74}_{-4.55}$ & $7.29$ & $0.64^{+0.18}_{-0.19}$ & $0.37$\\
Strong resonances + iso & \textsc{full} & mix & \checkmark & $-$ & $0.02^{+0.45}_{-0.02}$ & $0.47$ & $3.53^{+4.47}_{-2.10}$ & $6.57$ & $0.59^{+0.25}_{-0.18}$ & $0.43$\\
Strong resonances + iso & \textsc{full} & mix & $-$ & $-$ & $0.00^{+0.48}_{-0.00}$ & $0.48$ & $4.64^{+3.36}_{-3.12}$ & $6.48$ & $0.60^{+0.27}_{-0.17}$ & $0.44$\\
Strong resonances + iso & \textsc{simple} & NRSur7dq4 & \checkmark & $-$ & $0.00^{+0.76}_{-0.00}$ & $0.76$ & $7.48^{+0.52}_{-5.93}$ & $6.45$ & $0.60^{+0.31}_{-0.20}$ & $0.51$\\
Strong resonances + iso & \textsc{simple} & NRSur7dq4 & $-$ & $-$ & $0.00^{+0.70}_{-0.00}$ & $0.70$ & $7.44^{+0.56}_{-5.85}$ & $6.41$ & $0.65^{+0.32}_{-0.18}$ & $0.49$\\
Strong resonances + iso & \textsc{full} & NRSur7dq4 & \checkmark & $-$ & $0.09^{+0.40}_{-0.09}$ & $0.49$ & $6.71^{+1.29}_{-4.94}$ & $6.23$ & $0.57^{+0.29}_{-0.21}$ & $0.49$\\
Strong resonances + iso & \textsc{full} & NRSur7dq4 & $-$ & $-$ & $0.14^{+0.37}_{-0.14}$ & $0.50$ & $7.00^{+1.00}_{-5.20}$ & $6.20$ & $0.61^{+0.31}_{-0.19}$ & $0.50$\\
Weak resonances & \textsc{simple} & mix & \checkmark & $-$ & $0.30^{+0.60}_{-0.30}$ & $0.89$ & $0.90^{+6.56}_{-0.90}$ & $7.46$ & $\mathbf{0.73}^{+0.25}_{-0.19}$ & $0.45$\\
Weak resonances & \textsc{simple} & mix & $-$ & \checkmark & $0.44^{+0.49}_{-0.44}$ & $0.93$ & $0.00^{+7.49}_{-0.00}$ & $7.49$ & $\mathbf{0.74}^{+0.21}_{-0.23}$ & $0.44$\\
Weak resonances & \textsc{simple} & mix & $-$ & $-$ & $0.23^{+0.58}_{-0.23}$ & $0.82$ & $0.60^{+6.69}_{-0.60}$ & $7.29$ & $\mathbf{0.71}^{+0.26}_{-0.21}$ & $0.46$\\
Weak resonances & \textsc{full} & mix & \checkmark & $-$ & $0.55^{+0.40}_{-0.34}$ & $0.74$ & $0.00^{+7.47}_{-0.00}$ & $7.47$ & $0.98^{+0.02}_{-0.44}$ & $0.46$\\
Weak resonances & \textsc{full} & mix & $-$ & \checkmark & $0.76^{+0.24}_{-0.50}$ & $0.74$ & $0.35^{+6.90}_{-0.35}$ & $7.26$ & $0.93^{+0.07}_{-0.43}$ & $0.50$\\
Weak resonances & \textsc{full} & mix & $-$ & $-$ & $0.50^{+0.36}_{-0.30}$ & $0.65$ & $0.00^{+7.39}_{-0.00}$ & $7.39$ & $0.90^{+0.10}_{-0.39}$ & $0.49$\\
Weak resonances & \textsc{simple} & NRSur7dq4 & \checkmark & $-$ & $0.26^{+0.65}_{-0.26}$ & $0.91$ & $0.72^{+6.65}_{-0.72}$ & $7.38$ & $0.67^{+0.33}_{-0.23}$ & $0.56$\\
Weak resonances & \textsc{simple} & NRSur7dq4 & $-$ & $-$ & $0.28^{+0.61}_{-0.28}$ & $0.89$ & $0.00^{+7.27}_{-0.00}$ & $7.27$ & $\mathbf{0.63}^{+0.33}_{-0.26}$ & $0.59$\\
Weak resonances & \textsc{full} & NRSur7dq4 & \checkmark & $-$ & $0.81^{+0.19}_{-0.63}$ & $0.82$ & $0.00^{+7.38}_{-0.00}$ & $7.38$ & $1.00^{+0.00}_{-0.63}$ & $0.63$\\
Weak resonances & \textsc{full} & NRSur7dq4 & $-$ & $-$ & $0.65^{+0.35}_{-0.49}$ & $0.84$ & $0.00^{+7.23}_{-0.00}$ & $7.23$ & $1.00^{+0.00}_{-0.66}$ & $0.66$\\
Isotropic & \textsc{simple} & mix & $-$ & \checkmark & $0.25^{+0.69}_{-0.25}$ & $0.94$ & $5.79^{+2.21}_{-5.26}$ & $7.47$ & $0.00^{+0.27}_{-0.00}$ & $0.27$\\
Isotropic & \textsc{simple} & mix & $-$ & $-$ & $0.14^{+0.79}_{-0.14}$ & $0.93$ & $3.92^{+4.08}_{-3.42}$ & $7.50$ & $0.01^{+0.26}_{-0.01}$ & $0.27$\\
Isotropic & \textsc{full} & mix & $-$ & \checkmark & $0.08^{+0.86}_{-0.08}$ & $0.93$ & $5.06^{+2.94}_{-4.62}$ & $7.56$ & $0.00^{+0.29}_{-0.00}$ & $0.29$\\
Isotropic & \textsc{full} & mix & $-$ & $-$ & $0.00^{+0.91}_{-0.00}$ & $0.91$ & $4.32^{+3.68}_{-3.86}$ & $7.54$ & $0.02^{+0.31}_{-0.02}$ & $0.32$\\
Isotropic & \textsc{simple} & NRSur7dq4 & $-$ & \checkmark & $0.59^{+0.35}_{-0.59}$ & $0.94$ & $1.72^{+5.80}_{-1.72}$ & $7.52$ & $0.09^{+0.23}_{-0.09}$ & $0.33$\\
Isotropic & \textsc{simple} & NRSur7dq4 & $-$ & $-$ & $0.67^{+0.33}_{-0.62}$ & $0.94$ & $5.53^{+2.47}_{-5.00}$ & $7.47$ & $0.09^{+0.26}_{-0.09}$ & $0.35$\\
Isotropic & \textsc{full} & NRSur7dq4 & $-$ & \checkmark & $0.00^{+0.89}_{-0.00}$ & $0.89$ & $6.88^{+0.90}_{-6.69}$ & $7.58$ & $0.12^{+0.42}_{-0.12}$ & $0.55$\\
Isotropic & \textsc{full} & NRSur7dq4 & $-$ & $-$ & $0.06^{+0.80}_{-0.06}$ & $0.87$ & $3.67^{+4.33}_{-3.17}$ & $7.50$ & $0.14^{+0.45}_{-0.14}$ & $0.59$\\
\enddata
\tablenotetext{}{The table specifies the name of the analyzed population, the tilt and waveform models used, whether a separate mixture fraction between aligned and isotropic systems was introduced for the $\phi_{12}$ distribution ($\xi_{\mathrm{VM}}$), and whether the peaks of the $\phi_{12}$ \ac{VM} distributions were treated as free parameters ($\mu_{\mathrm{RMR/SMR}}$). For the three hyper-parameters governing the $\phi_{12}$ distribution common to all model variations ($f_{\mathrm{RMR}}, \kappa, \xi$), we report the maximum posterior value and the bounds and width of the 95\% credible interval calculated using the highest posterior density method. \textbf{Bold} values indicate cases where the true value (see Table~\ref{tab:sims}) lies outside the 95\% credible interval.}
\end{deluxetable*}

\end{document}